\documentclass[graybox,envcountchap,sectrefs]{svmono}

\usepackage{latexsym}
\usepackage{helvet}
\usepackage{courier}
\usepackage{type1cm}         

\usepackage{graphicx}        % standard LaTeX graphics tool
\usepackage[bottom]{footmisc}% places footnotes at page bottom

\usepackage{amsmath}
\newcommand{\nn}{\nonumber}
\newcommand{\be}{\begin{equation}}
\newcommand{\ee}{\end{equation}}
\newcommand{\bea}{\begin{eqnarray}}
\newcommand{\eea}{\end{eqnarray}}
\newcommand{\dotprod} {\raisebox{-.7mm}{\hspace{.4mm}\LARGE$\cdot$\hspace{.4mm}}}  %% large dot (vector dot product)
\newcommand{\crossprod} {\mbox{\boldmath{$\times$}} }
\newcommand{\vecnab}{\mbox{\boldmath{$\nabla$}}}
\newcommand{\bnabla}{\mbox{\boldmath{$\nabla$}}}
\newcommand{\bomega}{\mbox{\boldmath{$\omega$}}}
\newcommand{\bepsilon}{\mbox{\boldmath{$\epsilon$}}}
\newcommand{\deer}{(d{\bf r})\,}

\DeclareMathOperator{\tr}{tr}
\begin{document}

\author{Kimball A. Milton}
\title{Schwinger's Quantum Action Principle}
\subtitle{From Dirac's formulation through Feynman's path integrals, 
the Schwinger-Keldysh method, quantum field theory, to source theory}
\maketitle

\frontmatter%%%%%%%%%%%%%%%%%%%%%%%%%%%%%%%%%%%%%%%%%%%%%%%%%%%%%%

\abstract{Starting from the earlier notions of stationary action principles,
we show how Julian Schwinger's Quantum Action Principle descended from Dirac's 
formulation, which independently led Feynman to his path-integral formulation
of quantum mechanics.  The connection between the two is brought out, and
applications are discussed.  The Keldysh-Schwinger time-cycle method of 
extracting matrix elements in nonequilibrium situations is described. The 
variational formulation of quanum field theory and the development of
source theory constitute the latter part of this work.  In this document,
derived from Schwinger's lectures over four decades, the continuity of 
concepts, such as that of Green's functions, becomes apparent.}

\tableofcontents

\mainmatter%%%%%%%%%%%%%%%%%%%%%%%%%%%%%%%%%%%%%%%%%%%%%%%%%%%%%%%
\chapter{Historical Introduction}
\label{intro}
Variational principles for dynamical systems have a long history.  Although
precursors go back at least to Leibnitz (see for example \cite{wikisource})
and Euler \cite{euler}
the ``principle of least action'' was given modern form by de Maupertuis 
\cite{maupertuis1744,maupertuis1746}.  We will not attempt to trace the history
here; a brief useful account is given in Sommerfeld's lectures
 \cite{sommerfeld}.  The most important names in the history of the
development of dynamical systems, or at least those that will bear
most directly on the following discussion,   are those of Joseph-Louis Lagrange
\cite{lagrange} and William Rowan Hamilton \cite{Hamilton1834,Hamilton1835}.

Here we are concentrating on the work of Julian Schwinger (1918--1994), who
had profound and pervasive influence on 20th century physics, and whose
many students have become leaders in diverse fields.\footnote{For complex
reasons, Schwinger's influence on modern physics is not widely appreciated.
His contributions to our current understanding of nature are underrepresented
in textbooks, with some notable exceptions \cite{toms}.}
  For biographical
information about his life and work see \cite{mehra,milton}. Therefore,
we will take up the story in the modern era.
Shortly after Dirac's work with Fock and Podolsky \cite{dfp}, in which the
demonstration of the equivalence between his theory of quantum electrodynamics,
 and that of Heisenberg and Pauli \cite{heisenberg}, 
P. A. M. Dirac wrote a paper on
``The Lagrangian in Quantum Mechanics'' \cite{dirac}. This paper had a profound
influence on Richard Feynman's doctoral dissertation at Princeton on 
``The Principles of Least Action in Quantum Mechanics'' 
\cite{feynman}, and on his
later work on the formulations of the 
``Space-Time Approach to Quantum Electrodynamics'' 
\cite{feynmanst}.  Dirac's paper 
further formed the basis for Schwinger's development of the quantum
action principle, which first appeared in his final operator field
formulation of quantum field theory \cite{Schwinger1951}, 
which we will describe in Chapter 6.

The response of Feynman and Schwinger to Dirac's inspiring paper was completely
different.  Feynman was to give a global ``solution'' to the problem of
determining the transformation function, the probability amplitude connecting
the state of the system at one time to that at a later time, in terms of a sum
over classical trajectories, the famous path integral.  Schwinger, instead,
derived (initially postulated) a differential equation for that transformation
function in terms of a quantum action functional.  This differential equation
possessed Feynman's path integral as a formal solution, which remained
poorly defined; but Schwinger believed throughout his life that his approach
was ``more general, more elegant, more useful, and more tied to the historical
line of development as the quantum transcription of Hamilton's action
principle'' \cite{schwinger1973}.

Later, in a tribute to Feynman, Schwinger commented further.  Dirac, of course,
was the father of transformation theory \cite{Dirac1927}.  
The transformation function from a description of the system
at time $t_2$ to a description at time $t_1$ is ``the product
 of all the transformations functions associated with the successive
infinitesimal increments in time.''  Dirac said the latter, that is, the
transformation function from time $t$ to time $t+dt$ \textit{corresponds\/}
to $\exp[(i/ \hbar)dt\,L]$, where $L$ is the Lagrangian expressed in terms
of the coordinates at the two times.  For the transformation function between
$t_2$ and $t_1$ ``the integrand is $\exp[(i/\hbar)W]$. where 
$W=\int_{t_2}^{t_1}d t\,L$.''  
``Now we know, and Dirac surely knew, that to within
a constant factor the `correspondence,' for infinitesimal $d t$, 
is an equality
when we deal with a system of nonrelativistic particles possessing a
coordinate-dependent potential energy $V$ \dots.  Why then, did Dirac not make
a more precise, if less general  statement?  Because he was interested in a
general question: What, in quantum mechanics, corresponds to the classical
principle of stationary action?''

``Why, in the decade that followed, didn't someone pick up the computational
possibilities offered by this integral approach to the time transformation
function?  To answer this question bluntly, perhaps no one needed it---until
Feynman came along.'' \cite{Schwinger1989}.

But Schwinger followed the differential route, and starting in early 1950
began a new, his third, formulation of quantum electrodynamics, based on a
variational approach.  This was first published in 1951 \cite{Schwinger1951}. 
A bit later he started developing a new formulation of quantum kinematics,
which he called Measurement Algebra, which got its first public presentation
at \'Ecole de Physique des Houches in the summer of 1955.
There were several short notes in the Proceedings of the US
National Academy published in 1960, explaining both the quantum kinematical
approach and the dynamical action principle \cite{pnas1, pnas2, pnas3, pnas4}, 
but although  he often promised to write a book on the subject 
(as he also promised a book on quantum field theory)
nothing came of it.  Les Houches lectures, based on notes taken by Robert
Kohler, eventually appeared in 1970 \cite{leshouches}. 
Lectures based on a UCLA course on quantum mechanics
by Schwinger were eventually published
under Englert's editorship \cite{Schwinger2001}.  The incompleteness
of the written record may be partly alleviated by the present essay. 

We start on a classical footing.
\chapter{Review of Classical Action Principles}
\label{sec:1}
This section grew out of lectures
given by Schwinger at UCLA around 1974, which were
substantially transformed into Chapter 8 of {\it Classical Electrodynamics} 
\cite{Schwinger1998}.  (Remarkably, considering
his work on waveguide theory during World War II, now partially recorded
in Ref.~\cite{er}, he never gave
lectures on this subject at Harvard after 1947.)

\begin{figure}
\centering
\includegraphics{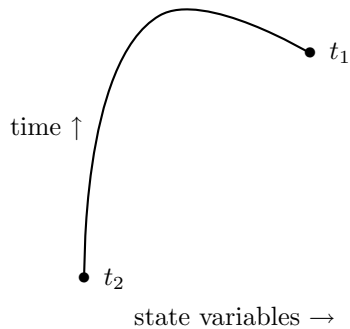}
%\begin{picture}(200,100)
%\put(10,10){\makebox(0,0){$\bullet$}}
%\put(100,100){\makebox(0,0){$\bullet$}}
%\qbezier(10,10)(10,100)(100,100)
%\put(10,10){\line(1,2){30}}
%\put(50,70){\oval(20,20)[tl]}
%\put(50,80){\line(5,2){50}}
%\put(15,10){$t_2$}
%\put(105,100){$t_1$}
%\put(80,0){state variables $\longrightarrow$}
%\put(-50,50){time $\uparrow$}
%\end{picture}
\caption{\label{fig8.1} A possible path from initial state to final state.}
%\label{fig8.1}
\end{figure}

      We start by reviewing and generalizing the Lagrange-Hamilton principle
 for a single particle. The action, $W_{12}$, is defined as the time
 integral of the Lagrangian, $L$, where the integration extends from an initial
 configuration or state at time $t_2$ to a final state at time $t_1$:
\begin{equation}
W_{12}=\int_{t_2}^{t_1} dt\, L. 
\label{8.1}
\end{equation}
The integral refers to any path, any line of time development, from the
initial to the final state, as shown in Fig.\ \ref{fig8.1}.
The actual time evolution of the system is selected by the principle
of stationary action: In response to infinitesimal variations of the
integration path, the action $W_{12}$ is stationary---does not have
a corresponding infinitesimal change---for variations about the correct
path, provided the initial and final configurations are held fixed,
\begin{equation}  
\delta W_{12} =0.  
\label{8.3}
\end{equation}
This means that, if we allow infinitesimal changes at the initial and
final times, including alterations of those times, the only contribution
to $\delta W_{12}$ then comes from the endpoint variations, or
 \begin{equation}
 \delta W_{12} = G_1 - G_2,
\label{8.2}
\end{equation}
where $G_a$, $a=1$ or $2$,  is a function, called the generator,
 depending on dynamical variables only at time $t_a$.
 In the following, we will consider three different realizations of
 the action principle, where, for simplicity, we will restrict our 
 attention to a single particle.

\section{Lagrangian Viewpoint}

      The nonrelativistic motion of a particle of mass $m$ moving in a 
potential $V({\bf r},t)$ is described by the Lagrangian
\begin{equation}
 L={1\over2}m\left({d{\bf r}\over dt}\right)^2-V({\bf r},t).
\label{8.4}
\end{equation}
 Here, the independent variables are $\bf r$ and $t$, so that two kinds
of variations can be considered.  First, a particular motion is
altered infinitesimally, that is,  the path is changed by
an amount $\delta{\bf r}$:
 \begin{equation}
{\bf r}(t) \to {\bf r}(t) + \delta {\bf r}(t).
\label{8.6}
\end{equation}
 Second, the final and initial times can be altered infinitesimally, by
$\delta t_1$ and $\delta t_2$, respectively.
It is more convenient, however, to think of these time displacements
as produced by a continuous variation of the time parameter, $\delta t(t)$,
\begin{equation}
 t \to t + \delta t(t),
\label{8.5}
\end{equation}
 so chosen that, at the endpoints,
\begin{equation}
 \delta t(t_1) = \delta t_1,\qquad    \delta t(t_2) = \delta t_2.
\label{1-18.5}
\end{equation}
 The corresponding change in the time differential is
\begin{equation}
 dt \to d (t + \delta t) = \left(   1 + {d\delta t\over dt }\right)dt,
\label{8.7}
\end{equation}
 which implies the transformation of the time derivative,
\begin{equation}
{ d\over dt}\to \left(1-{ d\delta t \over   dt}\right){d\over dt}.
\label{8.8}
\end{equation}
 Because of this redefinition of the time variable, the limits of integration
 in the action,
 \begin{equation}
 W_{12}=\int_2^1\left[{1\over2}m{(d{\bf r})^2\over dt}-dt\,V\right],
 \label{1-18.8}
 \end{equation}
   are {\it not\/} changed, the time displacement being produced through
   $\delta t(t)$ subject to (\ref{1-18.5}).
 The resulting variation in the action is now
\begin{eqnarray}
\delta W_{12}&=& \int_2^1dt\left\{m{ d{\bf r}\over dt}\dotprod {d \over dt}
\delta{\bf r}-\delta{\bf r}\dotprod\vecnab V   -{d\delta t\over dt}
\left[{1\over2}m\left({d{\bf r}\over dt}\right)^2+V\right]
-\delta t{\partial\over\partial t}V\right\}
\nonumber\\
&=&\int_2^1 dt\Bigg\{{d\over dt}\left[m{d{\bf r}\over dt}\dotprod \delta{\bf r}
-\left({1\over 2}m\left({d{\bf r}\over dt}\right)^2+V\right)\delta t\right]
\nonumber\\
&&\!\!\!\!\!\mbox{}+\delta{\bf r}\dotprod\left[-m{d^2\over dt^2}{\bf r}-\vecnab V
\right]+\delta t\left({d\over dt}
\left[{1\over2}m\left({d{\bf r}\over dt}\right)^2+V
\right]-{\partial\over\partial t}V\right)\!\!\!\Bigg\},\nn\\
\label{8.9}
\end{eqnarray}
 where, in the last form, we have integrated by parts in order to
 isolate $\delta{\bf r}$ and $\delta t$.
 
    Because $\delta{\bf r}$ and $\delta t$ are independent variations, 
the  principle of stationary action implies that the actual motion 
is governed by
\begin{subequations}
\begin{eqnarray}
m{ d^2\over dt^2}{\bf r}=&-&\vecnab V,
\label{8.10}\\
{d\over dt}\bigg[{1\over2} m \left({ d{\bf r}\over dt}\right)^2    &+& V
\bigg]={\partial\over\partial t}V,
\label{8.11}
\end{eqnarray}
 while the total time derivative gives the change at the endpoints,
\begin{equation}
      G = {\bf p\dotprod \delta r} - E\delta t,
\label{8.12}
\end{equation}
 with
\begin{equation}
\mbox{momentum}={\bf p}=m{d{\bf r}\over dt},\qquad
\mbox{energy}= E={1\over 2}m\left({ d{\bf r}\over dt}\right)^2+V.
\end{equation}
\end{subequations}
 Therefore, we have derived Newton's second law [the equation of motion in
 second-order form], (\ref{8.10}), 
and, for a static potential, $\partial V/\partial t=0$,
 the conservation of energy, (\ref{8.11}). The
 significance of (\ref{8.12}) will be discussed later in Section 
\ref{8.4}.

 \section{Hamiltonian Viewpoint}
      Using the above definition of the momentum, we can rewrite the Lagrangian
as
\begin{equation}
 L={\bf p}\dotprod{d{\bf r}\over   dt}- H({\bf r}, {\bf p},t),
\label{8.13}
\end{equation}
 where we have introduced the Hamiltonian
\begin{equation}
 H={ p^2\over 2m} + V ({\bf r},t).
\label{8.14}
\end{equation}
 We are here to regard $\bf r$, $\bf p$, and $t$ as independent variables in
 \begin{equation}
 W_{12}=\int_2^1[{\bf p}\dotprod d{\bf r}-dt\, H].
 \label{1-18.18}
 \end{equation}
 The change in the action, when $\bf r$, $\bf p$, and $t$ are all varied, is
\begin{eqnarray}
 \delta W_{12}
&=&\int_2^1 dt\left[{\bf  p}\dotprod{d\over dt}\delta{\bf r}-
\delta{\bf r}\dotprod{\partial H\over\partial{\bf r}}+\delta{\bf p}
\dotprod{d{\bf r}\over dt}-\delta{\bf p}\dotprod{\partial H\over\partial{\bf p}}
-{d\delta t\over dt}H-\delta t{\partial H\over\partial t}\right]\nonumber\\
&=&\int_2^1dt\bigg[{d\over dt}({\bf p\dotprod\delta r}-H\delta t)
+\delta{\bf r}\dotprod\left(-{d{\bf p}\over dt}-{\partial H\over\partial{\bf r}}
\right)\nonumber\\
&&\quad+\delta{\bf p}\dotprod\left({d{\bf r}\over dt}-{\partial H\over 
\partial {\bf p}}\right)+\delta t\left({dH\over dt}-{\partial H
\over\partial t}\right)\bigg].
\label{8.15}
\end{eqnarray}
 The action principle then implies
\begin{subequations}
\begin{eqnarray}
{d{\bf r}\over dt}&=&{\partial H\over\partial{\bf p}}={{\bf p}\over m},
\label{8.16}\\
{d{\bf p}\over dt}&=&-{\partial H\over \partial {\bf r}}=-\vecnab V,
\label{8.17}\\
{d H\over dt}&=&{\partial H\over\partial t},\label{8.18}\\
G&=&{\bf p\dotprod\delta r}-H\delta t.
\label{8.19}
\end{eqnarray}
\end{subequations}
In contrast with the Lagrangian differential equations of motion, which
involve second derivatives, these Hamiltonian equations contain only
first derivatives; they are called first-order equations.
They describe the same physical system, because when 
 (\ref{8.16}) is substituted into (\ref{8.17}), we recover 
 the Lagrangian-Newtonian equation (\ref{8.10}).
 Furthermore, if we insert (\ref{8.16}) into the Hamiltonian
 (\ref{8.14}), we identify $H$ with $E$.  The third equation
 (\ref{8.18}) is then identical with (\ref{8.11}).  We also note
 the equivalence of the two versions of $G$.
 
 But probably the most direct way of seeing that the same physical
 system is involved comes by writing the Lagrangian in the Hamiltonian
 viewpoint as
 \begin{equation}
 L={m\over2}\left({d{\bf r}\over dt}\right)^2-V-{1\over2m}
 \left({\bf p}-m{d{\bf r}\over dt}\right)^2.
 \label{1-18.26}
 \end{equation}
 The result of varying $\bf p$ in the stationary action principle
 is to produce
 \begin{equation}
 {\bf p}=m{d{\bf r}\over dt}.
 \label{1-18.27}
 \end{equation}
 But, if we accept this as the {\em definition\/} of $\bf p$, the
 corresponding term in $L$ disappears and we explicitly regain the 
 Lagrangian description.  We are justified in completely omitting the
 last term on the right side of (\ref{1-18.26}), despite its
 dependence on the variables $\bf r$ and $t$, because of its quadratic
 structure.  Its explicit contribution to $\delta L$ is
 \begin{equation}
 -{1\over m}\left({\bf p}-m{d{\bf r}\over dt}\right)\dotprod
 \left(\delta{\bf p}-m{d\over dt}\delta {\bf r}+m{d{\bf r}\over dt}
 {d\delta t\over dt}\right),
 \label{1-18.28}
 \end{equation}
 and the equation supplied by the stationary action principle for
 $\bf p$ variations, (\ref{1-18.27}), also guarantees that there is no
 contribution here to the results of $\bf r$ and $t$ variations.
 
 \section{A Third, Schwingerian, Viewpoint}
\label{sec8.3}

      Here we take $\bf r$, $\bf p$, and the velocity,
$\bf v$, as independent variables, so that
 the Lagrangian is written in the form
\begin{equation}
    L =  {\bf p}\dotprod\left({d{\bf r}\over dt}-{\bf v}\right)
           + {1\over2}  mv^2 - V({\bf r},t)\equiv
{\bf p}\dotprod{d{\bf r}\over dt}-H({\bf r,p,v},t),
\label{8.20}
\end{equation}
where 
\begin{equation}
H({\bf r,p,v},t)={\bf p\dotprod v}-{1\over2}mv^2+V({\bf r},t).
\label{8.26}
\end{equation}
 The variation of the action is now
\begin{eqnarray}
\delta W_{12}&=&\delta\int_2^1[{\bf p}\dotprod d{\bf r}-H\,dt]\nonumber\\
&=&\int_2^1dt\bigg[\delta{\bf p}\dotprod{d{\bf r}\over dt}
+{\bf p}\dotprod{d\over dt}\delta{\bf r}
-\delta{\bf r}\dotprod{\partial H\over\partial{\bf r}}
-\delta{\bf p}\dotprod{\partial H\over\partial{\bf p}}
-\delta{\bf v}\dotprod{\partial H\over\partial {\bf v}}\nonumber\\
&&\qquad\mbox{}-\delta t{\partial H\over\partial t}-H{d\delta t\over dt}\bigg]
\nonumber\\
&=&\int_2^1dt\bigg[{d\over dt}({\bf p\dotprod\delta r}- H\delta t)
-\delta{\bf r}\dotprod\left({d{\bf p}\over dt}+{\partial H\over\partial{\bf r}}
\right)\nonumber\\
&&\mbox{}+\delta{\bf p}\dotprod\left({d{\bf r}\over dt}-{\partial H
\over\partial {\bf p}}\right)
-\delta{\bf v}\dotprod{\partial H\over \partial{\bf v}}
+\delta t\left({d H\over dt}
-{\partial H\over\partial t}\right)\bigg],
\label{8.21}
\end{eqnarray}
 so that the action principle implies
\begin{subequations}
\begin{eqnarray}
{d{\bf p}\over dt}&=&-{\partial H\over\partial{\bf r}}=-\vecnab V,
\label{8.22}\\
{d{\bf r}\over dt}&=&{\partial H\over\partial{\bf p}}={\bf v},
\label{8.23}\\
{\bf 0}&=&-{\partial H\over\partial {\bf v}}=-{\bf p}+m{\bf v},
\label{8.24}\\
{d H\over dt}&=&{\partial H\over\partial t},
\label{8.25}\\
G&=&{\bf p\dotprod\delta r}-H\delta t.
\label{8.27}
\end{eqnarray}
\end{subequations}
 Notice that there is no equation of motion for $\bf v$ since $d{\bf v}/dt$
 does not occur in the Lagrangian, nor is it multiplied by a time derivative.
 Consequently, (\ref{8.24}) refers to
 a single time and is an equation of constraint.

      From this third approach, we have the option of returning to either of
the other two viewpoints by imposing an appropriate restriction.  Thus, if
we write (\ref{8.26}) as
\begin{equation}
H({\bf r,p,v},t)={p^2\over 2m}+V({\bf r},t)-{1\over2m}({\bf p}-m{\bf v})^2,
\label{1-18.33}
\end{equation}
and we adopt
\begin{equation}
{\bf v}={1\over m}{\bf p}
\label{1-18.34}
\end{equation}
as the {\em definition\/} of $\bf v$, we recover the Hamiltonian
description, (\ref{8.13}) and (\ref{8.14}). 
 Alternatively, we can present the Lagrangian (\ref{8.20}) as
\begin{equation}
L={m\over2}\left(d{\bf r}\over dt\right)^2-V+({\bf p}-m{\bf v})
\dotprod\left({d{\bf r}\over dt}-{\bf v}\right)-{m\over2}
\left({d{\bf r}\over dt}-{\bf v}\right)^2.
\label{1-18.35}
\end{equation}
Then, if we adopt the following as {\em definitions},
\begin{equation}
{\bf v}  = {d{\bf r}\over dt},\quad {\bf p}=m{\bf v},
\label{18.36}
\end{equation}
the resultant form of $L$ is that of the Lagrangian viewpoint, (\ref{8.4}).
It might seem that only the definition ${\bf v}=d{\bf r}/dt$, inserted
in (\ref{1-18.35}), suffices to regain the Lagrangian description.
But then the next to last term in (\ref{1-18.35}) would give the
following additional contribution to $\delta L$, associated with
the variation $\delta {\bf r}$:
\begin{equation}
({\bf p}-m{\bf v})\dotprod{d\over dt}\delta {\bf r}.
\label{1-18.37}
\end{equation}

      In the next Chapter, where the action formulation of electrodynamics is
 considered, we will see the advantage of adopting this third approach, which
 is characterized by the introduction of additional variables,
 similar to $\bf v$, for which there
 are no equations of motion.

 \section{Invariance and Conservation Laws}
 \label{sec8.4}
 There is more content to the principle of stationary action than equations
 of motion.  Suppose one considers a variation such that
 \begin{equation}
 \delta W_{12}=0,
 \label{1-19.1}
 \end{equation}
 independently of the choice of initial and final times.  We say that the
 action, which is left unchanged, is {\em invariant\/} under this alteration
 of path.  Then the stationary action principle (\ref{8.2}) asserts that
 \begin{equation}
 \delta W_{12}=G_1-G_2=0,
 \label{1-19.2}
 \end{equation}
 or, there is a quantity $G(t)$ that has the same value for any choice of
 time $t$; it is conserved in time.  A differential statement of that
 is
 \begin{equation}
 {d\over dt}G(t)=0.
 \label{1-19.3}
 \end{equation}
 The $G$ functions, which are usually
 referred to as generators, express the interrelation between 
conservation laws and invariances of the system.  
 
 Invariance implies conservation, and vice versa. 
 A more precise statement is the following:
\begin{quote}
      If there is a conservation law, the action is stationary under an
      infinitesimal transformation in an appropriate variable.
\end{quote}
 The converse of this statement is also true.
\begin{quote}
 If the action $W$ is invariant under an infinitesimal transformation (that is,
$\delta W = 0$), then there is a corresponding conservation law.
\end{quote}
  This is the celebrated theorem of Amalie Emmy Noether %(1882--1935)
\cite{Noether1918}.
  
  Here are some examples.
 Suppose the Hamiltonian of (\ref{8.13}) does not depend explicitly
 on time, or
 \begin{equation}
 W_{12}=\int_2^1[{\bf p}\dotprod d{\bf r}-H({\bf r,p})dt].
 \label{1-19.4}
 \end{equation}
 Then the variation (which as a rigid displacement in time, amounts
 to a shift in the time origin)
 \begin{equation}
 \delta t=\mbox{constant}
 \label{1-19.5}
 \end{equation}
 will give $\delta W_{12}=0$ [see the first line of (\ref{8.15}),
 with $\delta{\bf r}=0$, $\delta{\bf p}=0$, $d\delta t/dt=0$, 
$ \partial H/\partial t=0$].  The conclusion is that $G$ in (\ref{8.19}),
which here is just
\begin{equation}
G_t=-H\delta t,
\label{1-19.6}
\end{equation}
is a conserved quantity, or that
\begin{equation}
{dH\over dt}=0.
\label{1-19.7}
\end{equation}
This inference, that the Hamiltonian---the energy---is conserved, if there is
no explicit time dependence in $H$, is already present in (\ref{8.18}).
But now a more general principle is at work.

\begin{figure}
\centering
\includegraphics{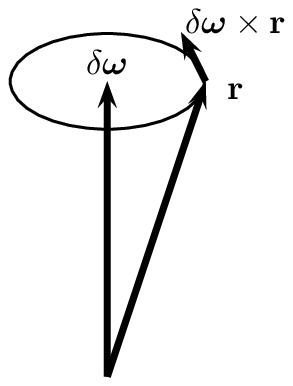}
%\begin{picture}(200,100)
%\put(100,0){\vector(0,1){90}}
%\put(100,100){\makebox(0,0){$\delta\mbox{\boldmath{$\omega$}}$}}
%\put(100,0){\vector(1,3){30}}
%\put(130,90){\vector(-1,1){10}}
%\put(130,100){$\delta\mbox{\boldmath{$\omega$}}\crossprod{\bf r}$}
%\put(130,80){$\bf r$}
%\qbezier[30](70,90)(100,70)(130,90)
%\qbezier[30](70,90)(100,110)(130,90)
%\end{picture}
\caption{$\delta\mbox{\boldmath{$\omega$}}\crossprod{\bf r}$ is perpendicular
to $\delta\mbox{\boldmath{$\omega$}}$ and $\bf r$, and represents an
infinitesimal rotation
of $\bf r$ about the $\delta\mbox{\boldmath{$\omega$}}$ axis.}
\label{fig8.2}
\end{figure}

Next, consider an infinitesimal, rigid rotation, one that maintains the
lengths and scalar products of all vectors.  Written explicitly for the
position vector $\bf r$, it is
\begin{equation}
 \delta{\bf r}  = \delta\vec{\omega}\crossprod{\bf r},
\end{equation}
where the constant vector $\delta\vec{\omega}$ gives 
the direction 
and magnitude of the rotation (see Fig.\ \ref{fig8.2}).
Now specialize (\ref{8.14}) to
\begin{equation}
H={p^2\over2m}+V(r),
\label{1-19.9}
\end{equation}
where $r=|{\bf r}|$, a rotationally invariant structure.  Then
\begin{equation}
W_{12}=\int_2^1[{\bf p}\dotprod d{\bf r}-H\, dt]
\label{1-19.10}
\end{equation}
is also invariant under the rigid rotation, implying the conservation of
\begin{equation}
G_{\delta\vec{\omega}}={\bf p\dotprod\delta r=
\delta\vec{\omega}\dotprod r\crossprod p.}
\label{1-19.11}
\end{equation}
This is the conservation of angular momentum,
\begin{equation}
{\bf L = r\crossprod p}, \quad {d\over dt}{\bf L=0}.
\label{1-19.12}
\end{equation}
 Of course, this is also contained within the equation of motion,
\begin{equation}
{ d \over dt}{\bf L}= -{\bf r} \crossprod\vecnab V=
-{\bf r\crossprod\hat r}{\partial V\over\partial r}={\bf 0},
\end{equation}
 since $V$ depends only on $|{\bf r}|$.

Conservation of linear momentum appears analogously when there is
invariance under a rigid translation.  For a single particle, (\ref{8.17})
tells us immediately that $\bf p$ is conserved if $V$ is a constant, say
zero.  Then, indeed, the action
\begin{equation}
W_{12}=\int_2^1\left[{\bf p}\dotprod d{\bf r}-{p^2\over2m}dt\right]
\label{1-19.13}
\end{equation}
is invariant under the displacement
\begin{equation}
\delta{\bf r}=\delta\mbox{\boldmath{$\epsilon$}}=\mbox{constant},
\label{1-19.14}
\end{equation}
and
\begin{equation}
G_{\delta\vec\epsilon} ={\bf p}\dotprod
\delta\mbox{\boldmath{$\epsilon$}}
\label{1-19.15}
\end{equation}
is conserved.  But the general principle acts just as easily for, say,
a system of two particles, $a$ and $b$, with Hamiltonian
\begin{equation}
H={p_a^2\over2m_a}+{p_b^2\over2m_b}+V({\bf r}_a-{\bf r}_b).
\label{1-19.16}
\end{equation}
This Hamiltonian and the associated action
\begin{equation}
W_{12}=\int_2^1[{\bf p}_a\dotprod d{\bf r}_a+{\bf p}_b\dotprod d{\bf r}_b
-H\,dt]
\label{1-19.17}
\end{equation}
are invariant under the rigid translation
\begin{equation}
\delta {\bf r}_a=\delta{\bf r}_b=\delta\mbox{\boldmath{$\epsilon$}},
\label{1-19.18}
\end{equation}
with the implication that
\begin{equation}
G_{\delta\mbox{\boldmath{$\epsilon$}}}={\bf p}_a\dotprod\delta{\bf r}_a
+{\bf p}_b\dotprod\delta{\bf r}_b=({\bf p}_a+{\bf p}_b)\dotprod
\delta\mbox{\boldmath{$\epsilon$}}
\label{1-19.19}
\end{equation}
is conserved.  This is the conservation of the total linear momentum,
\begin{equation}
{\bf P}={\bf p}_a+{\bf p}_b,\quad {d\over dt}{\bf P=0}.
\label{1-19.20}
\end{equation}

Something a bit more general appears when we consider a rigid translation
that grows linearly in time:
\begin{equation}
\delta{\bf r}_a=\delta{\bf r}_b=\delta{\bf v}\,t,
\label{1-19.21}
\end{equation}
using the example of two particles.  This gives each particle the common
additional velocity $\delta{\bf v}$, and therefore must also change
their momenta,
\begin{equation}
\delta{\bf p}_a=m_a\delta {\bf v},\quad\delta{\bf p}_b=m_b\delta {\bf v}.
\label{1-19.22}
\end{equation}
The response of the action (\ref{1-19.17}) to this variation is
\begin{eqnarray}
\delta W_{12}&=&\int_2^1[({\bf p}_a+{\bf p}_b)\dotprod\delta{\bf v}\,dt
+\delta{\bf v}\dotprod(m_a d{\bf r}_a+m_b d{\bf r}_b)-({\bf p}_a
+{\bf p}_b)\dotprod\delta{\bf v}\,dt]\nonumber\\
&=&\int_2^1d[(m_a{\bf r}_a+m_b{\bf r}_b)\dotprod\delta{\bf v}].
\label{1-19.23}
\end{eqnarray}
The action is {\em not\/} invariant; its variation has end-point contributions.
But there is still a conservation law, not of $G={\bf P\dotprod\delta v}t$,
but of ${\bf N}\dotprod\delta{\bf v}$, where
\begin{equation}
{\bf N=P}t-(m_a{\bf r}_a+m_b{\bf r}_b).
\label{1-19.24}
\end{equation}
Written in terms of the center-of-mass position vector
\begin{equation}
{\bf R}={m_a{\bf r}_a+m_b{\bf r}_b\over M},\quad M=m_a+m_b,
\label{1-19.25}
\end{equation}
the statement of conservation of
\begin{equation}
{\bf N=P}t-M{\bf R},
\label{1-19.26}
\end{equation}
namely
\begin{equation}
{\bf 0}={d{\bf N}\over dt}={\bf P}-M{d{\bf R}\over dt},
\label{1-19.27}
\end{equation}
is the familiar fact that the center of mass of an isolated system
moves at the constant velocity given by the ratio of the total momentum
to the total mass of that system.

\section{Nonconservation Laws. The Virial Theorem}

The action principle also supplies useful nonconservation laws.  Consider,
for constant $\delta\lambda$,
\begin{equation}
\delta{\bf r}=\delta\lambda{\bf r},\quad \delta {\bf p}=-\delta\lambda
{\bf p},
\label{1-19.28}
\end{equation}
which leaves ${\bf p}\dotprod d{\bf r}$ invariant,
\begin{equation}
\delta({\bf p}\dotprod d{\bf r})=(-\delta\lambda{\bf p})\dotprod d{\bf r}
+{\bf p}\dotprod(\delta\lambda d{\bf r})=0.
\label{1-19.29}
\end{equation}
But the response of the Hamiltonian
\begin{equation}
H=T(p)+V({\bf r}),\quad T(p)={p^2\over2m},
\label{1-19.30}
\end{equation}
is given by the noninvariant form
\begin{equation}
\delta H=\delta\lambda(-2T+{\bf r}\dotprod\vecnab V).
\label{1-19.31}
\end{equation}
Therefore we have, for an arbitrary time interval, for the
variation of the action (\ref{1-18.18}),
\begin{equation}
\delta W_{12}=\int_2^1 dt[ \delta\lambda(2T-{\bf r}\dotprod\vecnab V)]
=G_1-G_2=\int_2^1 dt{d\over dt}({\bf p\dotprod\delta\lambda r})
\label{1-19.32}
\end{equation}
or, the theorem
\begin{equation}
{d\over dt}{\bf r\dotprod p}=2T-{\bf r}\dotprod\vecnab V.
\label{1-19.33}
\end{equation}
%This is an example of the mechanical virial theorem to which we referred at 
%the end of Section 3.3.

For the particular situation of the Coulomb potential between charges,
$V=\mbox{constant}/r$, where
\begin{equation}
{\bf r}\dotprod\vecnab V=r{d\over dr}V=-V,
\label{1-19.34}
\end{equation}
the virial theorem asserts that
\begin{equation}
{d\over dt}({\bf r\dotprod p})=2T+V.
\label{1-19.35}
\end{equation}
We apply this to a {\em bound\/} system produced by a force of attraction.
On taking the time average of (\ref{1-19.35}) the time derivative term
disappears.  That is because, over an arbitrarily long time interval
$\tau=t_1-t_2$, the value of ${\bf r\dotprod p}(t_1)$ can differ by only
a finite amount from ${\bf r\dotprod p}(t_2)$, and
\begin{equation}
\overline{{d\over dt}({\bf r\dotprod p})}\equiv{1\over\tau}\int_{t_2}^{t_1}
dt{d\over dt}{\bf r\dotprod p}=
{{\bf r\dotprod p}(t_1)-{\bf r\dotprod  p}(t_2)\over\tau}\to0,
\label{1-19.36}
\end{equation}
as $\tau\to\infty$. The conclusion, for time averages,
\begin{equation}
2\overline T=-\overline V,
\label{1-19-37}
\end{equation}
is familiar in elementary discussions of motion in a $1/r$ potential.
%has been used qualitatively in Section 5.2.

Here is one more example of a nonconservation law:  Consider the
variations
\begin{subequations}
\begin{eqnarray}
\delta{\bf r}&=&\delta\lambda{{\bf r}\over r},\\
\delta{\bf p}&=&-\delta\lambda\left({{\bf p}\over r}-{{\bf r\, p\dotprod
r}\over r^3}\right)=\delta\lambda{{\bf r\crossprod(r\crossprod p)}
\over r^3}.
\label{1-19.38}
\end{eqnarray}
\end{subequations}
Again ${\bf p}\dotprod d{\bf r}$ is invariant:
\begin{equation}
\delta({\bf p}\dotprod d{\bf r})=-\delta\lambda\left({{\bf p}\over r}-
{{\bf r \,p\dotprod r}\over r^3}\right)\dotprod d{\bf r}
+{\bf p}\dotprod\left(\delta\lambda{d{\bf r}\over r}-\delta\lambda
{\bf r}{{\bf r}\dotprod d{\bf r}\over r^3}\right)=0,
\label{1-19.39}
\end{equation}
and the change of the Hamiltonian (\ref{1-19.30}) is now
\begin{equation}
\delta H=\delta\lambda\left[-{{\bf L}^2\over mr^3}+{{\bf r}\over r}\dotprod
\vecnab V\right].
\label{1-19.40}
\end{equation}
The resulting theorem, for $V=V(r)$, is
\begin{equation}
{d\over dt}\left({{\bf r}\over r}\dotprod {\bf p}\right)=
{{\bf L}^2\over mr^3}-{dV\over dr},
\label{1-19.41}
\end{equation}
which, when applied to the Coulomb potential, gives the bound-state time
average relation
\begin{equation}
{L^2\over m}\overline{\left({1\over r^3}\right)}=-\overline{\left(V\over r
\right)}.
\label{1-19.42}
\end{equation}
This relation is significant in hydrogen fine-structure calculations
(for example, see \cite{Schwinger2001}).

%\section{Problems for Chapter 8}

%\begin{enumerate}

%\item Suppose the system consists of $N$ particles interacting through
%a pairwise potential $V({\bf r}_a-{\bf r}_b)$.  Write down the Lagrangian
%and obtain the equations of motion.  What is the Hamiltonian, $H({\bf r}_a,
%{\bf p}_a)$?  Show that energy and total momentum are conserved.
%What is required for angular momentum to be conserved?

%\item For a free relativistic particle of rest mass $m_0$, the energy
%is $$E=\sqrt{p^2c^2+m_0^2c^4}.$$
%Use this as the Hamiltonian $H$, and from the Lagrangian
%$$L={\bf p}\dotprod{d{\bf r}\over dt}-H$$
%determine the relationship between the velocity 
%${\bf v}=d{\bf r}/dt$ and the momentum.
%Compute the energy in terms of the velocity.   
%Write the Lagrangian in terms of $\bf v$.

%\item Consider a particle bound by a potential of the form
%$$V=a r^b.$$
%Derive the time-averaged virial theorem relating $\overline T$
%to $\overline V$.  What is the smallest value of $b$ for which a
%bound state can occur?

%\end{enumerate}

 \chapter{Classical Field Theory---Electrodynamics}
\label{sec:2}
This Chapter again grew out of Schwinger's UCLA lectures. 
These evolved, torturously,  into Chapter 9 of 
\textit{Classical Electrodynamics} \cite{Schwinger1998}.  
Here we use Gaussian units.

\section{Action of Particle in Field}

It was stated in our review of mechanical action principles in the
previous Chapter that the
third viewpoint, which employs the variables $\bf r$, $\bf p$, and $\bf v$,
was particularly convenient for describing electromagnetic forces on
charged particles.  With the explicit, and linear, appearance of
$\bf v$ in what plays the role of the potential function 
when magnetic fields are present, %(\ref{6.7}),
we begin to see the basis for that remark.  Indeed, we have only to consult
(\ref{8.20}) to find the appropriate Lagrangian:
\begin{equation}
L={\bf p}\dotprod\left({d{\bf r}\over dt}-{\bf v}\right)+{1\over2}mv^2-e\phi
+{e\over c}{\bf v\dotprod A},
\label{1-20.16}
\end{equation}
where $\phi$ and $\mathbf{A}$ are the scalar and vector potentials,
respectively.
To recapitulate, the equations resulting from variations of $\bf p$, $\bf r$,
and $\bf v$ are, respectively,
\begin{subequations}
\begin{eqnarray}
{d{\bf r}\over dt}&=&{\bf v},\label{1-20.17a}\\
{d\over dt}{\bf p}&=&-e\vecnab \left[\phi-{1\over c}{\bf v\dotprod A}\right],
\label{1-17.20b}\\
{\bf p}&=&m{\bf v}+{e\over c}{\bf A}.
\label{1-20.17c}
\end{eqnarray}
\end{subequations}

We can now move to either the Lagrangian or the Hamiltonian formulation.
For the first, we simply adopt ${\bf v}=d{\bf r}/dt$ as a definition
(but see the discussion in Sec.~\ref{sec8.3}) and get
\begin{equation}
L={1\over2}m\left(d{\bf r}\over dt\right)^2-e\phi+{e\over c}{d{\bf r}\over dt}
\dotprod{\bf A}.
\label{1-20.18}
\end{equation}
Alternatively, we use (\ref{1-20.17c}) to define
\begin{equation}
{\bf v}={1\over m}\left({\bf p}-{e\over c}{\bf A}\right),
\label{1-20.19}
\end{equation}
and find
\begin{subequations}
\begin{eqnarray}
L&=&{\bf p}\dotprod{d{\bf r}\over dt}-H,\\
H&=&{1\over2m}\left({\bf p}-{e\over c}{\bf A}\right)^2+e\phi.
\label{1-20.20}
\end{eqnarray}
\end{subequations}
%Here we make contact with the energy considerations of Sec.~6.1; in particular%,
%$H$ coincides with the form of the energy given in (\ref{6.11}).

\section{Electrodynamic Action}

The electromagnetic field is a mechanical system.  It contributes its variables to the
action, to the Lagrangian of the whole system of charges and fields.  
In contrast with the point charges, the field is distributed in space.  
Its Lagrangian should therefore be, not a summation over discrete points, 
but an integration over all spatial volume elements,
\begin{equation}
L_{\rm field}=\int\deer{\cal L}_{\rm field};
\label{1-21.1}
\end{equation}
this introduces the Lagrange function, or Lagrangian density, ${\cal L}$.
The total Lagrangian must be the sum of  the particle part, (\ref{1-20.16}),
 and the field part, (\ref{1-21.1}), where the latter must be chosen so as 
 to give the Maxwell equations, in Gaussian units: %(\ref{1.27}):
\begin{subequations}
 \begin{eqnarray}
\vecnab\crossprod{\bf B}&=&{1\over c}{\partial\over\partial t}{\bf E}+
{4\pi\over c}{\bf j},\quad \vecnab\dotprod{\bf E}=4\pi\rho,\\
-\vecnab\crossprod{\bf E}&=&{1\over c}{\partial\over \partial t}{\bf B},
\!\quad\quad\qquad\vecnab\dotprod{\bf B}=0.
%\label{1.27}
\end{eqnarray}
\end{subequations}
 The homogeneous equations here are
equivalent to the construction of the electromagnetic
field in term of potentials, or, %(\ref{6.4}) and (\ref{6.2}),
\begin{subequations}
\begin{eqnarray}
{1\over c}{\partial\over\partial t}{\bf A}&=&-{\bf E}-\vecnab\phi,\\
{\bf B}&=&\vecnab\crossprod{\bf A}.
\label{1-21.2}
\end{eqnarray}
\end{subequations}
Thus, we recognize that  ${\bf A(r},t)$, ${\bf E(r},t)$, in analogy with
${\bf r}(t)$, ${\bf p}(t)$, obey equations of motion while $\phi({\bf r},t)$,
${\bf B(r},t)$, as analogues of ${\bf v}(t)$, do not.  There are enough clues
here to give the structure of ${\cal L}_{{\rm field}}$, apart from an overall
factor.  The anticipated complete Lagrangian for microscopic
 electrodynamics is
\begin{eqnarray}
L &=&\sum_a \left[{\bf p}_a\dotprod\left({d{\bf r}_a\over dt} - {\bf v}_a
\right)   + {1\over2} m_av_a^2-e_a\phi({\bf r}_a)
+{e_a\over c}{\bf v}_a\dotprod{\bf A(r}_a)\right]\nonumber\\
&&\!\!\!\!\!\!\!\!\!\!\mbox{}+{1\over4\pi}\int(d{\bf r})\,
\left[{\bf E}\dotprod
\left(-{1\over c}{\partial\over\partial t}{\bf A}-\vecnab\phi\right)
-{\bf B\dotprod\vecnab \crossprod A} + {1\over2} (B^2- E^2) \right].
\label{9.1}
\end{eqnarray}

 The terms that are summed in (\ref{9.1}) describe the behavior of charged 
 particles under the influence of the fields, while the terms that are integrated 
 describe the field behavior.  The independent variables are
\begin{equation} 
{\bf r}_a(t), \quad {\bf v}_a(t),\quad {\bf p}_a(t),\quad  \phi({\bf r},t),
\quad {\bf A}({\bf r},t),\quad {\bf E}({\bf r},t),\quad {\bf  B}({\bf r},t),
\quad t.
\end{equation}
 We now look at the response of the Lagrangian to variations in each of these
 variables separately, starting with the particle part:
\begin{subequations}
\begin{eqnarray}
 \delta{\bf r}_a:\quad \delta L &=& {d\over dt} (\delta{\bf r}_a\dotprod{\bf p}_a)
   + \delta{\bf r}_a\dotprod\left[ - {d{\bf p}_a\over dt}-\vecnab_ae_a\left(
 \phi({\bf r}_a) -{{\bf v}_a\over c}\dotprod{\bf A(r}_a) \right)\right],\nonumber\\
\label{9.2}\\
 \delta{\bf v}_a:\quad \delta L &=& \delta{\bf v}_a\dotprod\left[
-{\bf p}_a+m_a{\bf v}_a+{e_a\over c}{\bf A(r}_a)\right],
\label{9.3}\\
 \delta{\bf p}_a: \quad\delta L&=& \delta{\bf p}_a \dotprod\left(
{ d{\bf r}_a\over  dt} - {\bf v}_a\right).
\label{9.4}
\end{eqnarray}
\end{subequations}
 The stationary action principle now implies the equations of motion
\begin{subequations}
\begin{eqnarray}
{ d{\bf p}_a\over dt}&=&-e_a\vecnab_a\left(\phi({\bf r}_a)-{{\bf v}_a\over c}
\dotprod {\bf A(r}_a)\right),
\label{9.5a}\\
 m_a{\bf v}_a&=&{\bf p}_a- { e_a\over c}{\bf A (r}_a),
\label{9.5b}\\
{\bf v}_a&=&{d{\bf r}_a\over dt},
\label{9.5c}
\end{eqnarray}
\end{subequations}
 which are the known results, (\ref{1-20.17a})--(\ref{1-20.17c}).

The real work now lies in deriving the equations of motion for the fields.
 In order to cast all the field-dependent terms into integral form, we 
introduce charge and current densities,
\begin{subequations}
\begin{eqnarray}
\rho({\bf r},t)&=&\sum_a e_a\delta({\bf r-r}_a(t)),\\
{\bf j(r},t)&=&\sum_a e_a {\bf v}_a(t)\delta({\bf r-r}_a(t)),
\label{1-22.3}
\end{eqnarray}
\end{subequations}
 so that
\begin{equation}
\sum_a\left[-e_a\phi({\bf r}_a)+{e_a\over c}{\bf v}_a\dotprod{\bf A(r}_a)
\right]=\int(d{\bf r})\,\left[-\rho\phi+{1\over c}{\bf j\dotprod A}\right].
\label{9.6}
\end{equation}
 The volume integrals extend over sufficiently large regions to contain all
 the fields of interest.  Consequently, we can integrate by parts and ignore
 the surface terms.  The responses of the Lagrangian (\ref{9.1}) 
 to field variations, and the corresponding equations of motion deduced 
 from the action principle are
\begin{subequations}
\label{edvareqn}
\begin{eqnarray}
\delta\phi:\qquad \delta L&=&{1\over4\pi}\int(d{\bf r})\,\delta\phi
(\vecnab\dotprod{\bf E}-4\pi\rho),
\label{9.7a}\\
 \vecnab\dotprod{\bf E} &=& 4\pi\rho,
\label{9.7b}\\
 \delta{\bf A}:\qquad \delta L&=&-{1\over4\pi c}{d\over dt}\int
(d{\bf r})\, \delta{\bf A}\dotprod {\bf E} \nonumber\\
&&\mbox{} + {1\over4\pi}
\int\deer\delta{\bf A}\dotprod\left({1\over c}{\partial{\bf E}\over \partial t}
+{4\pi\over c}{\bf j}-\vecnab \crossprod{\bf B}\right),
\label{9.8a}\\
\vecnab\crossprod {\bf B}&=&{1\over c}{\partial\over\partial t}{\bf E}+
{4\pi\over c}{\bf j},
\label{9.8b}\\
 \delta{\bf E}:\qquad \delta L&=&{1\over4\pi}\int\deer\delta{\bf E}\dotprod\left(
-{1\over c}{\partial\over\partial
 t}{\bf A}-\vecnab\phi-{\bf E}\right),
\label{9.9a}\\
{\bf E}&=&-{1\over c}{\partial\over\partial t}{\bf A}-\vecnab\phi,
\label{9.9b}\\
 \delta{\bf B}:\qquad \delta L&=&{1\over4\pi}\int\deer\delta{\bf B}
\dotprod(-\vecnab\crossprod {\bf A+B}),
\label{9.10a}\\
{\bf B}&=&\vecnab   \crossprod {\bf A}.
\label{9.10b}
\end{eqnarray}
\end{subequations}
 We therefore recover Maxwell's equations, two of which are implicit in the
 construction of ${\bf E}$ and $\bf B$ in terms of potentials.  
By making a time variation of the action [variations due to the time dependence
 of the fields vanish by virtue of the stationary action principle---that is,
 they are already subsumed in Eqs.~(\ref{edvareqn})],
%(\ref{9.7a})--(\ref{9.10b})],
\begin{equation}
\delta t:\qquad\delta W=\int dt\,\left[
 {d \over dt}  (-H\delta t) + \delta t {dH\over dt} \right],
\label{9.11}
\end{equation}
 we identify the Hamiltonian of the system to be
\begin{eqnarray}
 H &=& \sum_a\left[\left({\bf  p}_a -{e_a\over c}{\bf A(r}_a)\right)\dotprod 
 {\bf v}_a - {1\over2}m_av_a^2+ e_a\phi({\bf r}_a)\right]\nonumber\\
&&+{1\over4\pi}\int\deer\left[{\bf E}\dotprod\vecnab\phi + {\bf B}\dotprod
\vecnab\crossprod {\bf A}+{1\over2}(E^2-B^2)\right],
\label{9.12}
\end{eqnarray}
 which is a constant of the motion,
$dH/dt=0$.
 The generators are inferred from
 the total time derivative terms in (\ref{9.2}), (\ref{9.8a}), and 
(\ref{9.11}),
\begin{subequations}
\begin{equation}
\delta W_{12}=G_1-G_2,
\end{equation}
 to be
\begin{equation}
 G = \sum_a\delta{\bf r}_a\dotprod
{\bf p}_a-{1\over4\pi c}\int\deer{\bf E \dotprod\delta A }- H\delta t.
\label{9.13}
\end{equation}
\end{subequations}

\section{Energy}
Notice that the total Lagrangian (\ref{9.1}) can be presented as
\begin{equation}
L=\sum_a{\bf p}_a\dotprod{d{\bf r}_a\over dt}-{1\over4\pi c}\int\deer
{\bf E}\dotprod{\partial\over\partial t}{\bf A}-H,
\label{1-23.1}
\end{equation}
where the Hamiltonian is given by (\ref{9.12}).
The narrower, Hamiltonian, description is reached by eliminating all
variables that do not obey equations of motion, and, correspondingly,
do not appear in $G$.  Those ``superfluous'' variables are the ${\bf v}_a$
and the fields $\phi$ and $\bf B$, which are eliminated by using
  (\ref{9.5b}), (\ref{9.7b}), and (\ref{9.10b}), 
 the equations without time derivatives, resulting, first,
  in the intermediate form
\begin{equation}
 H = \sum_a\left({1\over2m_a}\left({\bf p}_a-{e_a\over c}{\bf A}({\bf r}_a)
 \right)^2 
+e_a\phi({\bf r}_a)\right)+ \int\deer \left[ {E^2+B^2\over8\pi}-\rho\phi\right].
\label{9.14}
\end{equation}
 The first term here is the energy of the particles moving in the field 
[particle energy---see (\ref{1-20.20})], so we might call the second term the 
field energy.  
The ambiguity of these terms (whether the potential energy of particles 
is attributed to them or to the fields, or to both) is evident from 
the existence of a simpler form of the Hamiltonian
\begin{equation}
 H = \sum_a{1\over2 m_a}\left({\bf p}_a-{e_a\over c}{\bf A}({\bf r}_a)\right)^2
+\int\deer {E^2 + B^2\over8\pi},\quad \mathbf{B}=\vec{\nabla}\times \mathbf{A},
\label{9.15}
\end{equation}
 where we have used the equivalence of the two terms involving 
$\phi$, given in (\ref{9.6}). 

This apparently startling result suggests that the scalar potential has 
disappeared from the dynamical description.  But, in fact, it has not.  
If we vary the Lagrangian (\ref{1-23.1}), where $H$ is given by 
(\ref{9.15}), with respect to $\bf E$ we find
\begin{equation}
\delta L=-{1\over4\pi}\int\deer\delta {\bf E}\dotprod\left({1\over c}
{\partial\over\partial t}{\bf A+E}\right)=0.
\label{1-23.9}
\end{equation}
Do we conclude that ${1\over c}{\partial\over\partial t}{\bf A+E=0}$?
That would be true if the $\delta{\bf E(r},t)$ were arbitrary.  They are not;
${\bf E}$  is  subject to the restriction---the constraint---(\ref{9.7b}), which
means that any change in $\bf E$ must obey
\begin{equation}
\vecnab\dotprod\delta{\bf E}=0.
\label{1-23.10}
\end{equation}
The proper conclusion is that the vector multiplying $\delta{\bf E}$ in
(\ref{1-23.9}) is the gradient of a scalar function, just as in (\ref{9.9b}),
\begin{equation}
{1\over c}{\partial\over\partial t}{\bf A+E}=-\vecnab\phi,
\label{1-23.11}
\end{equation}
for that leads to
\begin{equation}
\delta L=-{1\over4\pi}\int\deer(\vecnab\dotprod\delta{\bf E})\phi=0,
\label{1-23.12}
\end{equation}
as required.

The fact that the energy is conserved,
\begin{equation}
      {dH\over dt} = 0,
\label{9.16}
\end{equation}
where
\begin{equation}
H=\sum_a{1\over2}m_av_a^2+\int\deer U,\quad U={E^2+B^2\over8\pi},
\label{1-24.2}
\end{equation}
is a simple sum of particle kinetic energy and integrated field energy
density,  can be verified directly by
 taking the time derivative of (\ref{9.14}). 
The time rate of change of the particle energy is computed directly:
%in (\ref{6.9}),
\begin{equation}
{d\over dt}\sum_a\left({1\over2}m_av_a^2+e_a\phi({\bf r}_a)\right)
=\sum_a{\partial\over\partial t}\left(e_a\phi({\bf r}_a)
-{e_a\over c}{\bf v}_a\dotprod{\bf A(r}_a)\right).
\label{9.17}
\end{equation}    
 We can compute the time derivative of the field energy by using the equation
 of energy conservation, %(\ref{3.6}),
\begin{equation}
{d\over dt}\int\deer U=-\int\deer {\bf j\dotprod E},
\label{9.18}
\end{equation}
 to be
\begin{eqnarray}
{d\over dt}\int\deer\left({ E^2 +B^2\over8\pi}-\rho\phi\right)&=&
\int\deer\left[-{\bf j \dotprod E}-\phi{\partial\over\partial t}\rho
-\rho{\partial\over\partial t}\phi\right]\nonumber\\
&=&-\int\deer\left[\rho{\partial\over\partial t}\phi-{1\over c}{\bf j}
\dotprod {\partial\over\partial t}{\bf A}\right]\nonumber\\
&=&-\sum_a e_a\left({\partial\over\partial t}\phi({\bf r}_a)-{1\over c}
{\bf v}_a\dotprod{\partial\over\partial t}{\bf A(r}_a)\right).\nonumber\\
\label{9.19}
\end{eqnarray}
 Here we have used (\ref{9.9b}), and have noted that
\begin{equation}
\int\deer\left[{\bf j}\dotprod\vecnab\phi-\phi{\partial\over\partial t}
\rho\right]=0
\end{equation}
 by charge conservation.  Observe that (\ref{9.17}) and 
(\ref{9.19}) are equal in magnitude
 and opposite in sign, so that their sum is zero.  
This proves the statement of energy conservation (\ref{9.16}).

\section{Momentum and Angular Momentum Conservation}
      The action principle not only provides us with the field equations,
 particle equations of motion, and expressions for the energy, but also with
 the generators (\ref{9.13}). The generators provide a connection between 
conservation laws and invariances of the action (recall Section \ref{sec8.4}). 
 Here we will further illustrate this connection by deriving momentum and 
angular momentum conservation from the invariance of the action under 
rigid coordinate translations and rotations, respectively. 
[In a similar way we could derive energy conservation, (\ref{9.16}), 
from the invariance under time displacements---see also Section \ref{sec9.7}].

\begin{figure}
\centering
\includegraphics{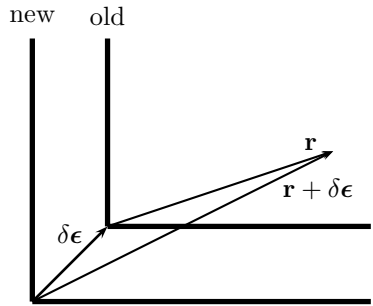}
%\begin{picture}(300,200)
%\put(0,0){\line(1,0){200}}
%\put(0,0){\line(0,1){150}}
%\put(30,40){\line(1,0){200}}
%\put(30,40){\line(0,1){150}}
%\put(0,0){\vector(3,4){30}}
%\put(30,40){\vector(2,1){100}}
%\qbezier(0,0)(65,45)(130,90)
%\put(10,30){$\delta{\mbox{\boldmath{$\epsilon$}}}$}
%\put(100,80){${\bf r}$}
%\put(115,70){${\bf r}+\delta {\mbox{\boldmath{$\epsilon$}}}$}
%\put(5,150){new}
%\put(35,190){old}
%\end{picture}
\caption{\label{fig9.1} Rigid coordinate displacement, where
the new coordinate system is displaced by a rigid translation 
$-\delta\mbox{\boldmath{$\epsilon$}}$ relative to the old coordinate system.}
\end{figure}

 Under an infinitesimal rigid coordinate displacement, 
 $\delta{\mbox{\boldmath{$\epsilon$}}}$, 
a given point which is described by $\bf r$ in the old coordinate 
system is described by ${\bf r}+\delta{\mbox{\boldmath{$\epsilon$}}}$
in the new one. (See Fig.\ \ref{fig9.1}.)
 The response of the particle term in (\ref{9.13}) is simple: 
$\delta {\mbox{\boldmath{$\epsilon$}}} \dotprod\sum_a{\bf p}_a$; for the
field part, we require the change, $\delta {\bf A}$, 
of the vector potential induced by
 the rigid coordinate displacement. The value of a field ${\cal F}$ 
at a physical point $P$ is unchanged under such a displacement,
so that if $\bf r$ and $\bf r+\delta \mbox{\boldmath{$\epsilon$}}$
 are the coordinates of $P$ in the two frames, there are corresponding
functions $F$ and $\overline F$ such that
\begin{equation}
{\cal F}(P) = F({\bf r})  =\overline{F}( {\bf r}  + 
\delta {\mbox{\boldmath{$\epsilon$}}}),
\label{9.20}
\end{equation}
 that is, the new function $\overline{F}$
 of the new coordinate equals the old function
 $F$ of the old coordinate.  The change in the function $F$ at the {\it same 
coordinate\/} is given by
\begin{equation}
\overline{F}({\bf r}) = F({\bf r}) + \delta F({\bf r}),
\end{equation}
 so that
\begin{equation}
 \delta F({\bf r}) = F({\bf r}-\delta {\mbox{\boldmath{$\epsilon$}}}) 
 - F({\bf r}) = 
-\delta{\mbox{\boldmath{$\epsilon$}}}\dotprod\vecnab F({\bf r}),
\label{9.21}
\end{equation}
 for a rigid translation (not a rotation).  
 
 As an example, consider the charge density
 \begin{equation}
 \rho({\bf r})=\sum_a e_a\delta({\bf r-r}_a).
 \label{1-24.7}
 \end{equation}
 If the positions of all the particles, the ${\bf r}_a$, are displaced by
 $\delta{\mbox{\boldmath{$\epsilon$}}}$, the charge density changes to
 \begin{equation}
 \rho({\bf r})+\delta\rho({\bf r})=\sum_a e_a\delta({\bf r-r}_a-
 \delta{\mbox{\boldmath{$\epsilon$}}}),
 \label{1-24.8}
 \end{equation}
 where
\begin{equation}
 \delta({\bf r-r}_a-\delta{\mbox{\boldmath{$\epsilon$}}})=\delta({\bf r-r}_a)
 -\delta{\mbox{\boldmath{$\epsilon$}}}\dotprod\vecnab_r\delta({\bf r-r}_a),
 \label{1-24.9}
 \end{equation}
 and therefore
 \begin{equation}
 \delta\rho({\bf r})=-\delta{\mbox{\boldmath{$\epsilon$}}}
 \dotprod\vecnab\rho({\bf r}),
 \label{1-24.10}
 \end{equation}
 in agreement with (\ref{9.21}).
 
  So the field part of $G$ in (\ref{9.13}) is
\begin{eqnarray}
-\int(d{\bf r})\,{1\over4\pi c} {\bf E \dotprod \delta A}&=&{1\over4\pi c}
\int\deer E_i(\delta{\mbox{\boldmath{$\epsilon$}}}
\dotprod \vecnab)A_i\nonumber\\
=&-&{1\over c}\sum_a e_a\delta {\mbox{\boldmath{$\epsilon$}}}
\dotprod{\bf A(r}_a)+
{1\over4\pi c}\int\deer({\bf E\crossprod B)}\dotprod\delta 
{\mbox{\boldmath{$\epsilon$}}},\nn\\
\label{9.22}
\end{eqnarray}
 where the last rearrangement makes use of (\ref{9.7b}) and (\ref{9.10b}),
 and the vector identity
 \begin{equation}
 \delta {\mbox{\boldmath{$\epsilon$}}}\crossprod(\vecnab\crossprod {\bf A})
 =\vecnab(\delta {\mbox{\boldmath{$\epsilon$}}}\dotprod{\bf A})
 -(\delta {\mbox{\boldmath{$\epsilon$}}}\dotprod\vecnab){\bf A}.
 \end{equation}
Including the particle part from (\ref{9.13}) we find the generator
corresponding to a rigid coordinate displacement can be written as
\begin{equation}
G=\delta{\bf {\mbox{\boldmath{$\epsilon$}}}\dotprod P},
\label{9.23}
\end{equation}
 where
\begin{equation}
{\bf P}=\sum_a\left({\bf p}_a-{e_a\over c}{\bf A(r}_a)\right)
+{1\over4\pi c}\int\deer{\bf E \crossprod B}\equiv\sum_a m_a{\bf v}_a+
\int\deer {\bf G},
\label{9.24}
\end{equation}
 with $\bf G$ the momentum density. %(\ref{3.10}). 
Since the action is invariant under a rigid displacement,
\begin{equation}
 0 = \delta W = G_1 - G_2 = ({\bf P}_1 -{\bf P}_2)\dotprod\delta{\bf r},
\end{equation}
 we see that
\begin{equation}
 {\bf P}_1 =  {\bf P}_2,
\end{equation}
 that is, the total momentum, $\bf P$, is conserved.  
This, of course, can also be verified by explicit calculation:
\begin{eqnarray}
{d\over dt}\int\deer {1\over4\pi c}{\bf E \crossprod B}
&=&-\int\deer \left[\rho{\bf E}+{1\over c}{\bf j\crossprod B}\right]\nonumber\\
&=&-\sum_ae_a\left({\bf E(r}_a)+{1\over c}{\bf v}_a\crossprod{\bf B(r}_a)\right),
\end{eqnarray}
% which restates (\ref{3.12}), 
from which the constancy of $\bf P$ follows.
% from (\ref{6.1}).

      Similar arguments can be carried out for a rigid rotation for which the
 change in the coordinate vector is
\def\bomega{\mbox{\boldmath{$\omega$}}}
\begin{equation}
\delta{\bf r} = \delta\bomega \crossprod{\bf r},
\end{equation}
 with $\delta\bomega$ 
 constant.  The corresponding change in a vector function is
\begin{equation}
{\bf\overline{A}(r + \delta r) = A(r) + \delta\bomega \crossprod A(r)}
\end{equation}
 since a vector transforms in the same way as $\bf r$, 
so the new function at the initial numerical values of the coordinates is
\begin{equation}
{\bf \overline{A} (r) =A (r) - (\delta r\dotprod\vecnab) A(r) + 
\delta\bomega \crossprod A(r)}.
\end{equation}
%(See Problem 9.\ref{prob9trans}.)
 The change in the vector potential is
\begin{equation}
{\bf \delta A = -(\delta r\dotprod \vecnab) A+ \delta\bomega \crossprod A}.
\end{equation}
 The generator can now be written in the form
\begin{equation}
 G = \delta\bomega\dotprod{\bf J},
\end{equation}
 where the total angular momentum, $\bf J$, is found to be 
% (see Problem 9.\ref{prob9.1})
\begin{equation}
{\bf J}=\sum_a{\bf r}_a\crossprod m_a{\bf v}_a+\int\deer{\bf r}
\crossprod\left({1\over4\pi c}{\bf E\crossprod B}\right),
\label{9.26}
\end{equation}
 which again is a constant of the motion.

\section{Gauge Invariance and the Conservation of Charge}
\label{sec9.6}
    An electromagnetic system possesses a conservation law, that of
    electric charge, which has no place in the usual mechanical framework.
    It is connected to a further invariance of the
    electromagnetic fields---the potentials
 are not uniquely defined in that if we let
\begin{equation}
{\bf A\to A }+ \vecnab\lambda,\qquad\phi\to\phi-{1\over c}{\partial
\over\partial t}\lambda,
\label{9.27}
\end{equation}
 the electric and magnetic fields defined by (\ref{9.9b}) and 
(\ref{9.10b}) remain unaltered, for an arbitrary function $\lambda$. 
This is called gauge invariance; the corresponding substitution 
(\ref{9.27}) is a gauge transformation.  [The term has its origin in
a now obsolete theory of Hermann Weyl (1885--1955) \cite{weyl1919}.] 

This invariance of
 the action must imply a corresponding conservation law.  To determine what is
 conserved, we compute the change in the Lagrangian, (\ref{9.1}), 
explicitly.  Trivially, the field part of $L$ remains unchanged.  
In considering the change of the particle part, we recognize that (\ref{9.27})
 is incomplete; since $\bf v$ is a physical quantity, 
${\bf p} - (e/c){\bf A}$
 must be invariant under a gauge transformation,
 which will only be true if (\ref{9.27}) is supplemented by
\begin{equation}
{\bf p} \to{\bf p} +{e\over c}\vecnab\lambda. 
\label{9.28}
\end{equation}
 Under the transformation (\ref{9.27}) and (\ref{9.28}), 
the Lagrangian becomes
\begin{eqnarray}
 L \to\overline{L} &\equiv& L + \sum_a\left[{e_a\over c}\vecnab\lambda
\dotprod\left({d{\bf r}_a\over dt}-{\bf v}_a\right)
+{e_a\over c} {\partial\over\partial t}\lambda+{e_a\over c}{\bf v}_a\dotprod
\vecnab\lambda\right]\nonumber\\
&=&L+\sum_a{e_a\over c}\left({\partial\over\partial t}\lambda
+{ d{\bf r}_a\over dt}\dotprod\vecnab\lambda\right)\nonumber\\
&=&L+{d\over dt}w,
\label{9.29}
\end{eqnarray}
where
\begin{equation}
w=\sum_a{e_a\over c}\lambda({\bf r}_a,t).
\label{9.30}
\end{equation}
 What is the physical consequence of adding a total time derivative to a
 Lagrangian?  It does not change the equations of motion, so the system is 
unaltered.  Since the entire change is in the end point behavior,
\begin{equation}
     \overline{W}_{12} = W_{12} +(w_1-w_2),  
\label{9.31}
\end{equation}
 the whole effect is a redefinition of the generators, $G$,
\begin{equation}
\overline{G} = G + \delta w.
\end{equation}
 This alteration reflects the fact that the Lagrangian itself is ambiguous up
 to a total time derivative term.  [This term may also be seen as arising
from the field term in the generator (\ref{9.13}).]
   
   To ascertain the implication of gauge invariance,
   we rewrite the change in the Lagrangian given in the first line 
   of (\ref{9.29}) by use of (\ref{9.5c}),
\begin{equation}
 \overline{ L}-L ={ 1\over c}\int\deer \left[\rho{\partial\over\partial t}
 \lambda+{\bf j  \dotprod\vecnab\lambda}\right],
\label{9.32}
\end{equation}
 and apply this result to an infinitesimal gauge transformation,
   $\lambda\to\delta\lambda$.
   The change in the action is then
   \begin{equation}
   \delta W_{12}=G_{\delta\lambda_1}-G_{\delta\lambda_2}-\int_{t_2}^{t_1}
   dt\int\deer{1\over c}\delta\lambda\left({\partial\over\partial t}\rho
   +\vecnab\dotprod {\bf j}\right),
   \label{1-25.13}
   \end{equation}
   with the generator being
   \begin{equation}
   G_{\delta\lambda}=\int\deer{1\over c}\rho\,\delta\lambda.
   \label{1-25.14}
   \end{equation}
   In view of the arbitrary nature of $\delta\lambda({\bf r},t)$,
   the stationary action principle now demands that, at every point,
\begin{equation}
{\partial\over\partial t}\rho+\vecnab\dotprod{\bf j}=0,
\label{9.33}
\end{equation}
that is, gauge invariance implies local charge conservation.
(Of course, this same result follows from Maxwell's equations.)
Then, the special situation $\delta\lambda=\mbox{constant}$, where
$\delta{\bf A}=\delta\phi=0$, and $W_{12}$ is certainly invariant,
implies a conservation law, that of
\begin{equation}
G_{\delta\lambda}={1\over c}\delta\lambda \,Q,
\label{1-25.16}
\end{equation}
in which 
\begin{equation}
Q=\int\deer\rho
\label{1-25.17}
\end{equation}
is the conserved total charge. 

\section{Gauge Invariance and Local Conservation Laws}
\label{sec9.7}
We have just derived the local conservation law of electric charge.
Electric charge is a property carried only by the particles, 
not by the electromagnetic
field.  In contrast, the mechanical properties of energy, linear momentum, and
angular momentum are attributes of both particles and fields.  For
these we have conservation laws of total quantities.  What about local
conservation laws?  The usual development of electrodynamics refers
to
%Early in this development (Chapter 3) we produced
local {\it non}-conservation laws; they concentrated on the fields
and characterized the charged particles as sources (or sinks) of field
mechanical properties.  It is natural to ask for a more even-handed
treatment of both charges and fields.  We shall supply it, in the framework
of a particular example.  The property of gauge invariance will be both
a valuable guide, and an aid to simplifying the calculations.

The time displacement of a complete physical system identifies its total
energy.  This suggests that time displacement of a part of the system
provides energetic information about that portion.  The ultimate limit of 
this spatial subdivision, a local description, should appear in response
to an (infinitesimal) time displacement that varies arbitrarily in space
as well as in time, $\delta t({\bf r},t)$.

Now we need a clue.  How do fields, and potentials, respond to such 
coordi\-nate-dependent displacements?  This is where the freedom of gauge
transformations enters:  The change of the vector and scalar potentials,
by $\vecnab\lambda({\bf r},t)$, $-(1/c)(\partial/\partial t)\lambda
({\bf r},t)$, respectively, serves as a model for the potentials themselves.
The advantage here is that the response of the scalar $\lambda({\bf r},t)$
to the time displacement can be reasonably taken to be
\begin{subequations}
\begin{equation}
(\lambda+\delta\lambda)({\bf r},t+\delta t)=\lambda({\bf r},t),
\label{1-26.1}
\end{equation}
or
\begin{equation}
\delta\lambda({\bf r},t)=-\delta t({\bf r},t){\partial\over\partial t}
\lambda({\bf r},t).
\label{1-26.2}
\end{equation}
\end{subequations}
Then we derive
\begin{subequations}
\begin{eqnarray}
\delta(\vecnab\lambda)&=&-\delta t{\partial\over\partial t}(\vecnab\lambda)
+\left(-{1\over c}{\partial\over\partial t}\lambda\right)c\vecnab\delta t,\\
\delta\left(-{1\over c}{\partial\over\partial t}\lambda\right)
&=&-\delta t\left(-{1\over c}{\partial^2\over\partial t^2}\lambda\right)
-\left(-{1\over c}{\partial\over\partial t}\lambda\right){\partial
\over\partial t}\delta t,
\label{1-26.3}
\end{eqnarray}
\end{subequations}
which is immediately generalized to
\begin{subequations}
\begin{eqnarray}
\delta {\bf A}&=&-\delta t{\partial\over\partial t}{\bf A}
+\phi c\vecnab\delta t,\\
\delta \phi&=&-\delta t{\partial\over\partial t}\phi-\phi{\partial
\over\partial t}\delta t,
\label{1-26.4}
\end{eqnarray}
\end{subequations}
or, equivalently,
\begin{subequations}
\begin{eqnarray}
\delta {\bf A}&=&c\delta t{\bf E}+\vecnab(\phi c\delta t),\\
\delta\phi&=&-{1\over c}{\partial\over\partial t}(\phi c\delta t).
\label{1-26.5}
\end{eqnarray}
\end{subequations}
In the latter form we recognize a gauge transformation, produced by the
scalar $\phi c\delta t$, which will not contribute to the changes of
field strengths.  Accordingly, for that calculation we have,
effectively, $\delta{\bf A}=c\delta t{\bf E}$, $\delta \phi=0$, leading
to
\begin{subequations}
\begin{eqnarray}
\delta{\bf E}&=&-{1\over c}{\partial\over\partial t}(c\delta t {\bf E})
=-\delta t{\partial\over\partial t}{\bf E}-{\bf E}{\partial\over
\partial t}\delta t,\\
\delta{\bf B}&=&\vecnab\crossprod(c\delta t{\bf E})=-\delta t
{\partial\over\partial t}{\bf B}-{\bf E}\crossprod\vecnab
c\delta t;
\label{1-26.6}
\end{eqnarray}
\end{subequations}
the last line employs the field equation $\vecnab\crossprod{\bf E}
=-(1/c)(\partial{\bf B}/\partial t)$.

In the following we adopt a viewpoint in which such homogeneous field
equations are accepted as consequences of the definition of the fields
in terms of potentials.  That permits the field Lagrange function
(\ref{9.1}) to be simplified:
\begin{equation}
{\cal L}_{\rm field}={1\over8\pi}(E^2-B^2).
\label{1-26.7}
\end{equation}
Then we can apply the field variation (\ref{1-26.6}) directly, and get
\begin{eqnarray}
\delta{\cal L}_{\rm field}&=&-\delta t{\partial\over\partial t}
{\cal L}_{\rm field}-{1\over4\pi}E^2{\partial\over\partial t}
\delta t-{c\over4\pi}{\bf E\crossprod B}\dotprod\vecnab\delta t\nonumber\\
&=&-{\partial\over\partial t}(\delta t{\cal L}_{\rm field})
-{1\over8\pi}(E^2+B^2){\partial\over\partial t}\delta t
-{c\over4\pi}{\bf E\crossprod B}\dotprod\vecnab\delta t.
\label{1-26.8}
\end{eqnarray}
Before commenting on these last, not unfamiliar, field structures, we turn
to the charged particles and put them on a somewhat similar footing
in terms of a continuous, rather than a discrete, description.

We therefore present the Lagrangian of the charges in (\ref{9.1}) in
terms of a corresponding Lagrange function,
\begin{subequations}
\begin{equation}
L_{\rm charges}=\int\deer{\cal L}_{\rm charges},
\label{1-26.9}
\end{equation}
where
\begin{equation}
{\cal L}_{\rm charges}=\sum_a{\cal L}_a
\label{1-26.10}
\end{equation}
and
\begin{equation}
{\cal L}_a=\delta({\bf r-r}_a(t))\left[{1\over2}m_av_a(t)^2
-e_a\phi({\bf r}_a,t)+{e_a\over c}{\bf v}_a(t)\dotprod{\bf A(r}_a,t)
\right];
\label{1-26.11}
\end{equation}
\end{subequations}
the latter adopts the Lagrangian viewpoint, with ${\bf v}_a=d{\bf r}_a/dt$
accepted as a definition.  Then, the effect of the time displacement
on the variables ${\bf r}_a(t)$, taken as
\begin{subequations}
\begin{eqnarray}
({\bf r}_a+\delta{\bf r}_a)(t+\delta t)&=&{\bf r}_a(t),\label{1-26.12}\\
\delta{\bf r}_a(t)&=&-\delta t({\bf r}_a,t){\bf v}_a(t),\label{1-26.13}
\end{eqnarray}
\end{subequations}
implies the velocity variation
\begin{equation}
\delta{\bf v}_a(t)=-\delta t({\bf r}_a,t){d\over dt}{\bf v}_a(t)
-{\bf v}_a(t)\left[{\partial\over\partial t}\delta t+{\bf v}_a\dotprod
\vecnab\delta t\right];
\label{1-26.14}
\end{equation}
the last step exhibits both the explicit and the implicit dependences
of $\delta t({\bf r}_a,t)$ on $t$.  In computing the variation of 
$\phi({\bf r}_a,t)$, for example, we combine the potential variation
given in (\ref{1-26.4}) with the effect of $\delta{\bf r}_a$:
\begin{subequations}
\begin{equation}
\delta\phi({\bf r}_a(t),t)=-\delta t{\partial\over\partial t}\phi
-\phi{\partial\over\partial t}\delta t-\delta t{\bf v}_a\dotprod
\vecnab_a\phi=-\delta t{d\over dt}\phi-\phi{\partial\over\partial t}
\delta t,
\label{1-26.15}
\end{equation}
and, similarly,
\begin{equation}
\delta{\bf A}({\bf r}_a(t),t)=-\delta t{\partial\over\partial t}{\bf A}
+\phi c\vecnab\delta t-\delta t{\bf v}_a\dotprod
\vecnab_a{\bf A}=-\delta t{d\over dt}{\bf A}+\phi c
\vecnab\delta t.
\label{1-26.16}
\end{equation}
\end{subequations}

The total effect of these variations on ${\cal L}_a$ is thus
\begin{subequations}
\begin{equation}
\delta{\cal L}_a=-\delta t{d\over dt}{\cal L}_a+\delta({\bf r-r}_a(t))
\left(-m_av_a^2-{e_a\over c}{\bf A\dotprod v}_a+e_a\phi\right)
\left({\partial\over\partial t}\delta t+{\bf v}_a\dotprod\vecnab\delta t
\right),
\label{1-26.17}
\end{equation}
or
\begin{equation}
\delta{\cal L}_a=-{d\over dt}(\delta t{\cal L}_a)-\delta({\bf r-r}_a(t))
E_a\left({\partial\over\partial t}\delta t+{\bf v}_a\dotprod\vecnab
\delta t\right),
\label{1-26.18}
\end{equation}
\end{subequations}
where we see the kinetic energy of the charged particle,
\begin{equation}
E_a={1\over2}m_av_a^2.
\label{1-26.19}
\end{equation}
We have retained the particle symbol $d/dt$ to the last, but now, 
being firmly back in the field, space-time viewpoint, it should be
written as $\partial/\partial t$, referring to all $t$ dependence,
with $\bf r$ being held fixed.  The union of these various 
contributions to the variation of the total Lagrange function is
\begin{equation}
\delta{\cal L}_{\rm tot}=-{\partial\over\partial t}(\delta t
{\cal L}_{\rm tot})-U_{\rm tot}{\partial\over\partial t}\delta t
-{\bf S}_{\rm tot}\dotprod\vecnab\delta t,
\label{1-26.20}
\end{equation}
where, from (\ref{1-26.8}) and (\ref{1-26.18}),
\begin{subequations}
\begin{equation}
U_{\rm tot}={1\over8\pi}(E^2+B^2)+\sum_a\delta({\bf r-r}_a(t))E_a
\label{1-26.21}
\end{equation}
and
\begin{equation}
{\bf S}_{\rm tot}={c\over4\pi}{\bf E\crossprod B}+\sum_a\delta
({\bf r-r}_a(t))E_a{\bf v}_a,
\label{1-26.22}
\end{equation}
\end{subequations}
are physically transparent forms for the total energy density and total
energy flux vector.

To focus on what is new in this development, we ignore boundary effects
in the stationary action principle, by setting the otherwise arbitrary
$\delta t({\bf r},t)$ equal to zero at $t_1$ and $t_2$.  Then, through
partial integration, we conclude that
\begin{equation}
\delta W_{12}=\int_{t_2}^{t_1}dt\int\deer\delta t\left({\partial\over
\partial t}U_{\rm tot}+\vecnab\dotprod{\bf S}_{\rm tot}\right)=0,
\label{1-26.23}
\end{equation}
from which follows the local statement of total energy conservation,
\begin{equation}
{\partial\over\partial t}U_{\rm tot}+\vecnab\dotprod{\bf S}_{\rm tot}=0,
\label{1-26.24}.
\end{equation}
%which generalizes (\ref{3.3}).

%\section{Problems for Chapter 9}
%\begin{enumerate}

%\item 
%\label{prob9.a}Consider the Lagrangian of a particle in a given
%electromagnetic field,  (\ref{1-20.16}):
%$$L({\bf r,p,v},t)={\bf p}\dotprod\left({d{\bf r}\over dt}-{\bf v}\right)
%+{1\over2}mv^2-e\phi({\bf r},t)+{e\over c}{\bf v\dotprod A(r},t).$$
%\begin{enumerate}
%\item Reduce to Lagrangian form and then derive the equations of motion.
%\item Reduce to Hamiltonian form and then derive the equations of motion.
%\item Show the equivalence between a) and b).
%\end{enumerate}

%\item \label{prob9trans} Verify the transformation laws under rigid
%rotations for scalars and vectors,
%$$\delta S({\bf r})=-(\delta\bomega\crossprod{\bf r})\dotprod\vecnab S({\bf r}%),$$
%$$\delta{\bf V(r})=-(\delta\bomega\crossprod{\bf r})\dotprod\vecnab{\bf V(r})
%+\delta\bomega\crossprod{\bf V(r}),$$
%by considering the transformations of the charge and current densities
%(\ref{1-22.3}).
%\item Derive the behavior of a vector field ${\bf V(r)}$ under
%infinitesimal rotations by considering the example of 
%the gradient of a scalar field $S({\bf r})$.

%\item \label{prob9.1}Fill in the details to derive (\ref{9.26}).

%\item Verify directly the local conservation law obeyed by
%$U_{\rm tot}$ and ${\bf S}_{\rm tot}$, (\ref{1-26.24}).

%\item\label{prob9.6}
% By considering $\delta{\bf r(r},t)$, analogous to $\delta t({\bf r},t)$
% in Section 9.6,
%in its effect on ${\cal L}_{\rm field}$, derive the field momentum
%density and stress dyadic.

%\end{enumerate} 
\chapter{Quantum Action Principle}
\label{sec:qap}

This Chapter, and the following one, are based on lectures given by
the author in quantum field theory courses at the University of Oklahoma
over several years, based in turn largely on lectures given by Schwinger at
Harvard in the late 1960s.

After the above reminder of classical variational principles,
we now turn to the dynamics of quantum mechanics.  We begin by considering
the transformation function $\langle a',t+dt|b',t\rangle$.  Here
$|b',t\rangle$ is a state specified by the values $b'=\{b'\}$ of a
complete set of dynamical variables $B(t)$, while $|a',t+dt\rangle$
is a state specified by values  $a'=\{a'\}$ of a (different) complete set
of dynamical variables $A(t+dt)$, defined at a slightly later time.\footnote{ 
Here Schwinger is using his standard notation, designating eigenvalues by 
primes.} 
We suppose that $A$ and $B$ do not possess
any explicit time dependence---that is, their definition does not depend
upon $t$.  Here
\be
\langle a',t+dt|=\langle a',t|U,
\ee
where the infinitesimal time translation operator is related to the generator
of time translations as follows,
\be
U=1+iG=1-i\,dt\,H.
\ee
The Hamiltonian $H$ is a function of dynamical variables, which we write 
generically as $\chi(t)$, and of $t$ explicitly.  Thus
\be
\langle a',t+dt|b',t\rangle=\langle a',t|1-i\,dt\,H(\chi(t),t)|b',t\rangle.
\ee

We next translate states and operators to time zero:
\begin{subequations}
\bea
\langle a',t|&=&\langle a'|U(t),\quad |b',t\rangle=U^{-1}(t)|b'\rangle,\\
\chi(t)&=&U^{-1}(t)\chi U(t),
\eea
\end{subequations}
where $\chi=\chi(0)$, etc.  Then,
\be
\langle a',t+dt|b',t\rangle=\langle a'|1-i\,dt\,H(\chi,t)|b'\rangle,\ee
or, as a differential equation
\bea
\delta_{\rm dyn}\langle a',t+dt|b',t\rangle&=&i\langle a'|\delta_{\rm dyn}[-dt H]
|b'\rangle\nonumber\\
&=&i\langle a',t+dt|\delta_{\rm dyn}[-dt\,H(\chi(t),t)]|b',t\rangle,
\label{dynvar}
\eea
where $\delta_{\rm dyn}$ corresponds to changes in initial and final times,
$\delta t_2$ and $\delta t_1$, and in the structure of $H$, $\delta H$.
[By reintroducing $dt$ in the state on the left in 
the second line, we make a negligible error of ${\cal O}(dt^2)$.]

However, we can also consider {\it kinematical\/} changes.  To understand
these, consider a system defined by coordinates and momenta, $\{q_a(t)\}$,
$\{p_a(t)\}$, $a=1,\dots,n$, which satisfy the canonical commutation relations,
\begin{subequations}
\bea[q_a(t),p_b(t)]&=&i\delta_{ab},\quad(\hbar=1)\\
\mbox{}[q_a(t),q_b(t)]&=&[p_a(t),p_b(t)]=0.
\eea
\end{subequations}
A spatial displacement $\delta q_a$ is induced by
\be
U=1+iG_q,\quad G_q=\sum_{a=1}^n p_a\delta q_a.
\ee
In fact ($\delta q_a$ is a number, not an operator), 
\bea
U^{-1}q_aU&=&q_a-\frac1i[q_a,G_q]\nonumber\\
&=&q_a-\delta q_a,
\eea
while
\be
U^{-1}p_aU=p_a-\frac1i[p_a,G_q]=p_a.
\ee

The (dual) symmetry between position and momentum,
\be
q\to p,\quad p\to-q,
\ee
gives us the form for the generator of a displacement in $p$:
\be
G_p=-\sum_aq_a\delta p_a.
\ee

A {\it kinematic\/} variation in the states is given by the generators
\begin{subequations}
\bea
\delta_{\rm kin}\langle\,\,\,|&=&\overline{\langle\,\,\,|}-\langle\,\,\,|
=\langle\,\,\,|iG,\\
\delta_{\rm kin}|\,\,\,\rangle&=&\overline{|\,\,\,\rangle}-|\,\,\,\rangle
=-iG|\,\,\,\rangle,
\eea\end{subequations}
so, for example, under a $\delta q$ variation, the transformation function
changes by
\be
\delta_q\langle a',t+dt|b',t\rangle=i\langle a',t+dt|\sum_a\left[p_a(t+dt)
\delta q_a(t+dt)-p_a(t)\delta q_a(t)\right]|b',t\rangle.
\ee
Now the dynamical variables at different times are related by Hamilton's 
equations,
\bea
\frac{d p_a(t)}{dt}&=&\frac1i[p_a(t),H(q(t),p(t),t)]\nonumber\\
&=&-\frac{\partial H}{\partial q_a}(t),
\eea
so
\be
p_a(t+dt)-p_a(t)=dt\frac{dp_a(t)}{dt}=-dt\frac{\partial H}{\partial q_a}(t).
\ee
Similarly, the other Hamilton's equation
\be
\frac{dq_a}{dt}=\frac{\partial H}{\partial p_a}
\ee
implies that
\be
q_a(t+dt)-q_a(t)=dt\frac{\partial H}{\partial p_a}(t).
\ee
From this we deduce first the $q$ variation of the transformation function,
\bea
&&\delta_q\langle a',t+dt|b',t\rangle\nonumber\\
&=&i\langle a',t+dt|\sum_ap_a(t)[\delta
q_a(t+dt)-\delta q_a(t)]-dt\frac{\partial H}{\partial q_a}\delta q_a(t)
+{\cal O}(dt^2)|b',t\rangle\nonumber\\
&=&i\langle a',t+dt|\delta_q\left[\sum_ap_a(t)\mbox{.}[q_a(t+dt)-q_a(t)]
-dt\,H(q(t),p(t),t)\right]|b',t\rangle,\nonumber\\
\eea
where the dot denotes symmetric multiplication of the $p$ and $q$ operators.

For $p$ variations we have a similar result:
\bea
&&\delta_p\langle a',t+dt|b',t\rangle\nonumber\\
&=&-i\langle a',t+dt|\sum_a[q_a(t+dt)
\delta p_a(t+dt)-q_a(t)\delta p_a(t)]|b',t\rangle\nonumber\\
&=&-i\langle a',t+dt|\sum_a q_a(t)[\delta p_a(t+dt)-\delta p_a(t)]+dt\,
\frac{\partial H}{\partial p_a}(t)\delta p_a(t)|b',t\rangle\nonumber\\
&=&i\langle a',t+dt|\delta_p\left[-\sum_aq_a(t)\mbox{.}(p_a(t+dt)-p_a(t))-dt\,
H(q(t),p(t),t)\right]|b',t\rangle.\nonumber\\
\eea
That is, for $q$ variations
\begin{subequations}
\be
\delta_q\langle a',t+dt|b',t\rangle=i\langle a',t+dt|\delta_q\left[dt L_q
\right]|b',t\rangle,
\ee
with the quantum Lagrangian
\be
L_q=\sum_ap_a\mbox{.}\dot q_a-H(q,p,t),
\ee
\end{subequations}
while for $p$ variations
\begin{subequations}
\be
\delta_p\langle a',t+dt|b',t\rangle=i\langle a',t+dt|\delta_p\left[dt L_p
\right]|b',t\rangle,
\ee
with the quantum Lagrangian
\be
L_p=-\sum_aq_a\mbox{.}\dot p_a-H(q,p,t).
\ee
\end{subequations}
We see here two alternative forms of the quantum Lagrangian.  Note that the
two forms differ by a total time derivative,
\be
L_q-L_p=\frac{d}{dt}\sum_a p_a\mbox{.} q_a.
\ee

We now can unite the kinematic transformations considered here with the 
dynamic ones considered earlier, in Eq.~(\ref{dynvar}):
\be
\delta=\delta_{\rm dyn}+\delta_{\rm kin}:\quad
\delta\langle a',t+dt|b',dt\rangle=i\langle a',t+dt|\delta[dt\,L]|b',t\rangle.
\label{infvp}
\ee

Suppose, for concreteness, that our states are defined by values of $q$, so
that
\be
\delta_p\langle a',t+dt|b't\rangle=0.
\ee
This is consistent, as a result of Hamilton's equations,
\be
\delta_pL_q=\sum_a\delta p_a\left(\dot q_a-\frac{\partial H}{\partial p_a}
\right)=0.
\ee
In the following we will use $L_q$.

It is immediately clear that we can iterate the infinitesimal version
(\ref{infvp})
of the quantum action principle by inserting at each time step a complete
set of intermediate states (to simplify the notation, we ignore their quantum
numbers):
\be
\langle t_1|t_2\rangle=\langle t_1|t_1-dt\rangle\langle t_1-dt|t_1-2dt\rangle
\cdots\langle t_2+2dt|t_2+dt\rangle\langle t_2+dt|t_2\rangle,
\ee
So in this way we deduce the general form of {\em Schwinger's quantum
action principle:}
\be
\delta\langle t_1|t_2\rangle=i\langle t_1|\delta\int_{t_2}^{t_1}dt\, L
|t_2\rangle.
\label{qap}
\ee
This summarizes all the properties of the system.

Suppose the dynamical system is given, that is, the structure of $H$ does not
change.  Then
\be
\delta\langle t_1|t_2\rangle=i\langle t_1|G_1-G_2|t_2\rangle,
\ee
where the generator $G_a$ depends on $p$ and $q$ at time $t_a$.  Comparing
with the action principle (\ref{qap}) we see
\be
\delta\int_{t_2}^{t_1}dt\,L=G_1-G_2,
\ee
which has exactly the form of the classical action principle 
(\ref{8.2}), except that
the Lagrangian $L$ and the generators $G$ are now operators.
If no changes occur at the endpoints, we have the {\it principle of
stationary action},
\be
\delta\int_{t_2}^{t_1}\left(\sum_a p_a.dq_a-H\,dt\right)=0.
\ee
As in the classical case, let us introduce a time parameter $\tau$, 
$t=t(\tau)$, such that $\tau_2$ and $\tau_1$ are fixed.  Calling the new
time parameter by the original name,  the above
variation reads
\bea
&&\sum_a\left[\delta p_a.d q_a+p_a.d\delta q_a-\delta H\,dt-H\,d\delta t\right]
\nonumber\\
&=&d\left[\sum_ap_a.\delta q_a-H\,\delta t\right]
+\sum_a\left[\delta p_a.d q_a-dp_a.\delta q_a\right]-\delta H\,dt+dH\,\delta t,
\nn\\
\eea
so the action principle says
\begin{subequations}
\bea
G&=&\sum_ap_a.\delta q_a-H\,\delta t,\\
\delta H&=&\frac{dH}{dt}\delta t+\sum_a\left(\delta p_a.\frac{dq_a}{dt}
-\delta q_a.\frac{dp_a}{dt}\right).
\eea
\end{subequations}
We will again assume $\delta p_a$, $\delta q_a$ are not operators (that is,
they are proportional to the unit operator); then we recover Hamilton's
equations,
\begin{subequations}
\bea
\frac{\partial H}{\partial t}&=&\frac{dH}{dt},\\
\frac{\partial H}{\partial p_a}&=&\frac{dq_a}{dt},\\
\frac{\partial H}{\partial q_a}&=&-\frac{dp_a}{dt}.
\eea
\end{subequations}
(Schwinger also explored the possibility of operator variations, see, for
example, his les Houches lectures
\cite{leshouches}.)
We learn from the generators,
\be
G_t=-H\,\delta t,\quad G_q=\sum_ap_a\delta q_a,
\ee
that the change in some function $F$ of the dynamical variable is
\be
\delta F=\frac{dF}{dt}\delta t+\frac1i[F,G],
\ee
so we deduce
\begin{subequations}
\bea
\frac{dF}{dt}&=&\frac{\partial F}{\partial t}+\frac1i[F,H],\\
\frac{\partial F}{\partial q_a}&=&\frac1i[F,p_a].
\eea
\end{subequations}
Note that from this the canonical commutation relations follow,
\be
[q_a,p_b]=i\delta_{ab},\quad[p_a,p_b]=0,
\ee
as well as Newton's law,
\be
\dot p_a=-\frac1i[H,p_a]=-\frac{\partial H}{\partial q_a}.
\ee

If we had used $L_p$ instead of $L_q$, we would have obtained the same
equations of motion, but in place of $G_q$, we would have obtained
\be
G_p=-\sum_aq_a\delta p_a,
\ee
which implies
\be
\frac{\partial F}{\partial p_a}=-\frac1i[F,q_a].
\ee
From this can be deduced the remaining canonical commutator,
\be
[q_a,q_b]=0,
\ee
as well as the remaining Hamilton equation,
\be
\dot q_a=\frac1i[q_a,H]=\frac{\partial H}{\partial p_a}.
\ee
It is easy to show that the effect of changing the Lagrangian
by a total time derivative (which is what is done in passing from $L_q$
to $L_p$) is to change the generators.

We now turn to examples.

\section{Harmonic Oscillator}
The harmonic oscillator is defined in terms of creation and annihilation
operators,\footnote{We follow Schwinger's usage of $y$ for the annihilation
operator, instead of the more usual $a$.}
 $y^\dagger$ and $y$, and the corresponding Hamiltonian $H$,
\begin{subequations}
\bea
[y,y^\dagger]&=&1,\\
H&=&\omega\left(y^\dagger y+\frac12\right).
\eea
\end{subequations}
The equations of motion are
\begin{subequations}
\label{hoeom}
\bea
\frac{dy}{dt}&=&\frac1i[y,H]=\frac1i\omega y,\label{hoeoma}\\
\frac{dy^\dagger}{dt}&=&\frac1i[y^\dagger,H]=-\frac1i\omega y^\dagger.
\label{hoeomb}
\eea
\end{subequations}
Eigenstates of $y$ and $y^\dagger$ exist, as right and left vectors,
respectively,
\begin{subequations}
\bea
y|y'\rangle&=&y'|y'\rangle,\\
\langle y^{\dagger\prime}|y^\dagger&=&y^{\dagger\prime}\langle 
y^{\dagger\prime}|,
\eea
\end{subequations}
while $\langle y'|$ and $|y^{\dagger\prime}\rangle$ do not exist.\footnote{If
$\langle y'|y=y'\langle y'|$ then we would have an evident contradiction:
\be
1=\langle y'|[y,y^\dagger]|y'\rangle=y'\langle y'|y^\dagger|y'\rangle
-\langle y'|y^\dagger|y'\rangle y'=0.
\ee}  These are the famous ``coherent states,'' to whom the name
Roy Glauber \cite{glauber} is invariably attached, although they were discovered
by Erwin Schr\"odinger \cite{Schrodinger1926}, and Glauber's approach, as he 
acknowledged, followed that of his mentor, Schwinger \cite{Schwinger1953}.

The transformation function we seek is therefore
\be
\langle y^{\dagger\prime},t_1|y'',t_2\rangle.
\ee
If we regard $y$ as a ``coordinate,'' the corresponding ``momentum'' is
$iy^\dagger$:
\be
\dot y=\frac1i\omega y=\frac{\partial H}{\partial iy^\dagger},\quad
i\dot y^\dagger=-\omega y^\dagger=-\frac{\partial H}{\partial y}.
\ee
The corresponding Lagrangian is therefore\footnote{\label{fn4}
We might note that
in terms of (dimensionless) position and momentum operators
\be
i y^\dagger.\dot y=\frac{i}2(q-ip).(\dot q+i\dot p)=\frac12(p.\dot q
-q.\dot p)+\frac{i}4\frac{d}{dt}(q^2+p^2),
\ee
where the first term in the final form is the average of the Legendre
transforms in $L_q$ and $L_p$.}
\be
L=iy^\dagger.\dot y-H.
\ee
Because we use $y$ as our state variable at the initial time, and $y^\dagger$
at the final time, we must exploit our freedom to redefine our generators
to write
\be
W_{12}=\int_2^1dt\,L-iy^\dagger(t_1).y(t_1).
\ee
Then the variation of the action is
\bea
\delta W_{12}&=&-i\delta(y_1^\dagger.y_1)+G_1-G_2\nonumber\\
&=&-i\delta y^\dagger_1.y_1-iy^\dagger_1.\delta y_1+iy^\dagger_1.\delta y_1
-iy^\dagger_2.\delta y_2-H\,\delta t_1+H\,\delta t_2\nonumber\\
&=&-i\delta y^\dagger_1.y_1-iy_2^\dagger.\delta y_2-H(\delta t_1-\delta t_2).
\eea
Then the quantum action principle says
\be
\delta\langle y^{\dagger\prime},t_1|y'',t_2\rangle=i\langle y^{\dagger\prime}
,t_1|-i\delta y_1^{\dagger\prime}y_1-iy_2^\dagger\delta y_2''
-\omega y_1^{\dagger\prime}y_1(\delta t_1-\delta t_2)|y'',t_2\rangle,
\ee
since by assumption the variations in the dynamical variables are numerical:
\be
[\delta y_1^\dagger,y_1]=[y_2^\dagger,\delta y_2],
\ee
and we have dropped the zero-point energy.
Now use the equations of motion (\ref{hoeoma}) and (\ref{hoeomb}) 
to deduce that
\be
y_1=e^{-i\omega(t_1-t_2)}y_2,\quad y_2^\dagger=e^{-i\omega(t_1-t_2)}y_1^\dagger
\ee
and hence
\bea
\delta\langle y^{\dagger\prime},t_1|y'',t_2\rangle
&=&\langle y^{\dagger\prime},t_1|\delta y^{\dagger\prime}e^{-i\omega(t_1-t_2)}
y''+y^{\dagger\prime}e^{-i\omega(t_1-t_2)}\delta y''\nonumber\\
&&\quad\mbox{}-i\omega y^{\dagger\prime}
e^{-i\omega(t_1-t_2)}(\delta t_1-\delta t_2)y''|y'',t_2\rangle\nonumber\\
&=&\langle y^{\dagger\prime},t_1|y'',t_2\rangle\delta\left[y^{\dagger\prime}
e^{-i\omega(t_1-t_2)}y''\right].
\eea
From this we can deduce that the transformation function has the exponential
form
\be
\langle y^{\dagger\prime},t_1|y'',t_2\rangle=\exp\left[y^{\dagger\prime}
e^{-i\omega(t_1-t_2)}y''\right],
\label{hosoln}
\ee
which has the correct boundary condition at $t_1=t_2$; and in particular,
$\langle 0|0\rangle=1$.

On the other hand, 
\be
\langle y^{\dagger\prime},t_1|y'',t_2\rangle=\langle y^{\dagger\prime}|
e^{-iH(t_1-t_2)}|y''\rangle,
\ee
where both states are expressed at the common time $t_2$, so, upon
inserting a complete set of energy eigenstates, we obtain ($t=t_1-t_2$)
\be
\sum_E\langle y^{\dagger\prime}|E\rangle e^{-iEt}\langle E|y''\rangle,
\ee
which we compare to the Taylor expansion of the previous formula,
\be
\sum_{n=0}^\infty \frac{(y^{\dagger\prime})^n}{\sqrt{n!}}e^{-in\omega t}
\frac{(y'')^n}{\sqrt{n!}}.
\ee
This gives all the eigenvectors and eigenvalues:
\begin{subequations}
\bea
E_n&=&n\omega,\quad n=0,1,2,\dots,\\
\langle y^{\dagger\prime}|E_n\rangle&=&\frac{(y^{\dagger\prime})^n}{\sqrt{n!}},
\label{cohstton}\\
\langle E_n|y''\rangle&=&\frac{(y'')^n}{\sqrt{n!}}.
\eea
\end{subequations}
These correspond to the usual construction of the eigenstates from the
ground state:
\be
|E_n\rangle=\frac{(y^\dagger)^n}{\sqrt{n!}}|0\rangle.
\ee

\section{Forced Harmonic Oscillator}
\label{sec:chap2}
Now we add a driving term to the Hamiltonian,
\be
H=\omega y^\dagger y+yK^*(t)+y^\dagger K(t),\label{forcedho}
\ee
where $K(t)$ is an external force ({\it Kraft\/} is force in German).
The equation of motion is
\be
i\frac{dy}{dt}=\frac{\partial H}{\partial y^\dagger}=[y,H]=\omega y+K(t),
\label{eom:fho}
\ee
while $y^\dagger$ satisfies the adjoint equation.  In the presence of $K(t)$,
we wish to compute the transformation function
$\langle y^{\dagger\prime},t_1|y'',t_2\rangle^K$.

Consider a variation of $K$.  According to the action principle
\bea
\delta_K\langle y^{\dagger\prime},t_1|y'',t_2\rangle^K&=&
\langle y^{\dagger\prime},t_1|i\delta_KW_{12}|y'',t_2\rangle^K
\nonumber\\
&=&-i\langle y^{\dagger\prime},t_1|\int_{t_2}^{t_1}dt[\delta K y^\dagger
+\delta K^* y]|y'',t_2\rangle^K.
\label{de:fho}
\eea
We can solve this differential equation by noting that the equation of
motion (\ref{eom:fho}) can be rewritten as
\be
i\frac{d}{dt}\left[e^{i\omega t}y(t)\right]=e^{i\omega t}K(t),
\ee
which is integrated to read
\be
e^{i\omega t}y(t)-e^{i\omega t_2}y(t_2)=-i\int_{t_2}^t dt'\,e^{i\omega t'}
K(t'),
\ee
or
\be
y(t)=e^{-i\omega(t-t_2)}y_2-i\int_{t_2}^t dt'\,e^{-i\omega(t-t')}K(t'),
\label{first}
\ee
and the adjoint\footnote{The consistency of these two equations
follows from
\be
e^{i\omega t_1}y_1=e^{i\omega t_2}y_2-i\int_{t_2}^{t_1}dt'\,e^{i\omega t'}
K(t'),\ee
so that the adjoint of Eq.~(\ref{first}) is
\bea
[y(t)]^\dagger&=&e^{i\omega t}\left[e^{-i\omega t_1}y_1^\dagger-i\int_{t_2}
^{t_1}dt'\,e^{-i\omega t'}K^*(t')\right]+i\int_{t_2}^tdt'\,e^{-i\omega(t'-t)}
K^*(t')\nonumber\\
&=&e^{i\omega(t-t_1)}y_1^\dagger+i\int_{t_1}^tdt'\,e^{-i\omega(t'-t)}K^*(t'),
\eea
which is Eq.~(\ref{second}).}
\be
y^\dagger(t)=e^{-i\omega(t_1-t)}y^\dagger_1
-i\int_{t}^{t_1} dt'\,e^{-i\omega(t'-t)}K^*(t').
\label{second}
\ee
Thus our differential equation (\ref{de:fho}) reads
\bea
&&\frac{\delta_K\langle y^{\dagger\prime},t_1|y'',t_2\rangle^K}
{\langle y^{\dagger\prime},t_1|y'',t_2\rangle^K}=
\delta_K\ln\langle y^{\dagger\prime},t_1|y'',t_2\rangle^K
\nonumber\\
&&\qquad=-i\int_{t_2}^{t_1}dt\,\delta K(t)\left[y^{\dagger\prime}
e^{-i\omega(t_1-t)}
-i\int_t^{t_1}dt'\,e^{-i\omega(t'-t)}K^*(t')\right]\nonumber\\
&&\qquad\quad\mbox{}-i\int_{t_2}^{t_1}dt\,\delta K^*(t)
\left[e^{-i\omega(t-t_2)}y''
-i\int_{t_2}^{t}dt'\,e^{-i\omega(t-t')}K(t')\right].\nn\\
\eea
Notice that in the terms bilinear in $K$ and $K^*$, $K$ always occurs 
earlier than $K^*$.  Therefore, these terms can be combined to read
\be
-\delta_K\int_{t_2}^{t_1}dt\,dt'\,K^*(t)\eta(t-t')e^{-i\omega(t-t')}K(t'),
\ee
where the step function is
\be
\eta(t)=\left\{\begin{array}{cc}
1,&t>0,\\
0,&t<0.
\end{array}\right.
\ee
Since we already know the $K=0$ value from Eq.~(\ref{hosoln}),
we may now immediately integrate our differential equation:
\bea
\langle y^{\dagger\prime},t_1|y'',t_2\rangle^K&=&\exp\bigg[y^{\dagger\prime}
e^{-i\omega(t_1-t_2)}y''\nonumber\\
&&\quad\mbox{}-iy^{\dagger\prime}\int_{t_2}^{t_1}dt\,e^{-i\omega(t_1-t)}K(t)
-i\int_{t_2}^{t_1}dt \,e^{-i\omega(t-t_2)}K^*(t)\,y''\nonumber\\
&&\quad\mbox{}-\int_{t_2}^{t_1}dt\,dt'\,K^*(t)\eta(t-t')e^{-i\omega(t-t')}
K(t')\bigg].
\label{fhosln}
\eea
The ground state is defined by $y''=y^{\dagger\prime}=0$, so
\be
\langle 0,t_1|0,t_2\rangle^K=\exp\left[-\int_{-\infty}^\infty dt\,dt'\,
K^*(t)\eta(t-t')e^{-i\omega(t-t')}K(t')\right],
\label{vpax}
\ee
where we now suppose that the forces turn off at the initial and final times,
$t_2$ and $t_1$, respectively.

A check of this result is obtained by computing the probability of the system
remaining in the ground state:
\bea
|\langle 0,t_1|0,t_2\rangle^K|^2&=&\exp\bigg\{-\int_{-\infty}^\infty dt\,dt'\,
K^*(t)e^{-i\omega(t-t')}\nn\\
&&\qquad\times[\eta(t-t')+\eta(t'-t)]K(t')\bigg\}\nonumber\\
&=&\exp\left[-\int_{-\infty}^\infty
 dt\,dt'\,K^*(t)e^{-i\omega(t-t')}K(t')\right]\nonumber\\
&=&\exp\left[-|K(\omega)|^2\right],
\eea
where the Fourier transform of the force is
\be
K(\omega)=\int_{-\infty}^\infty dt\,e^{i\omega t}K(t).
\ee
The probability requirement
\be
|\langle 0,t_1|0,t_2\rangle^K|^2\le1
\ee
is thus satisfied.  We see here a {\em resonance\/} effect: 
If the oscillator is
driven close to its natural frequency, so $K(\omega)$ is
large, there is a large probability of finding the system in an excited
state, and therefore of not remaining in the ground state.
Let us calculate this transition amplitude to an excited state.  By
setting $y''=0$ in Eq.~(\ref{fhosln}) we obtain
\bea
\langle y^{\dagger\prime},t_1|0,t_2\rangle^K&=&\exp\left[-iy^{\dagger\prime}
\int_{-\infty}^\infty dt\,e^{-i\omega(t_1-t)}K(t)\right]\langle 0,t_1|
0,t_2\rangle^K\nonumber\\
&=&\sum_n \langle y^{\dagger\prime},t_1|n,t_1\rangle\langle n,t_1|0,t_2
\rangle^K,
\label{vactocoh}
\eea
where we have inserted a sum over a complete set of energy eigenstates,
which possess the amplitude [see Eq.~(\ref{cohstton})]
\be
\langle y^{\dagger\prime}|n\rangle=\frac{(y^{\dagger\prime})^n}{\sqrt{n!}}.
\ee
If we expand the first line of Eq.~(\ref{vactocoh}) in powers of
 $y^{\dagger\prime}$, we find
\be
\langle n,t_1|0,t_2\rangle^K=\frac{(-i)^n}{\sqrt{n!}}e^{-in\omega t_1}
[K(\omega)]^n\langle0,t_1|0,t_2\rangle^K.
\ee
The corresponding probability is
\be
p(n,0)^K=|\langle n,t_1|0,t_2\rangle^K|^2=\frac{|K(\omega)|^{2n}}{n!}
e^{-|K(\omega)|^2},\label{pnok}
\ee
which is a Poisson distribution\footnote{A Poisson probability distribution
has the form
$p(n)=\lambda^ne^{-\lambda}/n!$. The mean value of $n$ for this distribution
is
\bea
\bar n&=&\sum_{n=0}^\infty n\,p(n)=\sum_{n=0}^\infty\frac{\lambda^n 
e^{-\lambda}}{(n-1)!}=\lambda\sum_{n=0}^\infty p(n)=\lambda.
\eea}
with mean $\bar n=|K(\omega)|^2$.

Finally, let us define the {\it Green's function\/} for this problem by
\be
G(t-t')=-i\eta(t-t')e^{-i\omega(t-t')}.
\label{hogf}
\ee
It satisfies the differential equation
\be
\left(i\frac{d}{dt}-\omega\right)G(t-t')=\delta(t-t'),\label{diffeq:gf}
\ee
as it must because [see Eq.~(\ref{eom:fho})]
\be
\left(i\frac{d}{dt}-\omega\right)y(t)=K(t),
\ee
where $y(t)$ is given by [see Eq.~(\ref{first})]
\be
y(t)=e^{-i\omega(t-t_2)}y_2+\int_{-\infty}^\infty dt'\,G(t-t')K(t').
\label{yoft}
\ee
Similarly, from Eq.~(\ref{second})
\be
y^\dagger(t)=e^{-i\omega(t_1-t)}y_1^\dagger
+\int_{-\infty}^\infty dt'\,G(t'-t)K^*(t').
\ee
We can now write the ground-state persistence amplitude (\ref{vpa}) as
\be
\langle 0,t_1|0,t_2\rangle^K=\exp\left[-i\int_{-\infty}^\infty dt\,dt'\,
K^*(t)G(t-t')K(t')\right],
\label{gspa}
\ee
and the general amplitude (\ref{fhosln}) as
\bea
\langle y^{\dagger\prime},t_1|y'',t_2\rangle^K&=&
\exp\bigg\{-i\int_{-\infty}^\infty dt\,dt'\left[K^*(t)+iy^{\dagger\prime}
\delta(t-t_1)\right]\nonumber\\
&&\quad\times G(t-t')\left[K(t')+iy''\delta(t'-t_2)\right]\bigg\},
\eea
which demonstrates that knowledge of $\langle 0,t_1|0,t_2\rangle^K$
for all $K$ determines everything:
\be
\langle y^{\dagger\prime},t_1|y'',t_2\rangle^K=\langle 0,t_1|0,t_2
\rangle^{K(t)+iy''\delta(t-t_2)+iy^{\dagger\prime}\delta(t-t_1)}.
\ee

\section{Feynman Path Integral Formulation}

Although much more familiar, the path integral formulation of quantum
mechanics \cite{feynman,feynmanst,Feynman1965}
 is rather vaguely defined.  We will here provide a formal 
``derivation'' based on the Schwinger principle, in the harmonic oscillator
context.  

Consider a forced oscillator, defined by the Lagrangian
(note in this section, $H$ does not include the source terms)
\be
L=iy^\dagger.\dot y-H(y,y^\dagger)-Ky^\dagger-K^*y.\label{holagrange}
\ee
As in the preceding section, the action principle says
\be
\delta_K\langle 0,t_1|0,t_2\rangle^K=-i\langle 0,t_1|\int_{t_2}^{t_1}dt\,
[\delta Ky^\dagger+\delta K^*y]|0,t_2\rangle^K,
\ee
or for $t_2<t<t_1$,
\begin{subequations}
\bea
i\frac{\delta}{\delta K(t)}\langle 0,t_1|0,t_2\rangle^K=\langle 0,t_1|
y^\dagger(t)|0,t_2\rangle^K,\\
i\frac{\delta}{\delta K^*(t)}\langle 0,t_1|0,t_2\rangle^K=\langle 0,t_1|
y(t)|0,t_2\rangle^K,
\eea
\end{subequations}
where we have introduced the concept of the functional derivative.
The equation of motion
\be
i\dot y-\frac{\partial H}{\partial y^\dagger}-K=0,
\quad -i\dot y^\dagger-\frac{\partial H}{\partial y}-K^*=0,
\label{rep:eom}
\ee
is thus equivalent to the functional differential equation,
\be
0=\left\{i\left[K(t),W\left[i\frac{\delta}{\delta K^*},i\frac{\delta}{\delta K}
\right]\right]-K(t)\right\}\langle0,t_1|0,t_2\rangle^K,
\label{fnalde}
\ee
where (the square brackets indicate functional dependence)
\be
W[y,y^\dagger]=\int_{t_2}^{t_1}dt\,
[iy^\dagger(t).\dot y(t)-H(y(t),y^\dagger(t))].
\ee
The reason Eq.~(\ref{fnalde}) holds is that by definition
\be
\frac{\delta}{\delta K(t)}K(t')=\delta(t-t'),
\ee
so
\bea
&&i\left[K(t),\int_{t_2}^{t_1}dt'\left(i\frac{i\delta}{\delta K(t')}.
\frac{d}{dt'}
\frac{i\delta}{\delta K^*(t')}-H\left(\frac{i\delta}{\delta K^*(t')},
\frac{i\delta}{\delta K(t')}\right)\right)\right]\nonumber\\ 
&&\qquad=i\frac{d}{dt}\frac{i\delta}{\delta K^*(t)}-\frac\partial{\partial
(i\delta /\delta K(t))}H\left(\frac{i\delta}{\delta K^*(t)},\frac{i\delta}
{\delta K(t)}\right),
\eea
which corresponds to the first two terms in the equation of motion
(\ref{rep:eom}), under
the correspondence
\be
y\leftrightarrow i\frac\delta{\delta K^*},\quad y^\dagger\leftrightarrow
i\frac\delta{\delta K}.
\ee
Since $[[K,W],W]=0$, we can write the functional equation (\ref{fnalde}) as
\be
0=e^{iW[i\delta/\delta K^*,i\delta/\delta K]}K
e^{-iW[i\delta/\delta K^*,i\delta/\delta K]}\langle0,t_1|0,t_2\rangle^K.
\ee
The above equation has a solution (up to a constant), because both
equations (\ref{rep:eom}) must hold,
\be
\langle0,t_1|0,t_2\rangle^K=e^{iW[i\delta/\delta K^*,i\delta/\delta K]}
\delta[K]\delta[K^*],
\ee
where $\delta[K]$, $\delta[K^*]$ are functional delta functions.  
The latter have functional Fourier decompositions (up to a multiplicative
constant),
\begin{subequations}
\bea
\delta[K]&=&\int[dy^\dagger]e^{-i\int dt \,K(t)y^\dagger(t)},\\
\delta[K^*]&=&\int[dy]e^{-i\int dt \,K^*(t)y(t)},
\eea
\end{subequations}
where $[dy]$ represents an element of integration over all (numerical-valued)
{\it functions\/} $y(t)$,
and so we finally have
\bea
&&\langle0,t_1|0,t_2\rangle^{K,K^*}\nonumber\\
&&\quad=\int[dy][dy^\dagger]\exp\left(-i\int_{t_2}
^{t_1}dt\left[K(t)y^\dagger (t)+K^*(t)y(t)\right]+iW[y,y^\dagger]\right)
\nonumber\\
&&\quad=
\int[dy][dy^\dagger]\exp\left(i\int_{t_2}^{t_1}dt\left[iy^\dagger \dot y
-H(y,y^\dagger)-Ky^\dagger-K^*y\right]\right),
\label{hopathint}
\eea
where $y$, $y^\dagger$ are now numerical, and the functional integration is
over all possible functions, over all possible ``paths.''  Of course, the
classical paths, the ones for which $W-\int dt(Ky^\dagger+K^*y)$ is
an extremum, receive the greatest weight, at least in the classical limit,
where $\hbar\to0$.

\subsection{Example} 
\label{sec:hogf}
Consider the harmonic oscillator Hamiltonian,
$H=\omega y^\dagger y$.  Suppose we wish to calculate, once again, the
ground state persistence amplitude, $\langle 0,t_1|0,t_2\rangle^K$.
It is perhaps easiest to perform a Fourier transform,
\be
y(\nu)=\int_{-\infty}^\infty dt\, e^{i\nu t}y(t),\quad
 y^*(-\nu)=\int_{-\infty}^\infty dt\, e^{-i\nu t}y^\dagger(t).
\ee
Then
\begin{subequations}
\bea
\int_{-\infty}^\infty dt\,y^\dagger(t)y(t)&=&\int_{-\infty}^\infty\frac{d\nu}
{2\pi}y(\nu)y^*(-\nu),\\
\int_{-\infty}^\infty dt\,iy^\dagger(t)\dot y(t)
&=&\int_{-\infty}^\infty\frac{d\nu}{2\pi}\nu y(\nu)y^*(-\nu).
\eea
\end{subequations}
Thus Eq.~(\ref{hopathint}) becomes
\bea
&&\langle0,t_1|0,t_2\rangle^{K,K^*}\nonumber\\
&=&\int[dy][dy^*]\exp\bigg\{i\int
\frac{d\nu}{2\pi}[y(\nu)(\nu-\omega)y^*(-\nu)\nn\\
&&\qquad\mbox{}-y^*(-\nu)K(\nu)
-y(\nu)K^*(-\nu)]\bigg\}\nonumber\\
&=&\int[dy][dy^*]\exp\bigg\{i\int\frac{d\nu}{2\pi}\left[y(\nu)-\frac{K(\nu)}
{\nu-\omega}\right](\nu-\omega)\left[y^*(-\nu)-\frac{K^*(-\nu)}{\nu-\omega}
\right]\nonumber\\
&&\qquad\mbox{}-i\int\frac{d\nu}{2\pi}K(\nu)\frac1{\nu-\omega}K^*(-\nu)\bigg\}
\nonumber\\
&=&\int[dy][dy^*]\exp\left\{i\int\frac{d\nu}{2\pi}y(\nu)(\nu-\omega)y^*(-\nu)
\right\}\nonumber\\
&&\qquad\times
\exp\left\{-i\int\frac{d\nu}{2\pi}K(\nu)\frac1{\nu-\omega}K^*(-\nu)
\right\}
\nonumber
\\
&=&\exp\left\{-i\int\frac{d\nu}{2\pi}K(\nu)\frac1{\nu-\omega}K^*(-\nu)\right\},
\eea
since the functional integral in the third equality,
obtained by shifting the integration variable,
\begin{subequations}
\bea
y(\nu)-\frac{K(\nu)}{\nu-\omega}&\to& y(\nu),\\
y^*(-\nu)-\frac{K^*(-\nu)}{\nu-\omega}&\to& y^*(-\nu),
\eea
\end{subequations}
 is $\langle0,t_1|0,t_2\rangle^{K=K^*=0}=1$.
How do we interpret the singularity
at $\nu=\omega$ in the remaining integral?  We should
have inserted a convergence factor in the original functional integral:
\be
\exp\left(i\int\frac{d\nu}{2\pi}\left[\dots\right]\right)\to\exp\left(
i\int\frac{d\nu}{2\pi}\left[\dots+i\epsilon y(\nu)y^*(-\nu)\right]\right),
\ee
where $\epsilon$ goes to zero through positive values.  Thus we have, in 
effect, $\nu-\omega\to\nu-\omega+i\epsilon$ and so we have for the ground-state
persistence amplitude
\be
\langle0,t_1|0,t_2\rangle^{K,K^*}=e^{-i\int dt\,dt'\,K^*(t)G(t-t')K(t')},
\label{hovpa}
\ee
which has the form of Eq.~(\ref{gspa}), with
\be
G(t-t')=\int_{-\infty}^\infty \frac{d\nu}{2\pi}\frac{e^{-i\nu(t-t')}}{\nu
-\omega+i\epsilon},
\ee
which is evaluated by closing the $\nu$ contour in the upper half plane
if $t-t'<0$, and in the lower half plane when $t-t'>0$.  
Since the pole is in the lower half plane we get
\be
G(t-t')=-i\eta(t-t')e^{-i\omega(t-t')},
\ee
which is exactly what we found in Eq.~(\ref{hogf}).

Now, let us rewrite the path integral
(\ref{hopathint}) in terms of co\"ordinates and momenta:
\begin{subequations}
\bea
q&=&\frac1{\sqrt{2\omega}}(y+y^\dagger),\quad p=\sqrt{\frac\omega2}\frac1i
(y-y^\dagger),\\
y&=&\sqrt{\frac\omega2}\left(q+\frac{ip}\omega\right),\quad
y^\dagger=\sqrt{\frac\omega2}\left(q-\frac{ip}\omega\right).
\eea
\end{subequations}
Then the numerical Lagrangian appearing in (\ref{hopathint})
 may be rewritten as (see footnote \ref{fn4} above)
\bea
L&=&iy^\dagger \dot y-\omega y^\dagger y-Ky^\dagger-K^*y\nonumber\\
&=&i\frac\omega2\left(q-i\frac{p}\omega\right)\left(\dot q+i\frac{\dot p}\omega
\right)-\frac{\omega^2}2\left(q^2+\frac{p^2}{\omega^2}\right)
\nonumber\\
&&\qquad\mbox{}
-\sqrt{\frac\omega2}K\left(q-\frac{ip}\omega\right)
-\sqrt{\frac\omega2}K^*\left(q+\frac{ip}\omega\right)\nonumber\\
&=&i\frac\omega4\frac{d}{dt}\left(q^2+\frac{p^2}{\omega^2}\right)+p\dot q
-\frac12\frac{d}{dt}(pq)-\frac12(p^2+\omega^2q^2)\nn\\
&&\qquad\mbox{}-\Re Kq
-\sqrt{\frac2\omega}\Im Kp\nonumber\\
&=&\frac{d}{dt}w+L(q,\dot q,t),
\label{transholag}
\eea
where, if we set $\dot q=p$, the Lagrangian is
\be
L(q,\dot q,t)=\frac12\dot q^2-\frac12\omega^2q^2+Fq,
\label{tradlag}
\ee
if
\be
\Im K=0,\quad F=-\sqrt{2\omega}\Re K.
\ee
In the path integral
\be
[dy][dy^\dagger]
=[dq][dp]\left|\frac{\partial(y,y^\dagger)}{\partial(q,p)}\right|,
\ee
where the Jacobian is
\be
\left|\frac{\partial(y,y^\dagger)}{\partial(q,p)}\right|
=\left|\begin{array}{cc}
\sqrt{\frac\omega2}\quad&\sqrt{\frac\omega2}\\
\\
\frac{i}{\sqrt{2\omega}}\quad&-\frac{i}{\sqrt{2\omega}}\end{array}
\right|=1,
\ee
and so from the penultimate line of Eq.~(\ref{transholag}), the path integral
(\ref{hopathint}) becomes
\bea
&&\langle 0,t_1|0,t_2\rangle^F
=\int[dy][dy^\dagger]
\exp\left[i\int_{t_2}^{t_1}dt\,L(y,y^\dagger)\right]\nonumber\\
&&\quad=\int[dq][dp]\exp\left[i\int_{t_2}^{t_1}dt\left(p\dot q-\frac12p^2
-\frac12\omega^2q^2+Fq\right)\right].
\eea
Now we can carry out the $p$ integration, since it is Gaussian:
\bea
\int[dp]e^{i\int dt\left[-\frac12p^2+p\dot q\right]}
&=&\int[dp]e^{i\int dt\left[-\frac12(p-\dot q)^2+\frac12\dot q^2\right]}
\nonumber\\
&=&e^{i\int dt\frac12\dot q^2}\prod_i\int_{-\infty}^\infty dp_i \,e^{-\frac12
ip_i^2\Delta t}.
\eea
Here we have discretized time so that $p(t_i)=p_i$, so the final functional
integral over $p$ is just an infinite product of constants, each one of which
equals $e^{-i\pi/4}\sqrt{2\pi/\Delta t}$.  
Thus we arrive at the form originally
written down by Feynman \cite{Feynman1965},
\be
\langle 0,t_1|0,t_2\rangle^F=\int[dq]\exp\left\{i\int_{t_2}^{t_1}dt\,
L(q,\dot q,t)\right\},
\label{feynpathint}
\ee
with the Lagrangian given by Eq.~(\ref{tradlag}),
where an infinite normalization constant has been absorbed into the measure.

\section{Toward Source Theory}
Let us return to the action principle.  Recall from Eq.~(\ref{hovpa})
\be
\langle 0t_1|0t_2\rangle^K=e^{-i\int dt\,dt'K^*(t)G(t-t')K(t')}.\label{vpa}
\ee
The action principle says
\be
\delta \langle t_1|t_2\rangle=i\langle t_1|\delta[W_1=\int dt\,L]|t_2\rangle.
\ee
In a general sense, the exponent in Eq.~(\ref{vpa}) is an integrated form
of the action.  In solving the equation of motion, we found in Eq.~(\ref{yoft})
\be
y(t)=e^{-i\omega(t-t_2)}y(t_2)+\int dt' G(t-t')K(t'),
\ee
where the first term is effectively zero here.  The net effect is to replace
an operator by a number:
\be
y'(t)=\int dt'G(t-t')K(t').
\ee
Then Eq.~(\ref{vpa}) can be written as
\be
\langle 0t_1|0t_2\rangle^K=e^{-i\int dt\,K^*(t)y'(t)}.\label{vpap}
\ee
Recall that the action was was the integral of the
Lagrangian (\ref{holagrange}), or
\be
W=\int dt\left[y^\dagger i\frac\partial{\partial t}y-\omega y^\dagger y
-y^\dagger K(t)-yK^*(t)\right],
\ee
so we see one term in Eq.~(\ref{vpap}) here, and the equation of motion 
(\ref{eom:fho}) cancels
out the rest!  So let's add something which gives the equation for $y'$:
\be
\langle 0t_1|0t_2\rangle^K=e^{i\int dt\left[y^{\dagger\prime}i\frac{d}{dt}y'
-\omega y^{\dagger\prime}y'-y^{\dagger\prime}K-y'K^*\right]}=e^{iW}.
\ee
Now insist that $W$ is stationary with respect to variations of $y'$, 
$y^{\dagger\prime}$, and we recover the equation of motion,
\be
\left(i\frac{d}{dt}-\omega\right)y'(t)=K(t).
\ee
This is the starting point for the development of source theory, which will
be treated in Chap.~\ref{ST}.

\chapter{Time-cycle or Schwinger-Keldysh formulation}
\label{sec:5}

A further utility of the action principle is the time-cycle or 
Schwinger-Keldysh formalism, which allows one to calculate matrix elements
and consider nonequilibrium systems.  Schwinger's original work on this
was his famous paper \cite{Schwinger1961}; Keldysh's paper appeared three
years later \cite{Keldysh1964}, and, rather mysteriously, 
cites the Martin-Schwinger equilibrium
paper \cite{Martin1959}, but not the nonequilibrium one \cite{Schwinger1961}.
The following was extracted from notes from Schwinger's lectures given in
1968 at Harvard, as taken by the author.

Consider the expectation value of some physical property $F(t)$ at a particular
time $t_1$ in a state $|b,t_2\rangle$:
\be
\langle F(t_1)\rangle_{b't_2}=\sum_{a'a''}\langle b't_2|a't_1\rangle
\langle a'|F|a''\rangle \langle a''t_1|b't_2\rangle,
\ee
which expresses the expectation value in terms of the matrix elements of the
operator $F$ in a complete set of states defined at time $t_1$, 
$\{|a't_1\rangle\}$.  Suppose the operator $F$ has no explicit time dependence.
Then we can use the action principle to write
\begin{subequations}
\be
\delta \langle a't_1|b't_2\rangle=i\langle a't_1|\delta\left[\int_{t_2}^{t_1}
dt\,L\right]|b't_2\rangle, 
\ee
and so
\be
\delta \langle b't_2|a't_1\rangle=-i\langle b't_2|\delta\left[\int_{t_2}^{t_1}
dt\,L\right]|a't_1\rangle, 
\ee
\end{subequations}
which can be obtained from the first equation by merely exchanging labels,
\be
\int_{t_2}^{t_1}=-\int_{t_1}^{t_2}.
\ee
If we consider
\be
\langle b't_2|b't_2\rangle=\sum_{a'}\langle b't_2|a't_1\rangle\langle a't_1|
b't_2\rangle,
\ee
the above variational equations indeed asserts that
\be
\delta \langle b't_2|b't_2\rangle=0.
\ee

\begin{figure}
\centering
\includegraphics{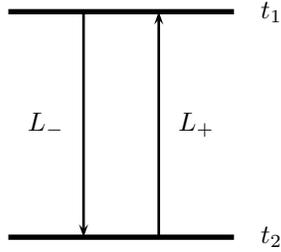}
\caption{\label{fig:tc1} A ``time-cycle,'' in which a system advances forward
in time from time $t_2$ to time $t_1$ under the influence of a Lagrangian
$L_+$, and then backward in time from time $t_1$ back to time $t_2$
under the influence of Lagrangian $L_-$.}
\end{figure}

We can interpret the above as a cycle in time, going from time $t_2$ to $t_1$
and then back again, as shown in Fig,~\ref{fig:tc1}.
  But, now imagine that the dynamics is different on the
forward and return trips, described by different Lagrangians $L_+$ and $L_-$.
Then
\be
\delta\langle b' t_2|b't_2\rangle=i\langle b't_2|\delta\left[\int_{t_2}^{t_1}
dt\,L_+-\int_{t_2}^{t_1}dt \,L_-\right]|b't_2\rangle.  
\ee
In particular, consider a perturbation of the form,
\be H=H_0+\lambda(t)F,\ee
where $\lambda(t)$ is some time-varying parameter.   If
we have an infinitesimal change, and, for example, $\delta
\lambda_+\ne0$, $\delta\lambda_-=0$, then
\be
\delta_{\lambda_+}\langle b't_2|b't_2\rangle^{\lambda_+\lambda_-}
=-i\langle b't_2|\int_{t_2}^{t_1} dt\,\delta \lambda_+F|b' t_2\rangle.
\ee
If we choose  $\delta\lambda_+$ to be an impulse,
\be
\delta\lambda_+=\delta \lambda\delta(t-t'), 
\ee
in this way we obtain the expectation value of $F(t')$.

\begin{figure}
\centering
\includegraphics{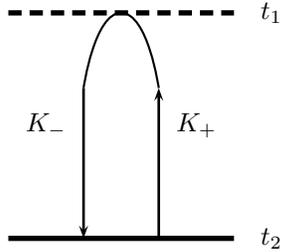}
\caption{\label{fig:tc2} A time cycle in which the harmonic oscillator
evolves from time $t_2$ to time $t_1$ under the influence of a force
$K_+$, and then from $t_1$ back to time $t_2$ under a force $K_-$.}
\end{figure}

Let's illustrate this with a driven harmonic oscillator, as described
by Eq.~(\ref{forcedho}), so now
\begin{subequations}
\bea
H_+&=&\omega y^\dagger y+K_+^*(t)y+K_+(t)y^\dagger,\\
H_-&=&\omega y^\dagger y+K_-^*(t)y+K_-(t)y^\dagger,
\eea
\end{subequations}
which describes the oscillator evolving forward in time from $t_2$ to $t_1$ 
under the influence of the force $K_+$, and backward in time from $t_1$
to $t_2$ under the influence of $K_-$, as shown in Fig.~\ref{fig:tc2}.
From the variational principle we can learn all about $y$ and 
$y^\dagger$.  We have already solved this problem by a more laborious method
above, in Section \ref{sec:chap2}.

It suffices to solve this problem with initial and final ground states.
If we consider only a $K^*$ variation,
\be
\delta_{K^*}\langle 0t_2|0t_2\rangle^{K_+,K_-}=-i\langle 0t_2|\int_{t_2}^{t_1}
dt\left[\delta K^*_+(t)y_+(t)-\delta K_-^*(t)y_-(t)\right]|0t_2\rangle.
\label{veho}
\ee
Now we must solve the equations of motion, so since effectively $y(t_2)\to0$, 
we have from Eq.~(\ref{first}), 
\begin{subequations}
\bea
y_+(t)&=&-i\int_{t_2}^{t} dt'\,e^{-i\omega(t-t')}K_+(t'),\\
y_-(t)&=&-i\int_{t_2}^{t_1} dt'\,e^{-i\omega(t-t')}K_+(t')
-i\int_{t_1}^{t}dt'\,e^{-i\omega(t-t')}K_-(t').\label{yminust}
\eea
\end{subequations}
The last term in the second equation is
\be
i \int_{t_2}^{t_1}dt'\,e^{-i\omega(t-t')}K_-(t')\eta(t'-t),
\ee
so naming the advanced and retarded Green's functions by extending
the definition in Eq.~(\ref{hogf}),
\be
G_{a,r}(t,t')=i e^{-i\omega(t-t')}\left\{\begin{array}{c}
\eta(t'-t)\\ -\eta(t-t')\end{array}\right\},
\ee
which satisfy the same differential equation (\ref{diffeq:gf}),
we effectively have
\begin{subequations}
\bea
y_+(t)&=&\int_{t_2}^{t_1}dt'\,G_r(t-t')K_+(t'),\\
y_-(t)&=&-i\int_{t_2}^{t_1} dt'\,e^{-i\omega(t-t')}K_+(t)
+\int_{t_2}^{t_1}dt'\,G_a(t-t')K_-(t'),
\eea
\end{subequations}
The solution to the variational equation (\ref{veho}) is now
\bea
\langle 0t_2|0t_2\rangle^{K_+,K_-}&=&
e^{-i\int dt\,dt'K_+^*(t)G_r(t-t')K_+(t')}
\nn\\
&&\times e^{i\int dt\,dt'K_-^*(t)G_a(t-t')K_-(t')}
e^{\int dt\,dt'K_-^*(t)e^{-i\omega(t-t')}K_+(t')}.\nn\\
\label{tcf}
\eea
This should reduce to 1 when $K_+=K_-=K$, so
\be
-iG_r(t-t')+iG_a(t-t')+e^{-i\omega(t-t')}=0,\label{identity}
\ee
which is, indeed, true.

\begin{figure}
\centering
\includegraphics{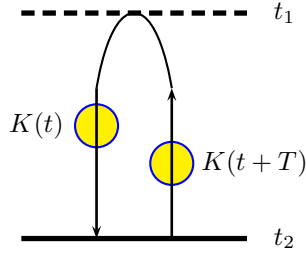}
\caption{\label{fig:tc3} Time cycle in which $K_-(t)=K(t)$, 
$K_+(t)=K(t+T)$, that is, the
forces are the same on the two legs, but displaced in time.}
\end{figure}

As an example, consider $K_-(t)=K(t)$, $K_+(t)=K(t+T)$, that is, the second
source is displaced forward by a time $T$.  
This is sketched in Fig.~\ref{fig:tc3}.
What does this mean?  From a
causal analysis, in terms of energy eigenstates, reading from right to left,
\be
\langle 0t_2|0t_2\rangle^{K_-,K_+}=\sum_n\langle 0t_2|nt_1\rangle^{K_-=K(t)}
\langle nt_1|0t_2\rangle^{K_+=K(t+T)}.
\ee
The effect on the second transformation function
is the same as moving the $n,t_1$ state to a later time,
\be
\langle nt_1|0t_2\rangle^{K(t+T)}=\langle n t_1+T|0t_2\rangle^{K(t)}
=e^{-in\omega T}\langle nt_1|0t_2\rangle^{K(t)},
\ee
so this says that
\be
\langle 0t_2|0t_2\rangle^{K_-K_+}=\sum_n e^{-in\omega T}p(n,0)^K,
\ee
which gives us the probabilities directly.  From the formula (\ref{tcf}) 
we have, using Eq.~(\ref{identity}), 
\bea
\langle 0t_2|0t_2\rangle^{K_-K_+}&=&
e^{\int dt\,dt'\,K^*(t)e^{-i\omega(t-t')}[K(t'+T)-K(t')]}\nn\\
&=&e^{\int dt\,dt'\,K^*(t)e^{-i\omega(t-t')}[e^{-i\omega T}-1]K(t')]}\nn\\
&=&e^{|\gamma|^2\left(e^{-i\omega T}-1\right)},\label{gfsimple}
\eea
where 
\be
\gamma=\int dt\,e^{i\omega t}K(t).
\ee
Thus we immediately obtain Eq.~(\ref{pnok}), or
\be
p(n,0)^K=e^{-|\gamma|^2}\frac{(|\gamma|^2)^n}{n!}.
\ee

The above Eq.~(\ref{gfsimple})
can be directly used to find certain average values.  For example,
\be
\langle e^{-in\omega T}\rangle^K_0=e^{|\gamma|^2\left(e^{-i\omega T}-1\right)}.
\ee
Expand this for small $\omega T$ and we find
\be
\langle n\rangle_0^K=|\gamma|^2.\label{meanofn}
\ee
In a bit more systematic way we obtain the dispersion:
\be
\langle e^{-i(n-\langle n\rangle)\omega T}\rangle=e^{|\gamma|^2(e^{-i\omega T}
-1+i\omega T)}.\label{expinwt}
\ee
Expanding this to second order in $\omega T$ we get
\be
\langle(n-\langle n\rangle)^2\rangle=\langle n^2\rangle-\langle n\rangle^2
\equiv (\Delta n)^2=|\gamma|^2=\langle n\rangle,\label{nminusn2}
\ee
or
\be
\frac{\Delta n}{\langle n\rangle}=\frac1{\sqrt{\langle n\rangle}}.
\ee
For large quantum numbers, which corresponds to the classical limit, the 
fluctuations become relatively small.

\begin{figure}
\centering
\includegraphics{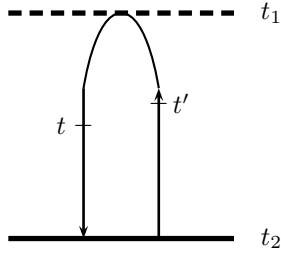}
\caption{\label{fig:tc4} Variational derivatives pick out operators
at definite times $t$ and $t'$.}
\end{figure}

Now consider a more general variational statement than in Eq.~(\ref{veho}),
\be
\delta\langle\,\,|\,\,\rangle^{K_-K_+}=
-i\langle\,\,|\int dt[\delta K_+^*y_+-\delta K_-^*y_-
+\delta K_+y_+^\dagger-\delta K_-y_-^\dagger|\,\,\rangle^{K_\pm},
\ee
%where the $\dots$ signify the omission of the other source variations, 
we see
that since we can change the source functions at will, and make very localized
changes,  it makes sense to define the variational derivatives
\begin{subequations}
\bea
&&i\frac\delta{\delta K_+^*(t)}\langle\,\,|\,\,\rangle^{K_\pm}=\langle\,\,|y_+(t)
|\,\,\rangle^{K_\pm},\\
&&-i\frac\delta{\delta K_-(t)}\langle\,\,|\,\,\rangle^{K_\pm}=\langle\,\,|
y_-^\dagger(t)|\,\,\rangle^{K_\pm}.
\eea
\end{subequations}
All expectation values of operator products at any time can be obtained
in this way---in particular, correlation functions.  Repeating this operation
we get
\bea
(-i)\frac\delta{\delta K_-(t)}i\frac\delta{\delta K_+^*(t')}\langle t_2|t_2
\rangle^{K_\pm}
&=&-i\frac\delta{\delta K_-(t)}\langle t_2|y_+(t')|t_2
\rangle^{K_\pm}\nn\\
&=&\langle t_2|y_-^\dagger(t)y_+(t')|t_2\rangle^{K_\pm}.
\eea
The operators are multiplied in the order of the time development.  The only 
place where $K_-$ appears is in the latter part of the time development.
See Fig.~\ref{fig:tc4}.

The distinction between $\pm$ disappears if we now set $K_+=K_-$:
\be
\frac\delta{\delta K_-(t)}\frac\delta{\delta K_+^*(t')}\langle0 t_2|0t_2
\rangle^{K_\pm}\bigg|_{K_+=K_-=K}=\langle 0t_2|y^\dagger(t)y(t')|0 t_2
\rangle^K.
\ee
As an example, set $t=t'=t_1$; then, from Eq.~(\ref{tcf}), 
this reads for the number operator $N(t)=y^\dagger(t)y(t)$,
\bea
\langle N(t_1)\rangle_0^K&=&\int dt \,K^*(t)G_a(t-t_1) \int dt'G_r(t_1-t')K(t')
\nn\\
&=&i \int dt\, e^{-i\omega(t-t_1)}K^*(t)(-i)\int dt'e^{-i\omega(t_1-t')}K(t')
=|\gamma|^2,\nn\\
\eea
as before, Eq.~(\ref{meanofn}).

\begin{figure}
\centering
\includegraphics{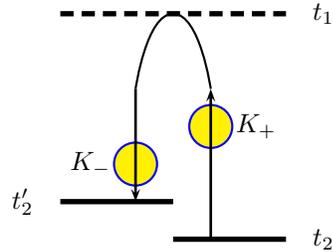}
\caption{\label{fig:tc5} Time cycle with different forces, $K_+$ and $K_-$.
on the forward and backward moving segments.  Now the initial time of the
time cycle, $t_2$, is different from the final time of the time cycle, $t_2'$,
with $\tau=t_2'-t_2$.  It is assumed that the time $t_1$ is later than
both $t_2$ and $t_2'$, and that the forces are localized as shown.}
\end{figure}

We would like to use more general starting and ending states than the ground
state.  We can obtain these by use of impulsive forces.  
It is convenient to deal
with all states at once, as in the generating function for $p(n,0)^K$ 
considered above.  Think of a time cycle starting at time $t_2$, advancing
forward to time $t_1$, during which time the force $K_+$ acts, then moving
back in time to a time $t'_2$, under the influence of the force $K_-$---See
Fig.~\ref{fig:tc5}.
  Let $t_2'=t_2+\tau$.  This displacement injects energy information.
Consider
\be
\sum_n\langle n t_2'|n t_2\rangle^{K_\pm}\equiv \tr \langle t_2'|t_2
\rangle^{K_\pm}=\sum_n e^{-in\omega\tau}\langle nt_2|nt_2\rangle^{K_\pm},
\ee
which uses (no force acts between times $t_2'$ and $t_2$)
\be
\langle n t_2'|=\langle n t_2|e^{-in\omega\tau}.
\ee
Analysis of this formula will yield individual transformation functions.

Now we must solve the dynamical equations subject to boundary conditions.
Let us compare $\tr \langle t_2'|y_+(t_2)|t_2\rangle$ with
$\tr \langle t_2'|y_-(t'_2)|t_2\rangle$.
\begin{subequations}
The first is
\be
\tr \langle t_2'|y_+(t_2)|t_2\rangle=\sum_n\langle n t_2'|y_+(t_2)|n t_2\rangle
=\sum_{nn'}\langle nt_2'|n't_2\rangle\langle n'|y|n\rangle,
\ee
while the second appears as
\be
\tr \langle t_2'|y_-(t_2')|t_2\rangle=\sum_{n'}\langle n' t_2'|y_-(t_2')|n' t_2
\rangle
=\sum_{nn'}\langle n'|y|n\rangle\langle nt_2'|n't_2\rangle.
\ee
\end{subequations}
Here, by introducing a complete set of states at the time of the operator, 
we have expressed the formula in terms 
of the matrix elements of stationary operators.
Remarkably, we see that the two expressions are equal; in effect, there
is a periodicity present here:
\be
y_+(t_2)=y_-(t_2'),\label{periodicity}
\ee
as far as traces are concerned.
Now, the equations of motion (\ref{eom:fho})  for the operators read
\be
\left(i\frac{d}{dt}-\omega\right)y(t)=K(t),
\ee
which has solution (\ref{yminust}) with the addition of the initial term, or
\bea
y_-(t)&=&e^{-i\omega(t-t_2)}y_+(t_2)-i\int_{t_2}^{t_1}dt'\,e^{-i\omega(t-t')}
K_+(t')\nn\\
&&\quad\mbox{}+i\int_t^{t_1}dt'\,e^{-i\omega(t-t')}K_-(t').
\eea
In particular,
\be
y_-(t_2')=e^{-i\omega\tau}y_+(t_2)-i\int dt'\,e^{-i\omega(t_2+\tau-t')}
(K_+-K_-)(t').\label{inpart}
\ee
Note that the integrals sweep over the full force history.  
Let us let $t_2=0$ for simplicity, although we will keep the label.
Because of the periodicity condition (\ref{periodicity}) this reads
\be
\left(e^{i\omega \tau}-1\right)y_+(t_2)=-i\int dt \, e^{i\omega t}(K_+-K_-)(t)
=-i(\gamma_+-\gamma_-),
\ee
or
\be
y_+(t_2)=\frac1{e^{i\omega \tau}-1}(-i)(\gamma_+-\gamma_-).
\ee

What we are interested in is
\be
\frac{\tr \langle t_2'|t_2\rangle^{K_2}}{\tr \langle t_2'|t_2\rangle},
\ee
The denominator, which refers to the free harmonic oscillator, is immediately
evaluated as
\be
\tr \langle t_2'|t_2\rangle=\sum_{n=0}^\infty e^{-in\omega\tau}
=\frac1{1-e^{-i\omega\tau}}.
\ee
(If $\tau$ be imaginary, we have thermodynamic utility.)  We have then
the variational equation
\be
\delta_{K_\pm^*}\left[\frac{\tr \langle t_2'|t_2\rangle^{K_2}}
{\tr \langle t_2'|t_2\rangle}\right]=
\frac{-i\tr \langle t_2'|\int dt\left(\delta K_+^*y_+-\delta K_-^*y_-\right)
|t_2\rangle^{K_\pm}}{\tr\langle t_2'|t_2\rangle},
\ee

Exactly as before, we get an equation for the logarithm---looking at the 
previous calculation leading to Eq.~(\ref{tcf}), 
we see an additional term, referring to the $y_+(t_2)$
boundary term in Eq.~(\ref{inpart}).  
The periodic boundary condition then gives 
\be
-\frac1{e^{i\omega\tau}-1}\delta(\gamma_+^*-\gamma_-^*)(\gamma_+-\gamma_-).
\ee
Therefore, to convert $\langle 0t_2|0 t_2\rangle^{K_\pm}$ in Eq.~(\ref{tcf}) to
\be
\frac{\tr \langle t_2'|t_2\rangle^{K_2}}{\tr \langle t_2'|t_2\rangle}=
\frac{\sum e^{-in\omega \tau}\langle n t_2|n t_2\rangle^{K_\pm}}{
\sum e^{-in\omega \tau}}
\ee
we must multiply by
\be
\exp[-\frac1{e^{i\omega\tau}-1}|\gamma_+-\gamma_-|^2].
\ee
This holds identically in $\tau$; in particular, in the limit where $\tau
\to -i\infty$, which corresponds to absolute zero temperature, we recover
$\langle 0t_2|0t_2\rangle^{K_\pm}$.

We find, generalizing Eq.~(\ref{tcf})
\bea
&&\frac{\sum_n e^{-in\omega\tau}\langle nt_2|n t_2\rangle^{K_\pm}}
{\sum_n e^{-in\omega\tau}}=e^{-i\int dt\,dt'\,K_+^*(t)G_r(t-t')K_+(t')}\nn\\
&&\times e^{i\int dt\,dt' K_-^*(t) G_0(t-t')K_-(t')}e^{\int dt\,dt'K_-^*(t)
e^{-i\omega(t-t')}K_+(t')}\nn\\
&&\times e^{-(e^{i\omega \tau}-1)^{-1}\int dt\,dt'(K_+^*-K_-^*)(t)e^{-i\omega
(t-t')}(K_+-K_-)(t')},\label{complexgf}
\eea
which is the exponential of a bilinear structure.  This is a generating
function for the amplitudes $\langle nt_2|nt_2\rangle^{K_\pm}$.  But it is
useful as it stands.

Put $\tau=-i\beta$; then this describes a thermodynamic average over a thermal
mixture at temperature $T$, where $\beta=1/kT$ in terms of Boltzmann's 
constant,
\be
\frac{\sum_n e^{-\beta n\omega}\langle\,\,|\,\,\rangle_n}{\sum_n e^{-\beta
n\omega}}
\ee
In terms of this replacement,
\be
\frac1{e^{i\omega\tau}-1}\to \frac1{e^{\beta\omega}-1}=\langle n\rangle_\beta,
\ee
because
\be
\frac{\sum_n n e^{-in\omega\tau}}{\sum_n e^{-in\omega\tau}}=\frac\partial
{\partial(-i\omega\tau)}\ln(\sum_n e^{-in\omega\tau})=
\frac\partial{\partial(-i\omega\tau)}\ln\frac1{1-e^{-i\omega\tau}}
=\frac1{e^{i\omega\tau}-1}.\label{thermal}
\ee

Now consider a time cycle with displacement $T$: the system evolves from time
$t_2$ to time $t_1$ under the influence of the force $K_+(t)$, and backwards
in time from $t_1$ to $t_2'$ under the force $K_-(t)$:
\be
K_-(t)=K(t), \quad K_+(t)=K(t+T).
\ee
This is again as illustrated in Fig.~\ref{fig:tc5}, with these replacements.
What is the physical meaning of this?  Insert in Eq.~(\ref{complexgf})
 a complete set of states at time
$t_1$:
\be
\langle n t_2|n t_2\rangle^{K_\pm}=\sum_{n'}\langle n t_2|n't_1\rangle^{K_-}
\langle n't_1|nt_2\rangle^{K_+}.
\ee
We did this before for the ground state.  The effect is the same as
moving the starting and ending times.
Appearing here is
\be
\langle n't_1|n t_2\rangle^{K(t+T)}=\langle n't_1+T|nt_2+T\rangle^{K(t)}
=e^{-in'\omega T}\langle n't_1|nt_2\rangle^{K(t)}e^{in\omega T}.
\ee
Therefore,
\be
\langle n t_2|nt_2\rangle^{K(t),K(t+T)}=\sum_{n'}e^{-i(n'-n)\omega T}p(n',n)^K
=\langle e^{-i(N-n)\omega T}\rangle_n^K.
\ee
Therefore, as a generalization for finite $\tau$ of 
Eq.~(\ref{expinwt}), we have from Eq.~(\ref{complexgf})
\bea
&&\left(\sum_{n'} e^{-in'\omega \tau}\right)^{-1}\sum_n
 e^{-in\omega\tau}\langle e^{-i(N-n)\omega T}\rangle_n^K\nn\\
&=&\exp\left[\left(e^{-i\omega T}-1\right)|\gamma|^2-\frac1{e^{i\omega\tau}-1}
\left(e^{i\omega T}-1\right)\left(e^{-i\omega T}-1\right)|\gamma|^2\right],
\label{gengf}
\eea
where $T$ gives the final state, and $\tau$ the initial state.  This used
the observation
\be
\int dt\,e^{i\omega t}K(t+T)=e^{-i\omega T}\int dt\,e^{i\omega t}K(t).
\ee
Expand both sides of Eq.~(\ref{gengf}) in powers of $T$, and we learn
\be
-i\omega T\sum_n\langle N-n\rangle^K_n \frac{e^{-in\omega \tau}}{\sum_{n'}
e^{-in'\omega\tau}}=-i\omega T|\gamma|^2,\label{angf}
\ee
or
\be
\langle N-n\rangle_\beta^K=|\gamma|^2,\ee
which generalizes the earlier result (\ref{meanofn}).  
Now apply Eq.~(\ref{angf}) as a generating function,
\be
\langle N-n\rangle_n^K=|\gamma|^2,
\ee
which reflects the linear nature of the system.

We can rewrite the above generating function more conveniently, by multiplying
by 
\be
e^{i\langle N-n\rangle \omega T}=e^{i\omega T|\gamma|^2},
\ee
that is, Eq,~(\ref{gengf}) can be written as
\bea
&&
\frac1{\sum e^{-in\omega\tau}}\sum e^{-in\omega\tau}\langle e^{-i(N-\langle
N\rangle)\omega T}\rangle_n^K\nn\\
&=&\exp\left[\left(e^{-i\omega T}-1+i\omega T\right)|\gamma|^2-
\frac1{e^{i\omega\tau}-1}\left(e^{-i\omega T}-1\right)\left(e^{i\omega T}-1
\right)|\gamma|^2\right].\nn\\
\label{gf2}
\eea
Now pick off the coefficient of $-(\omega T)^2/2$:
\be
\frac1{\sum e^{-in\omega \tau}}\sum e^{-in\omega \tau}\langle(N-\langle N
\rangle)^2\rangle_n^K=|\gamma|^2+2\frac1{e^{i\omega\tau}-1}|\gamma|^2,
\label{sogf}
\ee
or
\be
\langle (N-\langle N\rangle)^2\rangle_\beta^K=|\gamma|^2[1+2\langle 
n\rangle_\beta].
\ee
If, instead, we multiply Eq.~(\ref{sogf}) 
through by $\sum_n e^{-in\omega\tau}$,
we can use this as a generating function, and learn from Eq.~(\ref{thermal})
that
\be
\langle( N-\langle N\rangle)^2\rangle_n^K=|\gamma|^2(1+2n).
\ee
Note the simplicity of the derivation of this result, which does not involve
complicated functions like Laguerre polynomials.

\chapter{Relativistic Theory of Fields}
\label{sec:6}
This section is an adaptation of Chapter V of lectures given at Stanford
by Julian Schwinger in 1956 \cite{stanford1956}.

A state of a physical system is defined in terms of the maximum number of 
compatible measurements which can be made upon the system. If the state were 
defined on a space-like surface (one in which all points are in space-like 
relation: ($\Delta x)^2-(\Delta t)^2 >0$) then a measurement at any point is 
compatible with one made at any other point, since the disturbances introduced 
by the measurements cannot propagate faster than $c$, and hence cannot 
interfere. Thus, a state can be specified as an eigenvector of a complete set 
of commuting, Hermitian operators $\underline{a}$, associated with a definite 
space-like surface $\sigma:\,\, | a^\prime ,\sigma \rangle$. There always 
exists a coordinate system in which the space-like surface $\sigma$ is all of 
three-dimensional space at a given time; in this particular Lorentz frame the 
state vector is just: $| a^\prime ,t\rangle$. The 
problem of relativistic dynamics is to find the transformation 
function
\begin{equation}
\langle a^\prime_1 \sigma _1| a_2^{\prime \prime }\sigma _2\rangle.
\end{equation}
\begin{figure}
\centering
\includegraphics{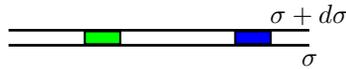}
\caption{\label{fig:v1} Spacetime volume bounded by two spacelike surfaces,
$\sigma$ and $\sigma+d\sigma$.  Points in the blue and green shaded regions
cannot interfere.}
\end{figure}

As in the non-relativistic case, we assume the existence of an action operator 
$W_{12}$ such that
\begin{equation}\label{5d1}
\delta \langle a_1^\prime \sigma _1| a_2^{\prime \prime }\sigma _2\rangle
=i\langle a_1^\prime \sigma _1| \delta [W_{12}]| a_2^{\prime \prime }\sigma _2\rangle.
\end{equation}
The contributions to the action operator are now given by
\begin{equation}
W_{12}=\sum_\sigma W_{\sigma +d\sigma ,\sigma }.
\end{equation}
Since measurements made at points in the space-like shell $d\sigma $ cannot 
interfere (e.g., in the two shaded areas, shown in Fig.~\ref{fig:v1}) their 
contribution to $W_{12}$ is additive,
\begin{equation}\label{5d2}
W_{12}= \int_{\sigma _2}^{\sigma _1}(dx)\mathcal{L}(x)=
\int_{\sigma _2}^{\sigma _1}(dx)\mathcal{L}[\chi _a(x)],
\end{equation}
where the $\chi _a(x)=\chi _a(x,y,z,t)$ are the dynamical variables of the 
system, necessarily Hermitian operators; 
the relativistic requirements automatically introduce the concept of 
fields. The relativistic notation used has the form
\begin{eqnarray}
x^0&=&t=-x_0, \quad x^k=x_k, \text{ where }  k=1,2,3,\nn\\
(dx)&=&dx^0dx_1dx_2dx_3, \quad \partial _\mu =\frac{\partial }{\partial x^\mu},
\end{eqnarray}
where the metric used is
\be
g_{\mu \nu }=
\left(\begin{matrix} -1 & 0 & 0 & 0\\
                   0 & 1 & 0 & 0 \\
                   0 & 0 & 1 & 0\\
                   0 & 0 & 0 & 1 \\
     \end{matrix}  \right).
\ee

The action principle again states that for a given dynamical system, the 
variations arise only from the end-point, that is,
\begin{equation}\label{5d3}
\delta W_{12}=G_1(\sigma _1)-G_2(\sigma _2)
\end{equation}
As before, from this requirement follow the equations of motion and the 
generators of infinitesimal transformations which yield the commutation 
relations of the field operators. The Lagrangian density $\mathcal{L}$ which 
will yield first order field equations is
\begin{equation}\label{5d4}
\mathcal{L} = \frac{1}{2}\left(\chi A^\mu \partial _\mu \chi 
-\partial _\mu \chi 
A^\mu \chi \right)-\mathcal{H}\left(\chi \right),%\label{folag}
\end{equation}
where the $A^\mu $ are a set of four numerical matrices, and space and time 
derivatives appear on a symmetric basis. The symmetrization of the kinematical 
term relates to the possibility of adding to $\mathcal{L}$ the 
relativistic analogue of our previous total time derivative, 
a four dimensional divergence. If
\begin{equation}\label{5d5}
\overline{\mathcal{L}}=\mathcal{L}-\partial _\mu f^\mu,
\end{equation}
then
\begin{equation}
\overline{W}_{12}=W_{12}-\int_{\sigma _2}^{\sigma _1}(dx) \partial _\mu f^\mu =
W_{12}-\int_{\sigma _1}d\sigma _\mu f^\mu +\int_{\sigma _2}d\sigma _\mu f^\mu,
\end{equation}
and
\begin{equation}
\overline{G}(\sigma _1)=G(\sigma _1)-\int_{\sigma _1}d\sigma _\mu f^\mu ,  
\qquad \overline{G}(\sigma _2)=G(\sigma _2)-\int_{\sigma _2}d\sigma _\mu 
f^\mu .
\end{equation}
As before, the equation of motion is unchanged. Since $\mathcal{L}$ is to be
Hermitian, we require
\begin{equation}\label{5d6}
A^{\mu \dagger }=-A^\mu,  \quad\mbox{so}\quad \mathcal{H}^\dagger =\mathcal{H}
\end{equation}

The rank of $A^\mu $is that of the number of independent fields. 
Note that the variation 
$\delta \langle a_1^\prime \sigma _1| a_2^{\prime \prime }
\sigma _2\rangle$  is independent of any coordinate system, since 
$\mathcal{L}$ is a Lorentz scalar.

We can now infer some fundamental properties from the requirement of invariance
of $\mathcal{L}$. Consider the coordinate transformation (Poincar\'e 
transformation)
\begin{equation}\label{5d7}
\overline{x}^\mu =\ell ^\mu{}_\nu  x^\nu -\ell ^\mu  
\end{equation}
where
\begin{equation}
g_{\mu \nu }\ell ^\mu {}_\lambda  \ell ^\nu {}_\kappa =g_{\lambda \kappa },\quad
\ell_\mu{}^\lambda\ell^\mu{}_\kappa=\delta^{\lambda}{}_\kappa.
\end{equation} 
We can divide the transformations into two subsets by considering the effect
of (\ref{5d7}) on $g_{00}$,
\begin{equation}
-g_{00}=1=\left(\ell ^0{}_0\right)^2-\sum_{k=1}^3\left(\ell ^k{}_0\right)^2 
\qquad \left(\ell ^0{}_0\right)^2=1+\sum_k\left(\ell ^k{}_0\right)^2\geq 1.
\end{equation} 
Since $\ell^0{}_0 = \frac{\partial \overline{x}^0}{\partial x^0}$
 it follows that we can never make a continuous change from a positive to a 
negative sense of time, i.e., generate an improper transformation continuously. 
We shall consider only the group of continuous proper Lorentz transformations.

Under such a coordinate change, the fields $\chi _a (x)$ change to 
new ones $\overline{\chi }_a(\overline{x})$  connected by a real linear 
transformation: 
\begin{equation}\label{5d8}
\overline{\chi }\left(\overline{x}\right)=L\chi (x).
\end{equation} 
Using $\overline{\partial }_\mu =\ell ^\nu{}_\mu \partial _\nu $, and writing the 
scalar $\mathcal{L}$ in the new frame we find
\begin{eqnarray}
\mathcal{L}&=&\frac{1}{2}\left(\overline{\chi }A^\mu \overline{\partial }_\mu 
\overline{\chi }-\overline{\partial}_\mu\overline{\chi}A^\mu\overline{\chi}
\right)-\mathcal{H}\left(\overline{\chi} \right)\nn\\
&=&\frac{1}{2}\left[\chi \left(
L^{\text{T}}A^\mu \ell ^\nu {}_\mu L\right) \partial _\nu \chi  
-\partial _\nu \chi \left( \ell ^\nu {}_\mu L^{\text{T}} A^\mu L 
\right)\chi \right] -\mathcal{H} \left(L\chi \right).
\end{eqnarray}
Thus
\begin{equation}\label{5d9}
L^{\text{T}}A^\mu L=\ell ^\mu {}_\nu A^\nu,  \qquad \mathcal{H}
\left(L\chi\right) =\mathcal{H}\left(\chi \right).
\end{equation}

If we choose for $\mathcal{H}$ the special form
\begin{equation}
\mathcal{H}=\chi B\chi =\chi L^{\text{T}}B L\chi, 
\end{equation}
where $B$ is Hermitian and non-singular, then 
$L^{\text{T}}=B L^{-1}B^{-1}$. 
From the first of equations (\ref{5d9}), we then obtain 
\begin{equation}\label{5d10}
L^{-1}\left(B^{-1}A^\mu \right)L=\ell ^\mu {}_\nu 
\left(B^{-1}A^\nu \right),
\end{equation}
showing that the combination $B^{-1}A^\mu $ transforms like a vector; this 
is implied by the required invariance of the kinematical term of $\mathcal{L}$. 

Consider now a general inhomogeneous infinitesimal Lorentz transformation, of 
the form
\begin{equation}
\overline{x}^\mu =x^\mu -\epsilon ^\mu +\epsilon ^\mu {}_\nu x^\nu,\label{gilt}
\end{equation}
where the $\epsilon ^\mu $ and $\epsilon ^\mu {}_\nu $ are infinitesimals, 
and the rotational nature of the $\epsilon ^{\mu \nu }$ is expressed by the 
relations $\epsilon _{\mu \nu }=-\epsilon _{\nu \mu }$. Then $L$ 
can be written as $L=1 + \frac{i}{2}\epsilon _{\mu \nu }S^{\mu \nu }$, where only 
the rotational $\epsilon _{\mu \nu }$ terms appear, since the translations 
$\epsilon ^\mu $ do not effect the gradient operators $\partial _\nu $, and no 
corresponding changes in the $\chi _\alpha $ are required to keep $\mathcal{L}$
 invariant. The $S^{\mu \nu }$ are (imaginary) operators, acting on the field 
variable, which will express the spin character of the fields. 

The variation $\delta W_\alpha $ allows us to change the field components at
each space-time point (call these changes $\delta _\alpha \chi _\alpha $), and
 to change the region of integration by displacing the boundary surfaces 
$\sigma _1$ and $\sigma _2$. In the previous non-relativistic treatment, 
instead of varying the end-point times $t_1$ and $t_2$ ,we used $t(\tau )$. 
Proceeding similarly here, we can express the variation of a space-like surface 
by varying the space-time coordinates under an infinitesimal Lorentz
transformation ($\delta x_\mu =\epsilon _\mu -\epsilon _{\mu \nu }x^\nu $) so chosen
that on $\sigma _1$ and $\sigma _2$ the required displacement is obtained.
The change in the action is
\begin{eqnarray}
\delta W_{12}&=&\delta _0\int_{\sigma _2}^{\sigma _1}\left(dx\right)
\left[\frac{1}{2}\left(\chi A^\mu \partial _\mu \chi 
-\partial _\mu \chi A^\mu \chi \right)-\mathcal{H}\right]\nn\\
&&\quad\mbox{}+\int_{\sigma _1}d\sigma _\mu \delta x^\mu \mathcal{L}
-\int_{\sigma _2}d\sigma _\mu \delta_0 x^\mu \mathcal{L} \nn\\
&=&\int_{\sigma _2}^{\sigma _1}\left(dx\right)\left[\delta _0\chi A^\mu 
\partial _\mu \chi -\partial _\mu \chi A^\mu \delta _0\chi 
-\delta _0\mathcal{H}\right] \nn\\
&&\quad\mbox{}+ \frac{1}{2}\int_{\sigma _2}^{\sigma _1}\left(dx\right)
\partial _\mu \left(\chi A^\mu \delta _0\chi 
-\delta _0\chi A^\mu \chi \right)
\nn\\
&&\quad\mbox{}+\int_{\sigma _1}d\sigma _\mu \delta x^\mu \mathcal{L} 
- \int_{\sigma _2} d\sigma _\mu \delta x^\mu \mathcal{L} \nn\\
&=&\int_{\sigma _2}^{\sigma _1}\left(dx\right)\delta _0\mathcal{L}
+\int_{\sigma _1}d\sigma_\mu\left\{\delta x^\mu \mathcal{L}+\frac{1}{2}
\left(\chi A^\mu \delta _0\chi -\delta _0\chi A^\mu \chi \right)\right\}\nn\\
&&\quad\mbox{}-\int_{\sigma _2}d\sigma _\mu \left\{\delta x^\mu \mathcal{L}
+\frac{1}{2}\left(\chi A^\mu \delta _0\chi -\delta _0\chi A^\mu \chi \right)
\right\}.\label{5d11} 
\end{eqnarray}

Applying the action principle, the interior variation $\delta _0 \mathcal{L}$ 
must vanish, giving the field equations of motion. What remains is the 
difference of two generators, $G(\sigma _1)-G(\sigma _2)$, where
\begin{equation}\label{5d111}
G(\sigma )=\int_\sigma d\sigma _\mu \left\{\delta x^\mu \mathcal{L}
+\frac{1}{2}\left(\chi A^\mu \delta _0 \chi -\delta _0 \chi A^\mu \chi \right)
\right\}.
\end{equation}
To re-write (\ref{5d111}) we recognize that the total change in the fields
is due to both the $\delta_0$ 
variation of the fields at a given space-time point on 
$\sigma $, and also to the variation induced by the infinitesimal Lorentz 
transformation of coordinates as $\sigma $ is displaced to $\sigma +d\sigma $; 
the latter is obtained from
\begin{equation}
\overline{\chi }\left(\overline{x}\right)=L\chi(x)=\chi(x)
+\frac{i}{2}\epsilon _{\mu \nu }S ^{\mu \nu }\chi (x),
\end{equation}
But 
\begin{equation}
\overline{\chi }\left(\overline{x}\right)=\overline{\chi }(x)
-\delta x^\mu \partial _\mu \chi (x),
\end{equation}
and therefore
\begin{equation}\label{5d12}
\overline{\chi }(x)-\chi (x)
=\delta x^\mu \partial _\mu \chi (x) +\frac{i}{2}\epsilon _{\mu \nu }S ^{\mu \nu }
\chi (x).
\end{equation}
The right hand side of (\ref{5d12}) is then the variation induced in the 
$\chi _\alpha $ by the coordinate transformation; the total variation of the 
fields is then
\begin{equation}\label{5d13}
\delta \chi (x)=\delta _0\chi (x)+\delta x^\mu \partial _\mu \chi (x)
+\frac{i}{2}\epsilon _{\mu \nu }S ^{\mu \nu }\chi (x).
\end{equation}
Solving for $\delta _0\chi $, and substituting into (\ref{5d111}), we obtain
for the generator
\begin{eqnarray}
G(\sigma )&=&\int d\sigma _\mu \left\{\delta x^\mu \mathcal{L} 
+\frac{1}{2}\left(\chi A^\mu \delta \chi -\delta \chi A^\mu \chi \right) 
\right\} \nn\\
&&\quad\mbox{}-\frac{1}{2} \int d\sigma _\mu \bigg\{\left(\chi A^\mu 
\partial ^\nu \chi -\partial ^\nu \chi A^\mu \chi \right)\delta x_\nu\nn\\
&&\qquad\mbox{}-\frac{i}{4}\epsilon _{\lambda \nu }\left(\chi A^\mu 
S^{\lambda \nu }\chi 
-S^{\lambda \nu }\chi A^\mu \chi \right)\bigg\}.
\end{eqnarray}
Adding a surface term, it is possible to bring $G(\sigma )$ into the form:
\begin{equation}\label{5d14}
G(\sigma )=\int d\sigma _\mu \frac{1}{2}\left(\chi A^\mu \delta \chi 
-\delta \chi A^\mu \chi \right) +\int d\sigma _\mu \delta x_\nu T^{\mu \nu },
\end{equation}
where $T^{\mu \nu }$ is the symmetric stress-tensor operator
\begin{eqnarray}
T^{\mu \nu }&=&g^{\mu \nu }\mathcal{L}-\frac{1}{4}\left[\chi A^{\text{\{}\mu }
\partial ^{\nu \text{\}}}\chi -\partial ^{\text{\{}\nu }\chi A^{\mu \text{\}}}
\chi \right]\nn\\
&&\quad\mbox{}-\frac{i}{4}\partial _\lambda \left[\chi A^{\text{\{}\nu }
S^{\mu \text{\}}\lambda }\chi -S^{\text{\{}\mu \lambda }A^{\nu \text{\}}}
\chi \right],\label{5d141}
\end{eqnarray}
and the brackets  $\{$ $\}$ represent symmetrization with respect to 
$\mu $ and $\nu $.

Applying the stationary action principle to the variation $\delta x_\nu $, taken
 as arbitrary, we note that the invariance of the action operator implies the 
conservation law:
\begin{equation}\label{5d15} 
\int_{\sigma _1} d\sigma _\mu T^{\mu \nu }=\int_{\sigma _2} d\sigma _\mu T^{\mu \nu },
\end{equation}
which, in turn, implies the corresponding differential conservation 
law
\begin{equation}\label{5d151}
\partial _\mu T^{\mu \nu }=0.
\end{equation}

The generator (\ref{5d14}) can be split into two parts, one representing 
changes induced by the coordinate variation, and the other giving the variation
induced by a change in the field variables,
\begin{subequations}
\begin{eqnarray} 
G_\chi &=&\int d\sigma _\mu\frac{1}{2}\left(\chi A^\mu\delta\chi
 -\delta \chi A^\mu \chi \right),
\\
G_x&=&\int d\sigma _\mu T^{\mu \nu }\delta x_\nu \
=\epsilon _\nu P^\nu  +\frac{1}{2}\epsilon _{\mu \nu } J^{\mu \nu },\label{5d16} 
\end{eqnarray}
\end{subequations}
where
\begin{equation}
P^\nu \equiv \int d\sigma _\mu  T^{\mu \nu },  
\qquad J^{\mu \nu }\equiv \int d\sigma _\lambda 
\left[x^\mu T^{\lambda \nu }-x^\nu T^{\lambda \mu }\right];
\end{equation}
$P^\nu $ and $J^{\mu \nu }$ are the generators for translations and rotations, 
respectively, and their commutation relations are determined by the group of 
transformations they represent. Specifically, $P^\nu $ is recognized as the 
4-momentum operator, and $J^{\mu \nu }$ as the relativistic generalization of the 
angular momentum operator.

The field equations are obtained by the vanishing of $\delta _0 \mathcal{L}$ in
(\ref{5d11}), 
\begin{equation}
2A^\mu \partial _\mu \chi =\frac{\partial _\ell \mathcal{H}}{\partial \chi } 
\qquad \text{ or } -2A^{\mu \text{T}}\partial _\mu \chi  
= \frac{\partial _r \mathcal{H}}{\partial \chi },\label{eomlandr}
\end{equation}
in terms of left and right derivatives, 
corresponding to the two equivalent ways of writing $G_\chi $:
\begin{equation}
G_\chi =\int d\sigma _\mu \chi A^\mu \delta \chi, \qquad  \text{ or }
\qquad G_\chi = 
-\int d\sigma _\mu \delta \chi A^\mu \chi. 
\end{equation}
If we continued with these two pairs of expressions,we would 
obtain two forms for the commutation rules of $\chi$; their equivalence 
then leads to the requirement that the $A^\mu $ and $\chi $ must decompose:
\begin{equation}
\chi =\phi +\psi,  \qquad \qquad A^\mu =a^\mu +s^\mu,\label{befd}  
\end{equation}
%as before, 
where the $a^\mu $ are anti-symmetric and real, the 
$s^\mu $ are symmetric and imaginary, and the $\phi $ and $\psi $ represent 
the kinematically independent fields of the Bose-Einstein and Fermi-Dirac 
types, respectively. The field equations are then
\begin{subequations}
\begin{eqnarray}
2a^\mu \partial _\mu \phi &=& \frac{\partial _\ell \mathcal{H}}{\partial \phi}  
= \frac{\partial _r\mathcal{H}}{\partial \phi },\\
2s^\mu \partial _\mu \psi &=&\frac{\partial _\ell \mathcal{H}}{\partial \psi }
=-\frac{\partial _r\mathcal{H}}{\partial \psi }.\label{5d17} 
\end{eqnarray}
\end{subequations}
Similarly, $G_\chi $ can be. divided into its $\phi $ and $\psi $ parts,
$G_\chi =G_\phi +G_\psi $,
\begin{eqnarray}
G_\phi &=&\int d \sigma _\mu \phi a^\mu \delta \phi  
= \int d \sigma _\mu  \left( a^\mu \delta \phi \right)\phi, \nn\\
G_\psi &=&\int d\sigma _\mu \psi s^\mu \delta \psi = \int d \sigma ^\mu 
\left(s^\mu \delta \psi \right)\psi. 
\end{eqnarray}
Again, we find (where $\{\,,\,\}$ denotes an anticommutator)
\begin{eqnarray} 
\left\{\delta \psi _\alpha (x),\psi _\beta (x^\prime \right\}&=&0,\nn\\
\left[\delta \phi _\alpha (x),\phi _k (x^\prime )\right]=
\left[\delta \phi _k(x),\psi _\alpha (x^\prime )\right]&=&
\left[\delta \psi _\alpha (x),\phi _k(x^\prime )\right]=0.\label{5d18}
\end{eqnarray}
The first of equations (\ref{5d18}) combined with the second of equations 
(\ref{5d17}) implies that $\mathcal{H}$ must be an even function of 
$\psi $.
 
The field equations may be written as equations of motion by singling out the 
time differentiation,
\begin{equation}\label{5d19}
2A^0\partial _0\chi =
\frac{\partial _\ell \mathcal{H}}{\partial \chi }-2A^k \partial _k \chi. 
\end{equation}
%Previously, 
If we took $A^0$ to be non-singular, we would be able to solve (\ref{5d19}) 
for $\partial 0\chi $. More generally, we now recognize the existence of the 
following possible situations:
\begin{enumerate}
\item $A^0$ is non-singular. In this case, all of the variables are 
kinematically independent. An example of this situation is the 
Dirac-Majorana spin-$\frac{1}{2}$ field.
\item  \label{case2}
$A^0$ is singular, but there are enough relations among the variables 
to determine all of them. Here, only those variables which 
possess equations of motion are kinematically independent, but the 
non-independent fields are determined from the independent fields. 
Examples of this are the spin zero and spin one fields. 
\item $A^0$ is singular, and there are not enough relations among the 
variables to determine all the fields, as in case \ref{case2}. 
The classic example of this is the spin 1, zero mass, electromagnetic field,
where the lack of determination corresponds to the possibility of introducing 
a gauge transformation. 
\end{enumerate}

\section{Inference of Particle Properties}
\label{sec:6.1}
We now consider the generators of infinitesimal coordinate 
(Lorentz) transformations $G_k$ and the commutation relations they 
imply. From (\ref{5d16}) we have 
\begin{equation}
G_x=\epsilon _\nu P^\nu +\frac{1}{2}\epsilon _{\mu \nu }J^{ \mu \nu },
\end{equation} 
which, when applied to the space-time coordinates $x_\mu $, generates 
the new $\overline{x}_\mu $,
\begin{equation}
\overline{x}^\prime =x^\mu -\epsilon ^\mu +\epsilon ^\mu {}_\nu x^\nu,
\end{equation} 
which is Eq.~(\ref{gilt}).
Accompanying this transformation, we have the apparent change in the 
fields given by (\ref{5d12}),
\begin{subequations}
\begin{equation}
-\delta \chi (x)=\overline{\chi }(x)-\chi (x)=
\delta x^\mu \partial _\mu \chi (x)+\frac{i}{2}\epsilon _{\mu \nu }S^{\mu \nu }
\chi (x),
\end{equation} 
or
\begin{equation}\label{5d20}
-\delta \chi (x)=\left[\epsilon ^\mu \partial _\mu 
+\frac{1}{2}\epsilon _{\mu \nu }\left(x^\mu \partial ^\nu -x^\nu \partial ^\mu 
-iS^{\mu \nu }\right)\right]\chi (x),
\end{equation}
\end{subequations}
using $\epsilon _{\mu \nu }= -\epsilon _{ \nu \mu }$.
Comparing with $\left[\chi ,G_x\right]=-i\delta \chi $, for arbitrary 
translations
($\epsilon _\mu $) and rotations ($\epsilon _{\mu \nu }$), we obtain
\begin{subequations}
\begin{eqnarray}\label{5d21}
\left[\chi ,P^\nu\right] &=&-i\partial ^\nu \chi,\\
\label{5d22}
\left[\chi ,J^{\mu \nu }\right]&=&-i\left(x^\mu \partial ^\nu -x^\nu 
\partial ^\mu +iS^{\mu \nu }\right)\chi.
\end{eqnarray}
\end{subequations}
From (\ref{5d22}) the identification of $S^{\mu \nu }$ with the intrinsic spin 
characteristics of the particle is evident. Considering the time 
component (\ref{5d21}) we obtain the standard commutator equation of 
motion
\begin{equation}
\left[\chi ,P^0\right]=-i\partial ^0\chi =i\partial _0\chi =
i\frac{\partial \chi }{\partial t}.
\end{equation}
To determine the manner in which the particle interpretation enters, we now 
represent the fields by Fourier integrals,
\begin{equation} 
\chi (x)=\int (dp)
e^{ix^\mu p_\mu }\chi (p),
\end{equation} 
where the $\chi (p)$ are operator functions of the numbers $p_\mu $. 
Substituting into (\ref{5d21}) and equating coefficients, we obtain
\begin{subequations}
\begin{equation}\label{5d23}
\left[\chi (p),P^\nu \right]=p^\nu \chi (p),
\end{equation}
or
\begin{equation}
P^\nu \chi (p)=\chi (p)\left(P^\nu -p^\nu\right).
\end{equation}
\end{subequations}
Since the 4 operators $P^\nu$ all commute, we can have simultaneous eigenstates
$| P^\prime \rangle$, where $P^\nu| p^\prime \rangle=P^{\nu \prime }| p^\prime
 \rangle$.
 Applying (\ref{5d23}) to these states 
\begin{equation}
P^\nu \chi | p^\prime \rangle=\chi \left(P^{\nu \prime }-p^\nu \right)
| P^\prime \rangle=\left(P^{\nu \prime }-p^\nu \right)\chi| P^\prime \rangle,
\end{equation} 
or writing $\chi | P^\prime \rangle$ as some new eigenvector 
$| p^\prime -p\rangle$,
\begin{equation}
P^\nu |P^\prime -p\rangle=\left(P^{\nu \prime }-p^\nu \right)| P^\prime -p\rangle,
\end{equation}
which shows that the effect of $\chi$  on $| P^\prime \rangle$ is to produce a 
state whose momentum eigenvalue has been changed by $-p^\nu $, indicating the 
capacity of the field to absorb or emit $p^\nu $. Specifically, considering the
time component $\nu =0$, we have two possible situations, depending on whether 
$p^0$ is greater than or less than zero. If $p^0>0$ , application of $\chi $ 
on $|P^\prime \rangle$ yields a state with lower energy $(P^{0\prime }-
p^{0\prime} $), 
and conversely if $p^0  < 0$. Since $\chi $ is Hermitian, $\chi $ can be 
split, in a Lorentz covariant way, into two parts, $\chi =\chi ^{(+)} 
+ \chi ^{(-)}$, where the superscripts indicate the positive and 
negative nature of the $p^0$ terms which enter in the respective
Fourier transforms, and where ($\chi ^{(+)})^\dagger =\chi ^{(-)}$. Then 
$\chi ^{(+)} $ applied to a state of definite energy acts as an energy 
annihilation operator, and $\chi ^{(-)}$ as an energy creation operator.

The $\chi ^{(+)}$ and $\chi ^{(-)}$ can then be used to create any physical 
state 
from the vacuum state, where we take the latter as the unique lowest-energy 
state. This vacuum state must necessarily correspond to the eigenvalue
$P^{k\prime} = 0$, ($k=1,2,3$), since if one of the $P^{k\prime }$ were not 
zero,
a rotation of the coordinate system could yield three non-zero momentum 
components, requiring (a super-position of) the corresponding eigenvectors for 
its description. But this precludes a description of the vacuum by a single 
non-degenerate state; we must therefore require that each $P^{k \prime} = 0$.
Since the vacuum is to have the lowest energy possible, and we may arbitrarily 
take this to be zero, we then characterize the vacuum state as that unique 
state for which $P^{\mu \prime }=0$, where $\chi ^{(+)}\mid 0>=0$.

So far we have considered $\chi $ as representing general fields; to introduce 
particle properties consider the Fourier transform of $\chi $, and imagine the 
numbers $p_\mu $ related by the relation:
\begin{equation}
-p_\mu p^\mu =m^2, \quad p^0 =\pm \sqrt{m^2+\mathbf{p}^2}.
\end{equation}
Then $\chi ^{(+)}$, for example, remains an operator which annihilates energy, 
but now is correlated with a momentum decrease. This is just the usual particle
interpretation; if $\chi ^{(+)}$ has this character it may be spoken of as a 
particle annihilation operator, and conversely for $\chi ^{(-)}$.
 
\section{The Connection Between Spin and Statistics}

We consider the simplest system, corresponding to linear field equations. Taking
 $\mathcal{H}=\chi B\chi $, this becomes for our two types of fields 
$\mathcal{H}=\phi B^{(1)}\phi +\psi B^{(2)}\psi $,  where $\phi $, $\psi $ 
represent fields of the $1^{st}$ and $2^{nd}$ kinds, respectively,
meaning bosonic and fermionic fields. (Note that 
$\mathcal{H}$ must be even in $\psi $,  and therefore, as noted below 
Eq.~(\ref{5d18}), no term of the form $\phi \psi $ can occur.) 
The matrix $B$ (and therefore $B^{(1)}$ and $B^{(2)}$) 
is Hermitian. Writing
\begin{equation}
\delta _\phi \mathcal{H}=\delta \phi B^{(1)}\phi +\phi B^{(1)}\delta \phi
  =\delta \phi \left(B^{(1)}\phi +B^{(1) \text{T}}\phi \right),
\end{equation}
and since
\begin{equation}
\frac{\partial _\ell \mathcal{H}}{\partial \phi }=\frac{\partial _r\mathcal{H}}
{\partial \phi },
\end{equation} 
for these variations of fields of the first kind, both terms in the above 
bracket must be equivalent, implying that $B^{(1)}=B^{(1)\text{T}}$. Since $B^{(1)}$
is also Hermitian, it is real. For $\delta \psi $ variations, the identical 
argument shows that $B^{(2)}=-B^{(2)\text{T}}$, i.e., $B^{(2)}$ is imaginary. 
It is 
precisely here that the connection between spin and statistics arises: We can 
construct matrices for $B^{-1}a^\mu$ only for particles of integral spin,
and for $B^{-1}s^\mu $ only when the particle spin is an integer plus one-half.
[See Eq.~(\ref{befd}).]

Writing the equations of motion (\ref{eomlandr}) for either field
\begin{equation}
2A^\mu \partial _\mu \chi =\frac{\partial _\ell \mathcal{H}}{\partial \chi} 
=2B\chi,
\end{equation}
and assuming---as the simplest case---that $B$ is non-singular, we obtain
\begin{equation}
B^{-1}A^\mu \partial _\mu \chi (x)=\chi (x).
\end{equation}
Going to the momentum representation $\chi (p)$ as before, this becomes 
\begin{equation}\label{5d24}
iB^{-1}A^\mu p_\mu \chi (p)=\chi (p).
\end{equation}
Now consider the matrix (of finite order) $iB^{-1}A^\mu p_\mu \equiv M$, which 
must satisfy its algebraic minimal characteristic equation
$M^n+a_1M^{n-1}+\cdot \cdot \cdot +a_n=0$, where the $a_n$ are numbers. From 
Eq.~(\ref{5d10}) we know that $L^{-1}\left(B^{-1}A^\mu \right)L
=\ell ^\mu {}_\nu \left(B^{-1}A^\nu \right)$. 
If we insert the proper combinations 
$L^{-1}L$ in each term, and use the relation
\begin{equation}
L^{-1}\left(B^{-1}A^\mu p_\mu \right)L=B^{-1}A^\nu \ell ^\mu {}_\nu p_\mu 
=B^{-1}A^\mu \overline{p}_\nu,
\end{equation}
then the result of a Lorentz transformation, giving us back the identical 
minimal equation (as it must), shows that the $a_j$ are Lorentz invariants, or 
functions of invariants. Since the only 4-vectors available are the $p_\mu $, 
we take $a_j=a_j(-p_\mu p^\mu )$. Furthermore, the minimal equation must be 
valid independently of the value of the numbers $p_\mu $, i.e., it must be an 
algebraic identity in $p_\mu $ ; the coefficients $a_j(-p^\mu p_\mu )$ must 
then 
be of the form $(-p^\mu p_\mu )^j$ times a numerical factor $c_j$, which is 
independent of the $p_\mu $. We can identify the two possible cases, 
corresponding to the degree of the minimal equation being either even ($n=2k$) 
or odd ($n=2k+1$); in either case the power to which $(B^{-1}A^\mu p_\mu )$ is 
raised must decrease in steps of two,
\begin{subequations}
\begin{eqnarray}
&&n=2k: \quad \left(iB^{-1}A^\mu p_\mu \right)^{2k}+ \left(-p_\mu p^\mu \right)
c_1 \left(iB^{-1}A^\mu p_\mu \right)^{2k-2}+ \dots\nn\\ 
&&\quad\qquad\mbox{}+ \left(-p_\mu p^\mu \right)^k c_k = 0, \label{5d25}\\ 
\label{5d26}
&&n=2k+1: \quad \left(iB^{-1}A^\mu p_\mu \right)^{2k+1}+ \left(-p_\mu 
p^\mu \right)c_1 \left(iB^{-1}A^\mu p_\mu \right)^{2k-1}+ \dots\nn\\
&&\quad\qquad
\mbox{}+\left(-p_\mu p^\mu \right)^k c_k \left(iB^{-1}A^\mu p_\mu \right)=0.
\end{eqnarray}
\end{subequations}
Note also that the numbers $c_j$ must be real, since 
\begin{subequations}
\begin{eqnarray}
B^\dagger &=&B, \quad \left(iA^\mu \right)^\dagger =iA^\mu ,\\
 \text{ and } \left(iB^{-1}A^\mu p_\mu \right)^\dagger &=&iA^\mu B^{-1}p_\mu  
=B \left(iB^{-1}A^\mu p_\mu \right)B^{-1}.
\end{eqnarray}
\end{subequations}
Taking the adjoint of the minimal equation then corresponds to making a
similarity transformation on the matrices and complex conjugating every $c_j$ 
term. Since each $B$ can be combined with a $B^{-1}$ term, and since we must
still have the same unique minimal equation, it follows that each 
$c_j=c_j^\ast $. All  of the above is a direct inference from the 
requirement of Lorentz invariance. 

If we now apply (\ref{5d25}) and (\ref{5d26}) to the field $\chi (p)$, and use 
(\ref{5d24}) we obtain for either case 
\begin{equation}\label{5d27}
\left[1+ \left(-p_\mu p^\mu \right)c_1+ \left(-p_\mu p^\mu \right)^2 c_2 +\cdot 
\cdot +\left(-p_\mu p^\mu \right)^k c_k\right]\chi (p)=0.
\end{equation}

\section{Fermi-Dirac Fields of Spin $1/2$}

Consider now
 only the simplest case of $k=1$, which corresponds to a minimal equation 
of degree 2. Then Eq.~(\ref{5d27}) becomes
\begin{equation}\label{5d28}
\left[1 + \left(-p_\mu p^\mu \right)c_1\right]\chi =0,
\end{equation}
and interpreting $c_1$ as $-m^{-2}$, for a non-vanishing $\chi $ 
we have the familiar relation $p_\mu p^\mu +m^2=0$.

We now go to the simplest case of all, that for which $n=2$ in 
Eq.~(\ref{5d25}); this then becomes
\begin{equation}\label{5d29}
\left(iB^{-1}A^\mu p_\mu \right)^2+\frac{p_\mu p^\mu }{m^2}=0.
\end{equation}
Writing
\begin{equation}
B=m\beta , \qquad A^\mu =i\alpha ^\mu , \text{ and } \beta ^{-1}\alpha ^\mu 
=\gamma ^\mu  
\end{equation}
then $\alpha ^\mu $ is Hermitian, and $\gamma ^{\mu \dagger} =
\beta \gamma ^\mu \beta ^{-1}$.
Eq.~(\ref{5d29}) becomes
\begin{equation}
\left(\gamma ^\mu p_\mu \right)^2=-p_\mu p^\mu =-g^{\mu \nu }p_\mu p_\nu 
=\gamma ^\mu \gamma ^\nu p_\mu p_\nu ,
\end{equation} 
and since only the symmetric combination $p_\mu p_\nu $ enters here, we have 
\begin{equation}\label{5d30}
\frac{1}{2} \left\{\gamma ^\mu ,\gamma ^\nu \right\}=-g^{\mu \nu }
\end{equation} 
as a necessary condition for Eq.~(\ref{5d29}) to be satisfied algebraically by 
$p_\mu $.

Before proceeding further, we remark that the construction of 5 
matrices, satisfying (\ref{5d30}), each of a definite symmetry (to represent 
the $\gamma ^\mu $ and $\gamma ^5=\gamma ^0\gamma ^1\gamma ^2\gamma ^3$),  
can be achieved in only one way: Three of the matrices must be symmetric, 
and the remaining two antisymmetric. This statement, which is easily verified, 
together with (\ref{5d29}) shows that we can construct matrices satisfying 
(\ref{5d30}) only for Fermi-DIrac fields, as follows:
Since $\alpha ^\mu $ is real, and $\beta $ imaginary, $\alpha ^\mu =
\beta \gamma ^\mu =-\beta \gamma ^{\mu \ast }$, 
i.e., all the $\gamma ^\mu $ are imaginary. From (\ref{5d30}) we note that 
\begin{equation}\label{5d301}
\left(\gamma ^0\right)^2=-g^{00}=+1 \qquad \left(\gamma ^k\right)^2=-g^{kk}=-1.
\end{equation} 
Defining $\gamma ^\mu =i\Gamma ^\mu $ where the $\Gamma ^\mu $are then real, 
it follows that
\begin{eqnarray}
\left(\Gamma ^0\right)^2&=&-1,  \qquad \left(\Gamma ^k\right)^2=+1,\nn\\
\sum_n\Gamma ^0_{jn}\Gamma ^0_{n\ell  }&=&- \delta _{j\ell },  \qquad \sum_n
\Gamma ^k_{jn}\Gamma ^k_{n\ell }=\delta _{j\ell }.\label{5d302}
\end{eqnarray} 
We are looking for matrices of definite symmetry; equating $j$ and $\ell $, 
(\ref{5d302}) can be satisfied only if $\Gamma ^0_{nj}=-\Gamma ^0_{jn}$, 
$\Gamma ^k_{nj}=+\Gamma ^k_{jn}$; i.e., $\gamma ^0$ is  antisymmetric 
(and imaginary, and therefore Hermitian), and the three $\gamma ^k $. are 
symmetric (and imaginary, and therefore skew-Hermitian). $\gamma ^5$ is then 
antisymmetric and real. It then follows that $\left\{\beta,\gamma ^k \right\}
=\left[\beta, \gamma ^0\right]=0$, and since $\beta $ and $\gamma ^0$ have the 
same properties, we may identify them: $\beta =\gamma ^0$.

If we now attempt to repeat this for Bose-Einstein fields, then the 
$\alpha ^\mu $ are anti-symmetric and imaginary, $\beta $ is symmetric and 
real,
 all the $\gamma ^\mu $ are the same as for the Fermi-Dirac case, and again 
$\left\{\beta, \gamma ^k \right\}=0$.  But this is a direct violation of the 
requirements that there be but three independent symmetric matrices, satisfying 
(\ref{5d301}). Equation (\ref{5d30}), from which the spin $\frac{1}{2}$ 
formalism is obtained, therefore refers only to Fermi-Dirac fields. 

The Lagrangian for the Fermi-Dirac spin $\frac{1}{2}$ field is then
\begin{equation}
\mathcal{L}=\frac{1}{2}\left[\psi ,i\alpha ^\mu \partial _\mu \psi \right] 
-\frac{m}{2}\left[\psi ,\beta\psi \right].
\end{equation}
More precisely, this is the Lagrangian for the uncharged Dirac~Majorana 
field---in order to represent charge we shall later have to double the number 
of components of $\psi $. For notation, we introduce the use of the dot 
$\cdot $ to symbolize the proper symmetrization brackets to be used for 
Fermi-Dirac or Bose-Einstein fields. Then
\begin{equation}
\mathcal{L}=\psi _\cdot i\alpha ^\mu \partial _\mu \psi -m\psi _\cdot 
\beta \psi.
\end{equation}
The field equations are then the familiar ones 
\begin{equation}
\left(\frac{1}{i}\alpha ^\mu \partial _\mu +m\beta \right)\psi =0,
\end{equation}
where, since $\psi $ is Hermitian, the adjoint of these equations must, and 
do, yield identical equations.

To obtain the commutation relations, we use: 
\begin{equation}
G_\psi = \int d\sigma _\mu \chi A^\mu \delta \chi \rightarrow \int d
\sigma _0\psi s^0\delta \psi,
\end{equation}
where $s^0=ia^0=i$ ($\beta=\gamma^0$), and form 
\begin{equation}
\left[\chi ,G_\chi \right]=\frac{i}{2}\delta \psi (x),
\end{equation}
or
\begin{equation}
\int d\sigma_0 \left\{\psi (x),\psi (x^\prime )s^0\right\}\delta\psi(x^\prime)
=\frac{i}{2}\delta \psi (x),
\end{equation}
where the factor $\frac{1}{2}$ enters because the $\psi $ are 
``non-canonical,'' and 
$[\,]\rightarrow  \{\, \}$ since $\delta \psi $ anti-commutes with $\psi $. 
This then yields
\begin{equation}
\left\{\psi (x),\psi (x^\prime )\right\}=
\frac{i}{2}\left(s^0\right)^{-1}\delta ^{(0)}(x-x_0),
\end{equation}
or
\begin{equation}\label{5d31}
\left\{\psi _\alpha (x),\psi _\beta (x^\prime )\right\}
=\frac{1}{2}\delta _{\alpha \beta }\delta ^{(0)}(x-x^\prime ),
\end{equation}
where $x$ and $x^\prime $ are points on the same space-like surface, 
and $(s^0)^{-1}=-i$.

\section{Spin Zero and One}
Let us return to the characteristic equation (\ref{5d26}) of the matrix
$\left(iB^{-1}A^\mu p_\mu \right)$. We will choose $k=1$, 
and examine the odd polynomial in $\left(iB^{-1}A^\mu p_\mu \right)$, 
which gives the possibility of describing a particle of zero mass. 
This is necessary to describe the electromagnetic field.

Define the four vector $\beta ^\mu \equiv imB^{-1}A^\mu $.
Then we have
\begin{eqnarray}
\left(\beta ^\mu p_\mu \right)^3 +p^\nu p_\nu \left(\beta ^\mu p_\mu \right)
&=&0, \nn\\
\mbox{or} \quad
\left(\beta ^\mu \beta ^\sigma \beta ^\nu +g^{\mu \nu }\beta ^\sigma \right)
p_\mu p_\sigma p_\nu &=&0.
\end{eqnarray}
From this equation, which is an identity in $p_\mu $, we can make statements 
only about the symmetric part of the matrix products. If we completely
symmetrize the matrix factor with respect to $\mu $, $\sigma $, and $\nu $, 
it must vanish. This is the sum of three terms of the form 
\begin{equation}
\beta ^\mu \beta ^\sigma \beta ^\nu +\beta ^\nu \beta ^\sigma \beta ^\mu 
=-g^{\mu \sigma  }\beta ^\nu -g^{\nu \sigma }\beta ^\mu,
\end{equation}
 which are the familiar Kemmer-Duffin commutation relations
\cite{petiau, duffin, kemmer} which are used to 
describe a particle of spin zero and one. All the $\beta $ matrices are 
singular. These matrices have 126 independent elements and hence are reducible 
to three sub-matrices of dimensionality 10, 5, and l; the sub-matrix of 
dimensionality 1 is trivially the null matrix; the one of dimensionality 5 and 
rank 2 describes a particle of spin 0; the matrix of dimensionality 10 and rank
6 describes a particle of spin 1.

The Lagrangian can be written, by choosing $B=m$, as 
\begin{equation}\label{5d32}
\mathcal{L}=\phi _\cdot i\beta ^\mu \partial _\mu \phi -m\phi _\cdot \phi, 
\end{equation}
which yields the equation of motion,
\begin{equation}\label{5d321}
\left(\beta ^\mu \frac{1}{i}\partial _\mu +m\right)\phi =0.
\end{equation}
Multiplying through by $\beta ^\sigma \beta ^\nu \frac{1}{i}\partial _\sigma 
\frac1i\partial_\nu$,
and symmetrizing with respect to $\sigma $ and $\mu $, using the commutation 
relations to reduce the triple matrix products and using the wave equation, 
we finally find 
\begin{equation}
\left[\frac{1}{i}\partial _\nu \frac{1}{i}\partial ^\nu +m\right]\phi =0,
\end{equation}
which verifies what has been put into the theory as the connection between 
energy and momentum. Thus each component of $\phi $ satisfies the Klein-Gordon 
equation.

The generator is
\begin{equation}\label{5d322}
G_\phi =-i\int d\sigma _0 \phi \beta ^0\delta \phi.
\end{equation}
The commutation relations are developed from
\begin{equation}
\left[\phi (x), \int d\sigma \phi (x^\prime )\beta ^0\delta \phi (x^\prime )
\right]=-\delta \phi (x),
\end{equation}
which yields 
\begin{equation}
\left[\phi _a(x),\left(\beta ^0\phi (x^\prime )\right)_b\right]
=-\delta ^\prime _{ab}\delta ^{(0)}(x-x^\prime ),
\end{equation}
where $\delta ^\prime _{ab}$ is a diagonal matrix having six ones and four 
zeros 
along the diagonal, which refer to the independent and dependent components, 
respectively. This equation cannot be solved for the commutation relations 
because $\beta ^0$ is a singular matrix. It eliminates the dependent components
of $\phi $ from the commutation relations.

If we multiply the field equation by $\left(1-(\beta ^0)^2\right)$  and use the
commutation relations, we find 
\begin{equation}
\left[\beta ^k \beta ^0 \beta ^0\partial _k +im\left(1-\beta ^0\beta ^0\right)
\right]\phi =0,
\end{equation} 
which is independent of time, and is the equation expressing the dependent
components of $\phi $ in terms of the independent components.

Let us re-examine the spin zero and spin one fields. Integral spin fields may 
be described in terms of ordinary tensors. They do not require the special 
spinor properties of the half-integral spin fields. We have seen that the spin 
zero representation of the Kenner-Duffin algebra has a dimensionality of five.
If the only tensor in addition to the field components which is introduced to 
form bilinear products in a scalar Lagrangian is the four-divergence, we must 
describe the field with tensors differing in rank by one. We shall construct the
 spin zero field with a scalar and four-vector as the necessary 5 components, 
and the spin one field with a four-vector and an anti-symmetrical second rank 
tensor as representing the 10 components.

\section{Spin Zero} 

The Lagrangian for a spin zero field is 
\begin{equation}\label{5d33}
\mathcal{L}=\frac{1}{2}\left(\phi _\cdot \partial _\mu \phi ^\mu -
\phi ^\mu _\cdot \partial _\mu \phi \right)
-\frac{m}{2}\left(\phi ^2-\phi _\mu \phi ^\mu \right).
\end{equation}
The field equations are determined by varying $\phi$ and $\phi ^\mu $,
\begin{equation}\label{5d331}
\partial _\mu \phi ^\mu =m\phi,  \qquad \qquad \partial _\mu \phi =m\phi _\mu, 
\end{equation}
which imply
\begin{equation}
\partial ^\mu \partial _\mu \phi =m\partial ^\mu \phi _\mu =m^2\phi,  \qquad 
\partial ^\mu \partial _\mu \phi ^\nu =\frac{1}{m}\partial ^\nu \partial ^\mu 
\partial _\mu \phi =m^2\phi ^\nu, 
\end{equation}
again yielding the Klein-Gordon equation.
 
The generator is 
\begin{equation}\label{5d333}
G= \int d \sigma _\mu \frac{1}{2}\left(\phi \delta \phi ^\mu -
\phi ^\mu \delta \phi \right)\rightarrow \int d\sigma _0 \frac{1}{2}
\left(\phi \delta \phi ^0-\phi ^0\delta \phi \right).
\end{equation}
Thus only $\phi$ and $\phi ^0$ are independent variables. This is reflected in 
the fact that the Klein-Gordon equation is second order, and hence we must 
specify both the wave function and its time derivative.\

The equations of motion have to be examined to see if the other components of 
the field are determined. We have from the field equations, as equations of 
motion,
\begin{equation}
\partial _0\phi =m\phi _0 \qquad  \partial _0\phi ^0=m\phi -\partial _k\phi ^k,
\end{equation}
and the following which is not an equation of motion, but does show that the 
$\phi ^k$ are determined in terms of the two independent components,
\begin{equation}
\partial _k\phi =m\phi _k.
\end{equation}
From the generator we see that we have one set of canonically conjugate
variables, $\phi $ and $\phi _0$. This means that the field has only one 
internal degree of freedom, and must be a spin zero field. This pair of 
conjugate variables must obey the canonical commutation relations (at
equal times), as is 
easily verified by using the generators for their respective change
\begin{equation}
\left[\phi (x),\phi (x^ \prime ) \right]=\left[\phi ^0(x),\phi ^0(x^ \prime )
\right]=0,\quad
 \left[\phi ^0(x),\phi (x^\prime )\right]=i \delta ^{(0)}(x-x^ \prime ),
\end{equation}
where $\delta ^{(0)}(x-x^\prime )$ means
$\delta (\mathbf{r},\mathbf{r}^\prime )$ on $\sigma_0 $.
The commutation relations obeyed by the dependent field components are derived 
from the field equations and these; for instance 
\begin{equation}
\frac{1}{i}\left[\phi ^0(x),\phi _k(x^\prime )\right]=-\frac{1}{m}\partial _k
\delta ^{(0)}(x-x^\prime ).
\end{equation}
These commutation relations can be written by inspection in a form not 
referring
to any particular coordinate system (where now $\delta ^{(\mu)}(x-x^\prime )$ 
means $\delta (\mathbf{r},\mathbf{r}^\prime )$ on $\sigma_\mu $), on the
spacelike surface $\sigma_\mu$,
\begin{eqnarray}
\left[\phi (x),\phi (x^\prime ))\right]&=&0 
\qquad \frac{1}{i}\left[\phi ^\mu(x) ,\phi (x^\prime )\right]
=\delta ^{(\mu )}(x-x^\prime ), \nn\\
\frac{1}{i}\left[\phi ^\mu(x) ,\phi ^\nu  (x^\prime )\right]&=&
-\frac{1}{m}\left[\partial ^\mu \delta ^{(\nu)} (x-x^\prime )
+\partial ^\nu \delta^{(\mu)}(x-x^\prime )\right].
\end{eqnarray}

\section{Spin One}
 
The description of a spin one field requires the use of a ten-component wave 
function. As we shall see, not all of these components are independent. We 
shall
 use a four-vector and an anti-symmetrical tensor of rank two. The Lagrangian 
is chosen to be
\begin{equation}\label{5d34}
\mathcal{L}=\frac{1}{2}\left(\phi _{\nu}{}_\cdot\partial _\mu G^{\mu \nu }
-G^{\mu\nu}{}_\cdot\partial _\mu \phi _\nu \right)
-\frac{m}{2}\left(\phi ^\mu \phi _\mu -\frac{1}{2}G^{\mu \nu }G_{\mu \nu }
\right). %\label{spin1lag}
\end{equation}
The factor of 1/2 in the last term is inserted because the 
unrestricted sum over
$\mu $ and $\nu $ counts each component twice. The field equations are
\begin{equation}\label{5d341}
\partial _\mu \phi _\nu -\partial _\nu \phi _\mu =mG_{\mu \nu }, \qquad 
\partial _\mu G^{\mu \nu }=m\phi ^\nu.
\end{equation}
The generator is
 \begin{equation}\label{5d342}
G= \int d\sigma _0 \frac{1}{2}\left(\phi_k\delta G^{0k}
-G^{0k}\delta \phi _k\right).
\end{equation}

 Thus only six of the ten components can be varied independently, $\phi _k $ 
and $G^{0k}$. Let us see if they have equations of motion and if the
other 4 components are determined in terms of the independent ones. The 
equations of motion are
\begin{equation}
\partial _0\phi _k=\partial _k\phi _0+mG_{0k}, \qquad 
\partial_0G^{0k}=m\phi^k-\partial_lG^{lk},
\end{equation}
for the independent field components. Also
\begin{equation}
\partial _k\phi _\ell -\partial _\ell \phi _k=mG_{k\ell }, \qquad 
m\phi ^0=\partial _kG^{k0}=-\partial _kG^{0k},
\end{equation}
which determine the four dependent field components. Thus all ten components of
the field are determined.

From the generator we see that we have 3 sets of canonically conjugate field 
variables; thus the field has 3 internal degrees of freedom. This corresponds 
to the three sub-states of a spin one field.

The commutation relations obeyed by the field components are obviously the 
canonically conjugate relations,
\begin{eqnarray}
\left[\phi ^k(x),\phi ^\ell (x^\prime )\right]&=&\left[G^{0k}(x),G^{0\ell }
(x^\prime )\right]=0 ,\nn\\
\frac{1}{i}\left[G^{0k}(x),\phi _\ell (x^\prime )\right]&=&\delta _\ell ^k
\delta ^{(0)}(x-x^\prime ).
\end{eqnarray}
The commutation relations obeyed by the dependent components can be realized 
from the above, by using their definitions in terms of the independent
components. For instance,
\begin{equation}
\frac{1}{i}\left[\phi ^0(x),\phi _\ell (x^\prime )\right]=-\frac{1}{m}
\delta ^k_\ell \partial _k\delta ^{(0)}(x-x^\prime )
=-\frac{1}{m}\partial _\ell \delta ^{(0)}(x-x^\prime).
\end{equation}

These relations can be generalized to refer to an arbitrary coordinate
system,
\begin{eqnarray}
\frac{1}{i}\left[\phi ^\mu (x),\phi ^\nu (x^\prime )\right]&=&-\frac{1}{m}
\left[\partial ^\mu \delta ^{(\nu)} (x-x^\prime )+\partial ^\nu \delta ^{(\mu)}
 (x-x^\prime )\right],\nn\\
\frac{1}{i}\left[G^{\mu \nu } (x),\phi_\lambda (x^\prime )\right]&=&
-\delta ^{\nu} _\lambda \delta ^{(\mu)} (x-x^\prime )
-\delta ^\mu _\lambda \delta ^{(\nu)} (x-x^\prime ),\nn\\
\frac{1}{i}\left[G^{\mu \nu } (x),G^{\lambda \kappa} (x^\prime )\right]&=&
-\frac{1}{m}
\left\{ g^{\mu \lambda }\left(\partial ^\nu \delta ^{(\kappa)}(x-x^\prime) 
+\partial ^\kappa\delta ^{(\nu)} (x-x^\prime )\right)\right\} \nn\\
    &&\qquad \text{ plus antisymmetrical terms} \dots.
\end{eqnarray}

\section{Electromagnetic Field}

The electromagnetic field is a spin one, massless field. A re-definition of the
field variables in the limit as the mass approaches zero will be made in the 
spin-one Lagrangian. Set
\begin{equation}
\sqrt{m}G^{\mu \nu }\equiv F^{\mu \nu },
\qquad \frac1{\sqrt{m}}\phi _\nu \equiv A_\nu.
\end{equation}
Then the Lagrangian (\ref{5d34})  becomes
\begin{equation}\label{5d35}
\mathcal{L}=\frac{1}{2}\left(A_{\nu \cdot }\partial _\mu 
F^{\mu \nu }-F^{\mu \nu }_\cdot \partial _\mu A_\nu \right)+ 
\frac{1}{4}F^{\mu \nu } F_{\mu \nu }.
\end{equation}
The field equations are
\begin{equation}\label{5d351}
\partial _\mu A_\nu -\partial _\nu A_\mu =F_{\mu \nu }, \qquad \partial _\mu 
F^{\mu \nu }=0,
\end{equation} 
and the generator becomes, in the local coordinate system, 
\begin{equation}\label{5d352}
G=\int d\sigma  \frac{1}{2}\left(A_k\delta F^{0 k }-F^{0k}\delta A_k \right).
\end{equation}
Thus, as in the case of a spin one non-zero mass field, only six of the field 
components can be varied independently. Their equations of motion are
\begin{equation}
\partial _0A_k=\partial _kA_0+F_{0k}, \qquad \partial _0F^{0k}=-\partial _\ell 
F^{\ell k}.
\end{equation}
 
We must now examine the rest of the field equations to see if the dependent 
components are determined,
\begin{equation}
\partial _kA_\ell -\partial _\ell A_k=F_{k\ell }, \qquad \partial _kF^{0k}=0.
\end{equation}
Thus $F^{kl}$  is determined, but because of the vanishing of the photon
mass, $A_0$ is not determined and a further restriction is placed on the
electric field, $F^{0k}$. This is exactly the freedom of making a gauge
transformation, which is allowed by the vanishing of the photon mass. 
That is,
\begin{equation}\label{5d36}
A_\mu \rightarrow A_\mu +\partial _\mu \Lambda . 
\end{equation}

It is convenient to split the vector part of $A_\mu $ into a longitudinal part,
which is the gradient of a scalar, and a transverse part, which is the curl of 
a vector. Thus we see that the gauge transformation affects only the 
longitudinal part of $A_k$. $F^{0k}$ is purely transverse because it has no 
divergence.

The equations of motion also break up into longitudinal and transverse 
components,  
\begin{equation}
\partial _0 A^{\text{L}}_k=\partial _kA_0
\qquad \partial _0A^{\text{T}}_k=F_{0k}.
\end{equation}
The first equation is entirely consistent with the possibility of making 
a gauge transformation which implies
\begin{equation}
 A_k^{\text{L}}=\partial _k \Lambda, \qquad A_0=\partial _0\Lambda.
\end{equation}   
By setting
\begin{equation}
A_k=\partial _k\Lambda +A_k^{\text{T}}
\end{equation}
in the generators and integrating by parts, or equivalently adding a surface 
term to the Lagrangian, we see that only the transverse components of $A_k$
appear in the generator,
\begin{equation}
G=\int d\sigma _0 \frac{1}{2}\left(A_k^{\text{T}}\delta F^{0k}-
F^{0k}\delta A_k^{\text{T}}\right).
\end{equation}

Since $F^{0k}$ and $A_k^{\text{T}}$ appear as canonical variables, we can 
impose the canonical commutation relations
\begin{subequations}
\begin{eqnarray}\label{5d36a}
\left[A_k^{\text{T}}(x),A_\ell ^{\text{T}}(x^\prime )\right]&=&
\left[F^{0k}(x),F^{0l}(x')\right]=0,\\
\label{5d361}
\left[F^{0k}(x),A^{\text{T}}_\ell (x^\prime )\right]&=&\delta ^k_\ell 
\left(\delta ^{(0)}(x-x^\prime )\right)^{\text{T}},
\end{eqnarray}
\end{subequations}
where only the transverse component of the term on the right side of 
(\ref{5d361}) can appear. If we define a term that subtracts off the 
longitudinal part of the delta function, the commutation relation can be
written as
\begin{equation}
\left[F^{0k}(x),A^{\text{T}}_\ell (x^\prime )\right]
=\delta ^k_\ell \delta ^{(0)}(x-x^\prime )-\partial ^k\partial ^\prime _\ell 
\mathcal{D}^0(x-x^\prime ).
\end{equation} 
The divergence of this purely transverse equation must vanish, hence
\begin{equation}
\delta ^k_\ell \partial _k\delta ^{(0)}(x-x^\prime )
=-\partial ^k\partial _k \partial _\ell \mathcal{D}^0(x-x^\prime ), 
\end{equation}
which can be satisfied if
\begin{equation}
\nabla ^2\mathcal{D}^0(x-x^\prime )=-\delta ^{(0)}(x-x^\prime ),
\end{equation}
or
\begin{equation}
\mathcal{D}^0(x-x^\prime )=\frac{1}{4\pi |\mathbf{x-x^\prime} | },
\end{equation}
which shows that the longitudinal part of $A_k$ is intimately connected with 
the Coulomb potential. %This connection will be clarified in a later paragraph.

\section{Introduction of Charge}
 
It was previously mentioned that the description of charge requires a doubling 
of the number of component fields $\chi $. 
%We develop the idea in this section, 
%reserving its generalization to arbitrary internal degrees of 
%freedom for the next chapter.

The simplest case to consider is the case of one internal degree of freedom. 
Calling the two basic fields $\chi _{(1)}$ and $\chi _{(2)}$, $\mathcal{L}$ 
[Eq.~(\ref{5d4})] becomes
\begin{equation}\label{5d37}
\mathcal{L}= \chi _{(1)\cdot }A^\mu \partial _\mu \chi _{(1)}+ 
\chi _{(2)\cdot }A^\mu \partial _\mu \chi _{(2)} -
\mathcal{H}(\chi _{(1)},\chi _{(2)}).
\end{equation}
Since both $\chi _{(1)}$ and $\chi _{(2)}$ have identical space-time 
properties, 
the kinematical portion of $\mathcal{L}$ is invariant under rotations and 
reflections in a new two-dimensional space composed of the components 
$\chi _{(1)}$ and $\chi _{(2)}$ in accordance with the sum-of-squares notation
of 
(\ref{5d37}). The rotations in 
this space, given by
\begin{equation}\label{5d38}
\overline{\chi }_{(1)}=\cos\lambda \chi _{(1)}+ \sin\lambda \chi_{(2)}, \qquad 
\overline{\chi }_{(2)}=-\sin\lambda \chi_{(1)} + \cos\lambda \chi _{(2)},
\end{equation}  
may be concisely expressed as 
\begin{equation}\label{5d39}
\overline{\chi }=e^{i\lambda q}\chi, 
\end{equation}
where
\begin{equation}\label{5d40}
\chi =\left(\begin{matrix} \chi _{(1)} \\ \chi _{(2)} \end{matrix}\right) 
\qquad q=\left(\begin{matrix} 
       0 & -i \\ i & 0 \\ \end{matrix}\right).
\end{equation}
The basic improper transformation (reflections) can be taken either as
\begin{equation}\label{5d41}
\overline{\chi }_{(1)}=\chi _{(2)}, \qquad \overline{\chi }_{(2)}=\chi _{(1)}
\end{equation}
or
\begin{equation}\label{5d411}
\overline{\chi }_{(1)}=\chi _{(2)}, \qquad \overline{\chi }_{(2)}=-\chi _{(1)}.
\end{equation}
Re-defining $A^\mu $ as
\begin{equation}
\left(\begin{matrix} A^\mu & 0 \\ 0 & A^\mu \\ \end{matrix}\right),
\end{equation}
allows us to write (\ref{5d37}) in matrix notation
\begin{equation}\label{5d42}
\mathcal{L}=\chi _\cdot A^\mu \partial _\mu \chi -\mathcal{H}(\chi ),
\end{equation}
where we have assumed that $\mathcal{H}$ has the required invariance under 
reflections and rotations in this space. Use of (\ref{5d39}) and 
$q^{\text{T}}=-q$ is then sufficient to guarantee the invariance of $\mathcal{L}$ 
under rotations. The reflection transformations in (\ref{5d41}) can be 
compactly written as
\begin{equation}\label{5d43}
\overline{\chi }=Ce^{i\frac{\pi }{2}q}\chi  \qquad \overline{\chi }=C\chi,
\end{equation}
where
\begin{equation}
C=\left( \begin{matrix} 0 & 1 \\ 1 & 0 \\ \end{matrix} \right)
\end{equation} 
Then the relations $C^{\text{T}}=C^{-1}=C$ show that $\mathcal{L}$ is invariant
under reflections.

The proper transformation (\ref{5d39}) can be evolved from a sequence of
infinitesimal transformations of the form
\begin{equation}\label{5d44}
\overline{\chi }=\left(1+i\delta \lambda q\right)\chi 
=\chi -\delta \chi  \qquad \delta\chi =-i\delta \lambda q\chi,
\end{equation}
and incorporating this infinitesimal variation into the action principle will 
yield a quantity which is conserved. The notation will suggest the
interpretation of this quantity as electrical charge, but until the manner in 
which it appears in interactions between the various fields is specified,
the nature of this charge is irrelevant.

If we now imagine that $\delta \lambda $ of [\ref{5d44}] is a continuous 
function of space~time, the finite change in $\lambda $ which builds up between
an initial $t(\sigma _2)$ and final $t(\sigma _1)$ times corresponds to 
different successive choices of $\chi $ The stipulated invariance of 
$\mathcal{L}$ then implies that $\delta _\lambda \mathcal{L}=0$. If $\delta 
\lambda $ were constant, $\delta _\lambda \mathcal{L}$ would vanish trivially,
in accordance with (\ref{5d39}) and (\ref{5d43}); we obtain something new since
the dependence of $\delta \lambda $ on $x^\mu $ introduces terms depending on 
$\partial _\mu \delta \lambda $:
\begin{equation}\label{5d45}
\delta _\lambda \mathcal{L}=-i\chi _\cdot A^\mu q\chi \partial _\mu \delta 
\lambda \equiv j^\mu \partial _\mu \delta \lambda =\partial _\mu \left[j^\mu 
\delta \lambda \right]-\delta \lambda \partial _\mu j^\mu,
\end{equation}
where
\begin{equation}
j^\mu \equiv -i\chi _\cdot A^\mu q\chi =-i\chi A^\mu q\chi 
=\chi _{(2)}A^\mu \chi _{(1)}-\chi _{(1)}A^\mu \chi _{(2)},\label{current6}
\end{equation}

The action principle now requires that the coefficient of $\delta\lambda $ 
vanish, yielding the differential conservation law
\begin{equation}\label{5d46}
\partial _\mu j^\mu =0.
\end{equation}
The term $\partial _\mu \left[j^\mu \delta \lambda \right]$ gives the generator
for changes on the boundary space-like surfaces $\sigma _1$. $\sigma _2$. 
Taking $\delta \lambda $ as constant over each surface, which corresponds to 
different, but definite, choices of the $\chi $ on $\sigma _{1,2}$, we have
\begin{equation}\label{5d47}
G_\lambda =\delta \lambda  \int d \sigma _\mu j^\mu \equiv Q\delta \lambda.
\end{equation} 
Further, if $\delta \lambda (\sigma _1)=\delta \lambda (\sigma _2)$  then it is
obvious that we get the same description as we would have obtained with 
$\delta \lambda =0$ in both surfaces, i.e.,
\begin{equation}\label{5d48}
\left[Q(\sigma _2)-Q(\sigma _1)\right]\delta \lambda =0, \quad 
\text{and \,therefore}\quad Q(\sigma _1)=Q(\sigma _2).
\end{equation}
This result also follows from (\ref{5d46}); the quantity $Q$, called the \
``charge", is therefore conserved.

Obtaining the commutation relations, from (\ref{5d44}) and (\ref{5d47}) we have
\begin{equation}\label{5d49}
\delta \chi =\frac{1}{i}\left[\chi ,G\right]=\frac{1}{i}\left[\chi ,Q 
\delta \lambda \right]=-iq\chi \delta \lambda,
\end{equation}
or 
\begin{subequations}
\begin{equation}
\left[\chi ,Q\right]=q \chi, 
\end{equation} 
or
\begin{equation}
\left[\chi _{(1)},Q\right]=-i\chi _{(2)}, \qquad \qquad 
\left[\chi _{(2)},Q \right]=-i \chi _{(1)}.
\end{equation} 
\end{subequations}
The generalization of (\ref{5d49}) for a finite rotation is
\begin{equation}\label{5d50}
e^{-i\lambda Q}\chi e^{i\lambda Q}=e^{i\lambda q} \chi,
\end{equation} 
which is analogous to translation induced by the momentum operator $P_\nu$,
\begin{eqnarray}
\left[\chi ,P_\nu \right]&=&\frac{1}{i}\partial _\nu \chi \nn\\
e^{-i\alpha \cdot P}\chi (x)e^{i\alpha \cdot P}&=&
e^{\alpha ^\nu \partial _\nu }\chi (x)=\chi (x+a).
\end{eqnarray}

There must also exist a unitary operator, in analogy with (\ref{5d50}), 
which accomplishes the improper transformations of equation (\ref{5d411}):
\begin{equation}\label{5d51}
\mathcal{C}^{-1}\chi \mathcal{C}= c\chi,\quad  \text{ or }\quad \begin{cases} 
\mathcal{C}^{-1} \chi _{(1)}\mathcal{C}=\chi _{(1)} \\
\mathcal{C}^{-1}\chi _{(2)}\mathcal{C}=-\chi _{(2)} 
\end{cases}.
\end{equation} 
From (\ref{5d47}) and (\ref{current6}) we then have 
\begin{equation}\label{5d52}
\mathcal{C}^{-1}Q\mathcal{C}=-Q.
\end{equation}
Thus, $\mathcal{C}$ has the interpretation of a charge reflection operator. 
Since $\mathcal{C}^2=+1$, its eigenvalues are $\mathcal{C} = \pm 1$. From 
(\ref{5d52}) we see that $\left[Q,\mathcal{C}\right] \not=0$ and the two 
operators cannot be simultaneously diagonalized, (in the representation of 
$\chi_{(1)}, \chi _{(2)}, \mathcal{C}$ is diagonal) \textit{except} for states
 of zero charge. If we arbitrarily assign the vacuum state the eigenvalue 
$\mathcal{C}^\prime  = +1$ , then this state of zero charge, $Q^\prime = 0$, 
has both a definite charge and charge symmetry.

To obtain a state (other than $Q^\prime =0$ ) of definite charge, where the
operator $Q$ is diagonalized, we must utilize non-Hermitian operators. Define
\begin{equation}\label{5d53}
\chi _{+}\equiv \chi _{(1)}-i\chi _{(2)}, \quad \chi _-=\chi _{(1)}+i\chi _{(2)},
\end{equation}
and then a simple calculation shows that 
\begin{equation}\label{5d54}
\left[\chi _+,Q\right]=\chi_ +, \qquad \left[\chi _-,Q\right]=-\chi _-,
\end{equation}
indicating that in this representation $Q$ is diagonal. However, $\mathcal{C}$ 
is no longer diagonal:
\begin{equation}
\mathcal{C}^{-1}\chi _{\pm}\mathcal{C}=\chi _{\mp}.
\end{equation}

From (\ref{5d54}) we see that if we have a state $| Q^\prime \rangle$
representing a definite charge, then
\begin{eqnarray}
Q \left(\chi _+| Q^\prime \rangle\right)&=&\left(Q^\prime -1\right)
\left(\chi _+| Q^\prime \rangle\right)\nn\\ 
Q \left(\chi _-| Q^\prime \rangle\right)&=&\left(Q^\prime +1\right)
\left(\chi _-| Q^\prime \rangle\right)
\end{eqnarray}
showing that $\chi _+$ creates a state for which the eigenvalue of $Q$ is 
$Q^\prime -1$, and conversely for $\chi _-$ These, then, represent charge 
annihilation and creation operators respectively. If we imagine that every 
physical state can be created by the action of these operators on the vacuum 
$| Q^\prime =0\rangle$ then the only possible values of $Q^\prime $ are $0$, 
$\pm 1$, $\pm 2$, \dots etc. (Note that charge annihilation represents the 
destruction of positive charge or the creation of negative charge, and vice 
versa.) To connect this with our previous particle description
in Sec.~\ref{sec:6.1}, where we had 
$\chi ^{(+)}$ and $\chi ^{(-)}$, we now obviously have the four possibilities: 
\begin{itemize}
\item $\chi ^{(+)}_+$ destroys a particle and decreases charge by 1, 
\item $\chi ^{(-)}_-$ creates a particle and increases charge by 1. 
\end{itemize}
These are inverse operations: $\left(\chi ^{(-)}_- \right)^\dagger 
=\chi ^{(+)}_+$.
\begin{itemize}
\item $\chi ^{(+)}_-$ destroys a particle but increases charge by 1. 
\item $\chi ^{(-)}_+$ creates a particle but decreases charge by 1.
\end{itemize} 
These are also inverse operations: $\left(\chi ^{(+)}_-\right)^\dagger 
=\chi ^{(-)}_+$.
This exhaustive description permits us to describe particles that carry charge.

In terms of the non-Hermitian variables defined by (\ref{5d53}), the 
Lagrangian (\ref{5d37}) becomes
\begin{equation}
\mathcal{L}=\frac{1}{2}\left[\chi _{-}{}_\cdot A^\mu \partial _\mu \chi _+ 
+\chi _{+}{}_\cdot A^\mu \partial _\mu \chi _-\right] - \mathcal{H},
\end{equation}
and the generator is
\begin{equation}
G = \int d\sigma _\mu \frac{1}{2}\left[\chi _-A^\mu \delta \chi _+
+\chi _+A^\mu \delta \chi _-\right],
\end{equation}
which shows that $\chi _+$ and $\chi _ -$ are a canonically conjugate set of 
field variables. The current (\ref{current6}) is
\begin{equation}
j^\mu =\frac{i}{2}\left[\chi _+A^\mu \chi _ - -\chi _ -A^\mu \chi _+\right].
\end{equation}
For variables of the second kind the commutation rules are
\begin{eqnarray}
\left\{\chi _+(x),\chi _-(x^\prime )\right\}A^0&=&i\delta ^{(0)}(x-x^\prime ),
\nn\\
\left\{\chi _{\pm }(x),\chi _{\pm }(x')\right\}&=&0.
\end{eqnarray}

\section{Quantum Electrodynamics}

As an illustration of the use of non-Hermitian fields, we now consider the 
ordinary charged Dirac field. We start with $\psi _{(1)}$ and $\psi _{(2)}$ as 
the two 4-component Hermitian fields which form the simplest charged field.
Let
\begin{subequations}
\begin{eqnarray}
\psi _{(1)}-i\psi _{(2)}=\psi _{+}&\equiv& \psi, \\
\psi _{(1)}+i\psi _{(2)}=\psi _{-}&\equiv&  \psi ^\dagger. 
\end{eqnarray}
\end{subequations}
There is an artificial asymmetry here depending on what is defined as $\psi $ 
and $\psi ^\dagger $. The commutation rules are now
\begin{eqnarray}
\left\{\psi _\alpha (x),\psi _\beta ^\dagger (x^\prime )\right\}&=&
\delta _{\alpha \beta }\delta ^{(0)}(x-x^\prime ), \nn\\
\left\{\psi _\alpha (x),\psi _\beta (x^\prime )\right\}&=&0 = 
\left\{\psi ^\dagger _\alpha (x),\psi ^\dagger _\beta (x^\prime )\right\}.
\end{eqnarray}
Since in this case, $A^0 = i$, the Lagrangian becomes:
\begin{equation}
\mathcal{L}=\frac{1}{2}\left[\psi ^\dagger _\cdot \alpha ^\mu i
\left(\partial _\mu \psi \right)-i\left(\partial _\mu \psi ^\dagger 
\right)_\cdot \alpha ^\mu \psi \right] -m\psi ^\dagger _\cdot 
\beta \psi.
\end{equation}
The current is then
\begin{equation}
j^\mu =\psi ^\dagger _\cdot \alpha ^\mu \psi.
\end{equation}
Setting
\begin{equation}
\alpha ^\mu =\beta \gamma ^\mu,  \qquad \psi ^\dagger \beta =\overline{\psi },
\end{equation}
the Lagrangian is
\begin{equation}
\mathcal{L}=\frac{1}{2}\left[\overline{\psi }_\cdot \gamma ^\mu 
i\partial _\mu \psi -i\partial _\mu \overline{\psi }_\cdot 
\gamma ^\mu \psi \right] - m\overline{\psi }_\cdot \psi.
\end{equation}

If we wish to interpret $j^\mu $ as the electric current, we can write down the
Lagrangian for the electromagnetic field plus the charged Dirac field and 
specify the coupling so that the usual field equations arise, in which $j^\mu $
acts as the source for the vector potential. The form of the coupling term is 
of course limited by Lorentz invariance and the spin $\frac{1}{2}$ and spin 1 
algebra.
Thus the most general Lagrangian which can be formed in this case is
\begin{eqnarray}
\mathcal{L}&=&\frac{1}{2}\left[\overline{\psi }_\cdot \gamma ^\mu 
i\partial _\mu \psi -i\partial _\mu \overline{\psi }_\cdot \gamma ^\mu \psi 
\right] - m\overline{\psi }_\cdot \psi\nn\\
&&\quad\mbox{}+ \left[A_{\nu \cdot}\partial _\mu F^{\mu \nu }-
F^{\mu \nu }_\cdot \partial _\mu A_\nu \right]\frac{1}{4}F^{\mu \nu }
F_{\mu \nu}\nn\\
&&\quad\mbox{}+eA^\mu _\cdot j_\mu +\frac{1}{2}\mu F^{\mu \nu }_\cdot 
\left(\overline{\psi }_\cdot \sigma _{\mu \nu }\psi \right).\label{5d55}
\end{eqnarray}
The symmetrization between $A_\mu $ and $j^\mu $ is necessary because in general,
the sources for the $A^\mu $ are partly the $j_\mu $ and hence these need not 
commute. The last term, a Pauli moment, even though it appears covariant, may, 
in fact, not be covariant due to its operator properties. These show up only in
higher order terms in the perturbation expansion.

The Lagrangian given by equation (\ref{5d55}) is invariant under the rotation 
and reflection of $\chi _{(1)}$ and $\chi _{(2)}$, which correspond to the 
replacements
\begin{equation}
\psi \rightarrow e^{i\lambda (x)}\psi, \quad\overline{\psi }\rightarrow 
e^{-i\lambda (x)}\overline{\psi },\quad
A_\mu \rightarrow A_\mu +\frac{1}{c}\partial _\mu \Lambda (x),
\end{equation}
and to the interchange of $\psi $ and $\overline{\psi}$, respectively. Note 
that in the second case---that of reflection---we must also replace 
$A_\mu \rightarrow -A_\mu $, $F_{\mu \nu }\rightarrow -F_{\mu \nu }$, in order to have
$\mathcal{L}$ unaltered.

We obtain the equations of motion and generators by variations
\begin{subequations}
\begin{eqnarray}
\delta \overline{\psi }:&&\qquad 
\left[\gamma ^\mu \left(\frac{1}{i}\partial _\mu -eA_\mu{}_\cdot \right) 
+ m\right]\psi
 =0, \label{5d56a}\\
\delta \psi:&&\qquad \overline{\psi }\left[\gamma ^\mu \left(i
\partial _\mu ^{\text{T}}-eA_{\mu \cdot }\right) +m \right]=0,\label{5d56b} \\
\delta F_{\mu \nu }:&&\quad \partial _\mu A_\nu -\partial _\nu A_\mu =
F_{\mu \nu },\label{5d56c}\\
\delta A_\mu:&& \quad \partial _\nu F^{\mu \nu }=e\overline{\psi }_\cdot 
\gamma ^\mu \psi \equiv j^\mu. \label{5d56d}
\end{eqnarray}
\end{subequations}
The generators are
\begin{equation}
G=\int d\sigma _\mu \frac{1}{2}\left[\overline{\psi} i\gamma ^\mu \delta \psi 
-\delta \overline{\psi }i\gamma ^\mu \psi -A_\nu \delta F^{\mu \nu }
-F^{\mu \nu }\delta A_\nu \right],
\end{equation}
and by adding appropriate 4-divergences, this becomes 
\begin{equation}\label{5d57}
G=\int d\sigma _\mu \left[\overline{\psi}i\gamma ^\mu \delta \psi 
-F^{\mu \nu }\delta A_\nu \right]=\int d\sigma _0
\left[\overline{\psi }i\gamma ^0\delta \psi  -F^{0k}\delta A_k\right],
\end{equation}
in a local coordinate system. In this form, the independent variables appear to
be only $\psi $ and the $A_k$. $\psi $ is definitely an independent variable; 
its equation of motion is Eq.~(\ref{5d56a}). To re-examine the 
electromagnetic field, we write Eqs.~(\ref{5d56c}) and (\ref{5d56d}) as
\begin{eqnarray}
\partial _0A_k&=&\partial _kA_0+F_{0k}, \qquad F_{k\ell }
=\partial _kA_\ell -\partial _\ell A_k, \nn\\
\partial _0F^{k0}&=&j^k-\partial _\ell F^{k0}, \qquad  \partial _kF^{0k}=j^0.
\label{5d58}
\end{eqnarray}
Since we have the freedom of making a gauge transformation, wherein the
longitudinal components of $A^\mu $ are then arbitrary, it is advantageous to 
re-write equations (\ref{5d58}) in terms of transverse (T) and longitudinal (L) 
components: 
\begin{eqnarray}
&&(1)\quad \partial _0A_k^{\text{T}} = F_{0k}^{\text{T}}, \qquad
(2)\quad \partial _0F^{k0 \text{T}}=j^{k\text{T}} - \partial _\ell F^{k\ell },
\nn\\
&&(3)\quad F_{k\ell }=F_{k\ell }^{\text{L}}
=\partial _k A_\ell -\partial _\ell A_k,
\qquad
(4)\quad \partial _0A^{\text{L}}_k = \partial _kA_0+F_{0k}^{\text{L}},\nn\\
&&(5)\quad \partial _0F^{k0\text{L}}=j^{k\text{L}},\qquad
(6)\quad \partial _kF^{0k\text{L}} = j^0.
\end{eqnarray}
Here, $F_{0k}^{\text{L}}\ne0$ 
since the electric field is no longer divergence-free. The 
solution of item (6) is immediately
\begin{equation}\label{5d60}
F^{0k \text{L}} = -\partial ^k \int d\sigma ^{(0)}\mathcal{D}_{(0)}
(x-x^\prime )j^0(x^\prime ),
\end{equation}
demonstrating that the longitudinal part of the electromagnetic field is 
not a kinematically independent quantity, but depends on the $j^0$ of all 
the charged particles present. Note that item (5) is not an independent 
statement,
since it is the result of the conservation law for $j^\mu (x)$. Since 
$A_k^{\text{L}}=\partial _k\Lambda(x)$, using item (5) shows that 
\begin{equation}
A_0=\partial _0\Lambda(x) + \int d\sigma ^\prime  \mathcal{D}^{(0)}(x-x^\prime )
j_0(x^\prime ),
\end{equation}
i.e., $A_0$ is completely arbitrary, corresponding to the freedom of choice of 
$\Lambda (x)$.

The generator in equation (\ref{5d57}) can now be written as 
\begin{equation}\label{5d61}
G=\int d\sigma \left[\overline{\psi } i\gamma ^0\delta \psi -F^{0k \text{T}}
\delta A^{\text{T}}_k-F^{0k\text{L}}\delta A_k^{\text{L}} \right].
\end{equation}
By the addition of a surface term, the last term in the
integrand in (\ref{5d61}) becomes equal to
$=j^0\delta \Lambda$,  which does not refer to the kinematically independent 
part of the electromagnetic field, and gives the generator for the change in 
the Dirac field when an infinitesimal gauge transformation is made---See
Eq.~(\ref{5d47}). The 
electromagnetic field commutation rules are again as given in equations
(\ref{5d36a}) and (\ref{5d361}); 
the anti-commutation relation for the Dirac fields works out to 
be
\begin{equation}
\left\{\psi (x),\overline{\psi }(x^\prime )\right\}
=\gamma ^0\delta ^{(0)}(x-x^\prime )
=-\gamma _0\delta ^{(0)}(x-x^\prime )\rightarrow 
-\gamma _\mu \delta ^{(\mu)} (x-x^\prime ).
\end{equation}

Using the non-Hermitian $\psi $, $\overline{\psi }$, the effect of the charge 
reflection operator $\mathcal{C}$ is to interchange the fields:
\begin{equation}\label{5d62}
\mathcal{C}^{-1}\psi \mathcal{C}=\psi ^\dagger,  \qquad \qquad 
\mathcal{C}^{-1}\psi ^\dagger  \mathcal{C}=\psi.
\end{equation}
In order to maintain the invariance of $\mathcal{L}$, we must now also require 
that this unitary operator reverse the sign of the $A_\mu $.
 \begin{equation}\label{5d63}
\mathcal{C}^{-1}A^\mu  \mathcal{C}=-A^\mu.
 \end{equation}
Again, $\mathcal{C}^2=+1$, and therefore its eigenvalues are 
$\mathcal{C}=\pm 1$.  From the arguments of the proceeding section, use of the 
transformation:
 \begin{equation}\label{5d62b}
\mathcal{C}^{-1}Q  \mathcal{C}=- Q
 \end{equation}
permits a classification, in terms of the quantum number $c^\prime $, 
for systems
of zero net charge. For example, consider the production of photons by repeated
application of the operator $A$ (actually  $A^{(-)}$, a photon
creation operator) to the vacuum state 
$| 0\rangle$, characterized by $c^\prime =+ 1$, and assume that no coupling 
terms
 are present in $\mathcal{L}$ (i.e., $e=0$). If $A$ is applied $n$ times to 
$| 0\rangle$ then by (\ref{5d63}) 
\begin{equation}
\mathcal{C}\left(A \cdot \cdot \cdot A \right)| 0\rangle = \left(-1\right)^n
\left(A\cdot \cdot \cdot A\right)| 0\rangle,
\end{equation} 
and we have a state $| (A^\prime \cdot \cdot\cdot A^\prime )\rangle$ whose 
eigenvalue 
of $\mathcal{C}$ depends on the (even or odd) number of photons present: 
$C^\prime =\left(-1\right)^n=\pm 1$.  Whichever +1 or -1 eigenvalue of 
$c^\prime $ 
we begin with persists as the coupling is turned on, even though what is now 
called the state of $n$ photons is a superposition of many different states.
 
As an example of this, let us consider the decay of positronium. In the absence
of the coupling term between $A_\mu $ and $\psi $ (i.e., $e=0$), the system of 
($e^+ + e^-$) is stable, and has a definite eigenvalue $c^\prime $. When the 
coupling is introduced, the only permitted final states will be those of the 
same eigenvalue. From our previous interpretation, $\psi $ is the operator 
which creates a particle $e^-$ and annihilates $ e^+$ , and conversely for 
$\psi ^\dagger $. To obtain a state with $e^-$ at $x$ and $e^+$ at $x^\prime $,
we form $\left(\psi ^\dagger (x)_{\cdot} \psi (x^\prime )\right)| 0\rangle$, 
and 
recognize that the actual state is a superposition of these with the amplitude 
wave function $\psi (x,x^\prime )$. Using (\ref{5d62}) we then have
\begin{eqnarray}
\mathcal{C}|e^+e^-\rangle&=&\mathcal{C} \int \int (dx) (dx^\prime) \psi 
(x,x^\prime )\left(\psi ^\dagger (x)_\cdot \psi (x^\prime )\right)| 0\rangle
\nn\\
&=& \mbox{}- \int \int (dx) (dx^\prime) \psi (x,x^\prime )\left(\psi ^\dagger 
(x^\prime )_\cdot \psi (x)\right)| 0\rangle\nn\\
&=&-\int \int (dx) (dx^\prime) \psi (x',x)\left(\psi ^\dagger (x)_\cdot
\psi (x^\prime )\right)| 0\rangle.
\end{eqnarray}

If the wave function $\psi (x,x^\prime )$ is symmetric, then $c^\prime =-1$; if
$\psi (x,x^\prime )$ is anti-symmetric, $c^\prime =+1$. As the coupling is now 
turned on, this no longer remains the exact state, but the $c^\prime $ value 
remains the same. Thus we have the selection rule for positronium: from a $^1S$
state it can decay only into an even number of photons ($c^\prime =+1$, 
$\psi (x,x^\prime )$ is antisymmetric), from a $^3S$ state it can decay only in
to an odd number of photons ($c^\prime =-1$, $\psi (x,x^\prime )$ is 
symmetric).
\chapter{Nonrelativistic Source Theory}
\label{ST}

The following is based on lectures given at Schwinger by Harvard in Spring 
1969, as transcribed by the author.  The goal was to construct a general theory
of particles, in a nonrelativistic context.  As such, this provides a 
transition between nonrelativistic quantum mechanics and source theory,
the general development of which was given later in Schwinger's three-volume
treatise \cite{Schwinger1970,Schwinger1973, Schwinger1989}.

\begin{figure}
\centering
\includegraphics{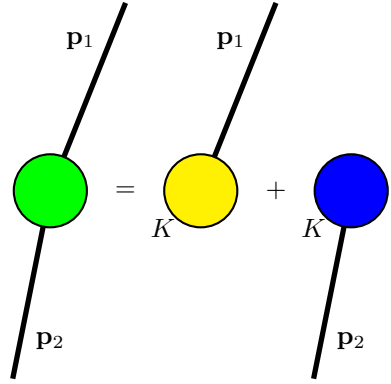}
\caption{\label{fig:s1} A scattering process in which a particle of momentum
$\mathbf{p}_2$ scatters into a state of a particle with momentum $\mathbf{p}_1$
can be thought of as a combination of two processes: One in which the particle 
is created by a source $K$ in the state $\mathbf{p}_1$ and a second in which
the particle of momentum $\mathbf{p}_2$ is absorbed by the source $K$.}
\end{figure}

The measurement symbol, or projection operator, which forms the
basis for Schwinger's approach to quantum mechanics \cite{Schwinger2001},
\be
|a',b'|=|a'\rangle\langle b'|
\ee
represents a idealized process in which the state $b'$ is annihilated and a new
state $a'$ is produced.  Wouldn't it be useful to separate these processes?
Recall how oscillator states, as discussed in    Chap.~\ref{sec:qap}, 
 were created by forces.  We generalize to sources
which can create and destroy particles.  A scattering process, which occurs
in a more or less localized scattering region, can be abstracted into a 
two-stage process, in which first an incoming particle is absorbed, and then a
new particle is created, as sketched in Fig.~\ref{fig:s1}.
In the individual processes the source $K$ acts to create or annihilate the
particle.

What does this mean quantum mechanically?  We must describe the processes
by probability amplitudes; we want to construct everything from the one
particle production mechanism, 
\be
\langle 1_{\mathbf{p}+}|0_-\rangle^K,\quad \langle 0_+|1_{\mathbf{p}-}
\rangle^K,
\ee
where the first amplitude represents the process in which a single particle
state of momentum $\mathbf{p}$ is created, where before the source acts only
the vacuum state is present, while the second represents the process in 
which a single particle of momentum $\mathbf{p}$ is absorbed after which
the vacuum state is present.

Because the processes occur in space and time, the source must be a function
$K(\mathbf{r},t)$, which exhibits a certain degree of localizability.  How
does the effectiveness of the source vary with different degrees of freedom?
The complementary measure is the corresponding function
in momentum space $K(\mathbf{p},E)$,
where, nonrelativistically,
 $E=p^2/(2m)$.  We expect the relationship between the production 
amplitude and the source function to be, at least for a weak source,
\be
\langle 1_{\mathbf{p}+}|0_-\rangle^K=\sqrt{\frac{(d\mathbf{p})}{(2\pi)^3}}(-i)
K(\mathbf{p}).
\ee
Here $(d\mathbf{p})=dp_1dp_2dp_3$.  The $-i$ factor is purely conventional
for later convenience.  The square root of the momentum-space element is
present to properly account for the density of states in the continuum picture.
To compute the annihilation amplitude $\langle 0_+|1_{\mathbf{p}-}\rangle$,
we can use orthogonality, $\langle 1_{\mathbf{p}-}|0_-\rangle=0.$  This must
be maintained by the dynamics.  The completeness relation
\be
1=|0_+\rangle\langle 0_+|+\sum_{\mathbf{p}}|1_{\mathbf{p}+}\rangle\langle
1_{\mathbf{p}+}|+\sum_{\mathbf{p,p'}}|1_{\mathbf{p}+}1_{\mathbf{p'}+}\rangle
\langle 1_{\mathbf{p}+}1_{\mathbf{p'}+}|+\dots,
\ee implies
\be
0=\langle 1_{\mathbf{p}-}|0_+\rangle^K\langle 0_+|0_-\rangle^K+
\sum_{\mathbf{p}'}\langle 1_{\mathbf{p-}}|1_{\mathbf{p'}+}\rangle\langle
1_{\mathbf{p'}+}|0_-\rangle^K+\dots,
\ee 
where we shall consider a weak source, so we will drop the higher terms.
To lowest order (in powers of the source)
\be
\langle 0_+|0_-\rangle^0=1,\quad \langle 1_{\mathbf{p}-}|1_{\mathbf{p}'+}
\rangle^0=\delta_{\mathbf{pp'}},
\ee
so
\be
0=\langle 1_{\mathbf{p}-}|0_+\rangle^K+\langle1_{\mathbf{p}+}|0_-\rangle^K,
\ee
or
\be
\langle 0_+|1_{\mathbf{p}-}\rangle^K=-\left[\langle 1_{\mathbf{p}+}|0_-
\rangle^{K}\right]^*.
\ee
Therefore, the effectiveness of producing or absorbing a particle by a weak
source is
\be
\langle 1_{\mathbf{p}+}|0_-\rangle^K=\sqrt{\frac{(d\mathbf{p})}{(2\pi)^3}}(-i)
K(\mathbf{p}),\quad
\langle 0_+| 1_{\mathbf{p}-}\rangle^K=\sqrt{\frac{(d\mathbf{p})}{(2\pi)^3}}(-i)
K(\mathbf{p})^*.\label{weak1p}
\ee
These equations, in fact, define what we mean by a source.

Now we need to seek the relation to the space-time description.  What happens
when a source is displaced,
\be
\bar K(\mathbf{r},t)=K(\mathbf{r+R},t+T).
\ee
Relativity (here Galilean) means that the same effect occurs by displacing
the space-time coordinate system to which the initial and final states are
referred, the generator of such an infinitesimal displacement being
\be
G=\mathbf{P}\cdot \delta\bepsilon-H\delta t,
\ee
which implies that the displacement operator is
\be
U=e^{i\mathbf{P\cdot R}-iHT}.
\ee
Thus the 1 particle states change according to
\be
\langle 1_{\mathbf{p}}|\to\langle 1_{\mathbf{p}}|e^{i\mathbf{P\cdot R}-iHT}
=e^{i\mathbf{p\cdot R}-iET}\langle 1_{\mathbf{p}}|,
\ee
while the vacuum state is unchanged,
\be
\langle 0_+|\to\langle 0_+|.
\ee
Thus if the one particle amplitude is proportional to $K(\mathbf{p})$,
\be
\langle 1_{\mathbf{p}+}|0_-\rangle^K\sim K(\mathbf{p}),
\ee
that of the displaced source is
\be
\langle 1_{\mathbf{p}+}|0_-\rangle^{\bar K}\sim e^{i\mathbf{p\cdot R}-iET}
K(\mathbf{p})\sim \bar K(\mathbf{p}).
\ee
This implies that $K(\mathbf{p})$ is obtained by Fourier transforming
$K(\mathbf{r},t)$,
\be
K(\mathbf{p})=K(\mathbf{p},E)=\int(d\mathbf{r})dt e^{-i\mathbf{p\cdot r}+iEt}
K(\mathbf{r},t).\label{ftsource}
\ee
In general, one wants to remove the connection between $E$ and $\mathbf{p}$.

\begin{figure}
\centering
\includegraphics{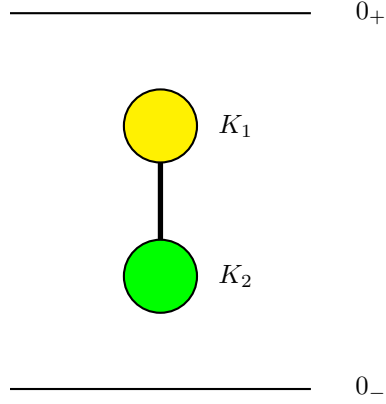}
\caption{\label{fig:s2} Exchange of a particle between a source $K_2$ and
a source $K_1$, where the latter is localized later than the former.  Before
the first source acts, the system is in the vacuum state, as it is after
the second source acts.  Time is imagined as plotted vertically in these 
``causal'' diagrams.}
\end{figure}

What about strong sources?  Remember for the oscillator, the most basic
object was the ground-state persistence amplitude $\langle 0_+|0_-\rangle^K$.
The latter contains the process of the exchange of particles between temporally
separated sources, as illustrated in Fig.~\ref{fig:s2}.
Here a single particle is emitted by source $K_2$ and absorbed later by the
source $K_1$.  Before and after either source acts, the system is in the
vacuum, no-particle, state.  We imagine the total source to be the sum of
the two components,
\be
K(x)=K_1(x)+K_2(x).
\ee
The decomposition shown is called a causal arrangement.  Because we are
so far considering weak sources, when we insert a complete set of states
at an intermediate time between the action of the two sources, the vacuum
persistence amplitude is
\be
\langle 0_+|0_-\rangle^K=\langle 0_+|0_-\rangle^{K_1}
\langle 0_+|0_-\rangle^{K_2}+\sum_{\mathbf{p}}\langle 0_+|1_{\mathbf{p}-}
\rangle^{K_1}\langle 1_{\mathbf{p}+}|0_-\rangle^{K_2}+\dots.
\ee
Now using the one-particle creation and annihilation amplitudes (\ref{weak1p}),
together with the Fourier transform (\ref{ftsource}) we see that the 
one-particle exchange term here is
\be
-i\int(d\mathbf{r})dt (d\mathbf{r}')dt' K_1^*(\mathbf{r},t)
\left[-i\int\frac{(d\mathbf{p})}{(2\pi)^3}e^{i\mathbf{p\cdot (r-r')}-iE(t-t')}
\right]K_2(\mathbf{r}',t').
\ee
But the source is a unitary whole---results can depend only on the total source
$K$ and not its parts.  This is a statement of the uniformity of space and 
time.  This will introduce terms that will refer to each component source
separately; what we don't want is a term that involves $K_2^*$ and $K_1$;
therefore, we define the retarded Green's function
\be
G(\mathbf{r-r'},t-t')=-i\eta(t-t')\int
\frac{(d\mathbf{p})}{(2\pi)^3}e^{i\mathbf{p\cdot (r-r')}-iE(t-t')},
\ee
in terms of which we infer
\be
\langle 0_+|0_-\rangle^K=1-i\int(d\mathbf{r})dt (d\mathbf{r}')dt' 
K^*(\mathbf{r},t)G(\mathbf{r-r'},t-t')K(\mathbf{r}',t').
\ee
The Green's function satisfies the differential equation
\be
\left[i\frac{\partial}{\partial t}-\frac{\left(\frac1i\bnabla\right)^2}{2m}
\right]G(\mathbf{r-r'},t-t')=\delta(t-t')\delta(\mathbf{r-r'}).\label{gfeqna}
\ee

\begin{figure}
\centering
\includegraphics{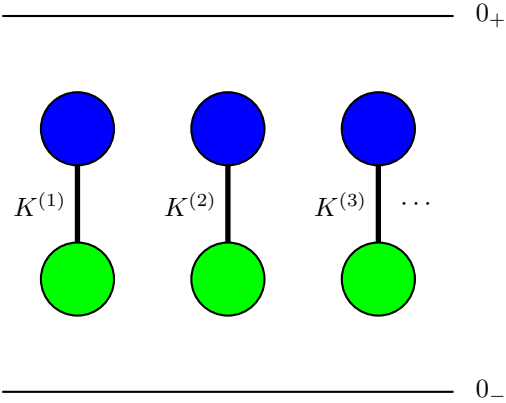}
\caption{\label{fig:beam} Exchange of noninteracting particles between 
spatially nonoverlapping sources.  Particles emitted by the lower source
of $K^{(1)}$ are only detected by the upper source of the same set.  It is
arranged that there is no cross-coupling.} 
\end{figure}

Now we want to remove the restriction to weak sources.  Suppose we have a
beam of noninteracting particles, detected for example by spatially separated
sources, as illustrated in Fig.~\ref{fig:beam}.
 Each particle is produced and
detected by a single pair of weak sources $K^{(\alpha)}$, $\alpha=1, 2, 3,
\dots$.  There is no interaction between different pairs of sources.  Thus
the vacuum persistence amplitude for this arrangement is
\be
\langle 0_+|0_-\rangle^K=\prod_\alpha\left[1-i\int(d\mathbf{r})dt(d\mathbf{r}')
dt' K^*(\mathbf{r},t)G(\mathbf{r-r}',t-t')K(\mathbf{r}',t')\right]^{(\alpha)},
\ee
because each source is weak.  But only $K$, not $K^{(\alpha)}$ should enter;
physics shouldn't depend on the channel.  So
\be
\langle 0_+|0_-\rangle^K=\prod_\alpha e^{-i\int (K^*GK)^{(\alpha)}}=
e^{-i\sum_\alpha \int K^{(\alpha)}{}^*GK^{(\alpha)}}=e^{-i\int K^*GK}, 
\ee
where the last step depends upon the arrangement that prohibits cross
coupling between the component sources.  This looks just like the structure
we saw for the harmonic oscillator (\ref{gspa}), 
except now the integrals are over space as well as time:
\be\langle 0_+|0_-\rangle^K=e^{-i\int(d\mathbf{r})dt(d\mathbf{r}')dt'
K^*(\mathbf{r},t)G(\mathbf{r-r'},t-t')K(\mathbf{r}',t')}.
\ee
We will apply this to generalized (not weak) 
emission and absorption processes, but
still assuming that the particles are not interacting.

\begin{figure}
\centering
\includegraphics{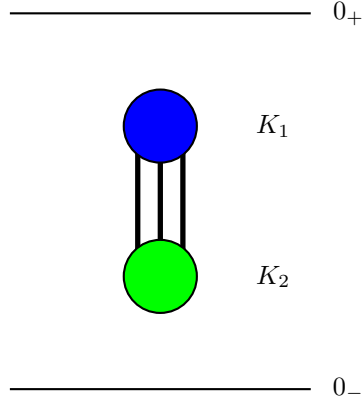}
\caption{\label{fig:s4} Exchange of noninteracting particles between causally
separated sources, $K_2$ and $K_1$.}
\end{figure}

Figure \ref{fig:s4} shows the exchange of noninteracting particles between
causally separated sources, $K=K_1+K_2$.  The total vacuum persistence 
amplitude is
\be
\langle 0_+|0_-\rangle^K=\langle 0_+|0_-\rangle^{K_1}
\langle 0_+|0_-\rangle^{K_2}e^{-i\int K_1^*GK_2}\label{causalex}.
\ee
The fact that there is a causal relation between the two sources means that
\be
G(\mathbf{r-r'},t-t')=-i\int\frac{(d\mathbf{p})}{(2\pi)^3}e^{i\mathbf{p\cdot
(r-r')}-iE(t-t')}.
\ee
Define the discrete specification of the source in momentum space as
\be
K_p=\sqrt{\frac{(d\mathbf{p})}{(2\pi)^3}}K(p).
\ee
Then the final exponential term in Eq.~(\ref{causalex}) is
\be
e^{\sum_p(-iK_{1p})^*(iK_{2p})}.
\ee
On the other hand, if we insert a complete set of multiparticle states at
an intermediate time, we have
\be
\langle 0_+|0_-\rangle^K=\sum_{\{n\}}\langle 0_+|\{n\}\rangle^{K_1}
\langle \{n\}|0_-\rangle^{K_2}.
\ee
Compare this with the expansion of Eq.~(\ref{causalex}):
\be
\langle 0_+|0_-\rangle^K=\langle 0_+|0_-\rangle^{K_1}
\prod_{\mathbf{p}}\sum_{n_{\mathbf{p}}=0}^\infty
\frac{(-iK_{1p}^*)^{n_{\mathbf{p}}}}{\sqrt{n_{\mathbf{p}}!}}
\frac{(-iK_{2p}^*)^{n_{\mathbf{p}}}}{\sqrt{n_{\mathbf{p}}!}}
\langle 0_+|0_-\rangle^{K_2},
\ee
where the occupation numbers in a given momentum cell are given by
$\{n\}=\{n_{\mathbf{p}}\}$.  From this we infer the probability amplitudes
for producing and absorbing particles by a strong source:
\begin{subequations}
\bea
\langle \{n\}|0_-\rangle^K&=&\prod_{\mathbf{p}}\frac{(-iK_{\mathbf{p}})^{n_
{\mathbf{p}}}}{\sqrt{n_{\mathbf{p}}!}}\langle 0_+|0_-\rangle^K,\label{mpprod}
\\
\langle 0_+|\{n\}\rangle^K&=&\prod_{\mathbf{p}}\frac{(-iK^*_{\mathbf{p}})^{n_
{\mathbf{p}}}}{\sqrt{n_{\mathbf{p}}!}}\langle 0_+|0_-\rangle^K.\label{mpdet}
\eea
\end{subequations}
As a check of this, we verify that the total probability must be unity:
\be
1=\sum_{\{n\}}p(\{n\},0)^K=|\langle 0_+|0_-\rangle^K|^2e^{\sum_{\mathbf{p}}
|K_{\mathbf{p}}|^2}.\label{probcond}
\ee
Independently,
\bea
|\langle 0_+|0_-\rangle^K|^2&=&\exp\bigg[-\int(d\mathbf{r})dt(\mathbf{r}')dt'
K^*(\mathbf{r},t)iG(\mathbf(\mathbf{r-r'},t-t')K(\mathbf{r}',t') \nn\\
&&\quad\mbox{}-\int(d\mathbf{r})dt(\mathbf{r}')dt'
K^*(\mathbf{r},t)[iG(\mathbf(\mathbf{r'-r},t'-t)]^*
K(\mathbf{r}',t')\bigg].\nn\\
\eea
But the combinations of Green's functions appearing here is
\be iG(\mathbf(\mathbf{r-r'},t-t')+[iG(\mathbf(\mathbf{r'-r},t'-t)]^*
=\int\frac{(d\mathbf{p})}{(2\pi)^3}e^{i\mathbf{p\cdot(r-r')}-iE(t-t')},
\ee
which is a solution of the homogeneous equation, so indeed
\be
|\langle 0_+|0_-\rangle^K|^2 = e^{-\sum_{\mathbf{p}}|K_{\mathbf{p}}|^2},
\ee
so the probability condition (\ref{probcond}) is satisfied.

The description so far of the exchange of particles between sources is a sort
of action at a distance picture.  We are often concerned with excitations 
produced by a source---a more local description.  A test source is used to
measure effects.  So let us add an additional infinitesimal source,
\be
K(\mathbf{r},t)\to
K(\mathbf{r},t)+\delta K(\mathbf{r},t),\ee
which results in the following infinitesimal change in the action,
\be
\delta W=-\int(d\mathbf{r})dt[\delta K^*(\mathbf{r},t)\psi(\mathbf{r},t)
+\delta K(\mathbf{r},t)\psi^*(\mathbf{r},t)],
\ee
which defines new objects which refer to the pre-existing situation.   Here
%\alpheqn
\bea
\psi(\mathbf{r},t)&=&\int(d\mathbf{r}')dt' G(\mathbf{r-r'},t-t')K(\mathbf{r'},t)
\label{deffield}\\
\psi^*(\mathbf{r},t)&=&\int(d\mathbf{r}')dt' K^*(\mathbf{r'},t) 
G(\mathbf{r'-r},t'-t)K(\mathbf{r}',t');
\eea
%\reseteqn
the latter is not the complex conjugate of $\psi$, because
\be
G(\mathbf{r-r'},t-t')^*\ne G(\mathbf{r'-r},t'-t),
\ee
$G$ being the retarded Green's function.  The differential equation 
satisfied by the Green's function (\ref{gfeqna}) is
\be
\left(i\frac\partial{\partial t}-T\right)G(\mathbf{r-r'},t-t')=
\delta(\mathbf{r-r'})\delta(t-t'),
\ee
where $T=-\nabla^2/(2m)$ is the kinetic energy differential operator.  
Therefore, the {\it field\/} $\psi$ satisfies
\be
\left(i\frac\partial{\partial t}-T\right)\psi(\mathbf{r},t)=K(\mathbf{r},t).
\label{fe1}
\ee  This is analogous to the equation satisfied by the harmonic
oscillator variable (\ref{eom:fho}), or
\be
\left(i\frac{d}{dt}-\omega\right)y(t)=K(t).
\ee
Now because
\be
\left(-i\frac\partial{\partial t}-T\right)G(\mathbf{r'-r},t'-t)=
\delta(\mathbf{r-r'})\delta(t-t'),
\ee
the field $\psi^*$ satisfies
\be
\left(-i\frac\partial{\partial t}-T\right)\psi^*(\mathbf{r},t)
=K^*(\mathbf{r},t),\label{fe2}
\ee
which is the complex conjugate equation.  The boundary conditions are
different in the two cases: $\psi$ is a retarded solution, while $\psi^*$ is
an advanced solution.

What does $W$ have to do with action?  Let us write the alternative forms
\bea
W&=&-\int(d\mathbf{r})dt\, K^*(\mathbf{r},t)\psi(\mathbf{r},t) =
-\int(d\mathbf{r})dt\, \psi^*(\mathbf{r},t) K(\mathbf{r},t)\nn\\
&=&
-\int(d\mathbf{r})dt\, \psi^*(\mathbf{r},t)\left(i\frac\partial{\partial t}
-T\right) \psi(\mathbf{r},t).
\eea
Combining these forms appropriately, we can write
\be
W=\int(d\mathbf{r})dt\left[\psi^*\left(i\frac\partial{\partial t}-T\right)\psi
-K^*\psi-\psi^*K\right].  
\ee
Think of this last as a functional of $K$, $K^*$, $\psi$, and $\psi^*$,
so
\be
\delta W=\int (d\mathbf{r})dt\left[-\delta K^*\psi-\psi^*\delta K\right]
+\delta_{\psi,
\psi^*}W.
\ee
But the definition (\ref{deffield})
of $\psi$ and $\psi^*$ shows that the last variation is 
zero, which is a statement of the stationary action principle:
\bea
\delta_{\psi,\psi^*}W&=&\int(d\mathbf{r})dt\, \delta\psi^*\left[\left(i\frac
\partial{\partial t}-T\right)\psi-K\right]\nn\\
&&\quad\mbox{}+\int(d\mathbf{r})dt\, \delta\psi\left[\left(-i\frac
\partial{\partial t}-T\right)\psi^*-K^*\right]=0.
\eea
That is, the stationary action principle, that $W$ is unchanged under
infinitesimal field variations, supplies the equations of motion
(\ref{fe1}) and (\ref{fe2}).

Fundamental to physics is the notion of the uniformity of space and time, 
that the laws of physics are independent of the locale.  This is reflected
in the indistinguishability of identical particles, which, in turn, 
is reflected in the
probability amplitude of a source producing a multi-particle distribution,
Eq.~(\ref{mpprod}), which says all that is possible.  Experimentally, we know
of two kinds of statistics.  Here $n_{\mathbf{p}}$ is unlimited, so this cannot
refer to Fermi-Dirac statistics.  In the Bose-Einstein case we have stimulated
emission.  Let us see this.

\begin{figure}
\centering
\includegraphics{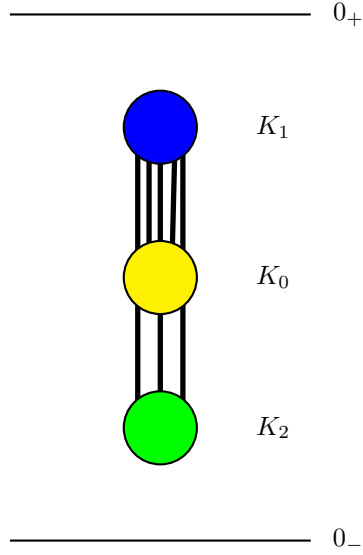}
\caption{\label{fig:s5} Effect of a weak source $K_0$ acting at an
intermediate time between strong sources $K_2$ and $K_1$.}
\end{figure}

Figure \ref{fig:s5} shows the interchange of noninteracting particles between
sources $K_2$ and $K_1$, but now with a {\it weak\/} source
$K_0$ in between.
The total source is composed of three causally separated pieces,
\be
K=K_1+K_0+K_2.
\ee
Using this causal arrangement, the vacuum persistence amplitude is
\bea
\langle 0_+|0_-\rangle^K&=&e^{-i\int K^*GK}\nn\\
&=&\langle 0_+|0_-\rangle^{K_1+K_2}\langle 0_+|0_-\rangle^{K_0}
e^{-i[\int K_1^*GK_0+\int K_0^*GK_2]}.
\eea
Here, because the disturbance by $K_0$ is regarded as weak, we approximate
$\langle 0_+|0_-\rangle^{K_0}\approx1$, and the exponential term is expanded to
first order in $K_0$:
\bea
\langle 0_+|0_-\rangle^K&\approx&\sum_{\{n\}}\langle 0_+|\{n\}\rangle^{K_1}
\langle \{n\}|0_-\rangle^{K_2}\nn\\
&&\times\left\{1+\sum_{\mathbf{p}}[(-iK_{1\mathbf{p}}^*)
(-iK_{0\mathbf{p}})+(-iK_{0\mathbf{p}}^*)(-iK_{2\mathbf{p}})]\right\}.
\eea
Compare this with the multi-particle exchange description between the three
sources,
\be
\langle 0_+|0_-\rangle^K=\sum_{\{n\}\{n\}'}\langle 0_+|\{n\}_-\rangle^{K_1}
\langle\{n\}_+|\{n\}'_-\rangle^{K_0}\langle \{n\}'_+|0_-\rangle^{K_2}.
\ee
Now recall the connection between $K_p$ and $\langle \{n\}|0_-\rangle^K$, 
Eq.~(\ref{mpdet}), so we infer
\be
\langle 0_+|\{n\}\rangle^{K_1}(-iK_{1\mathbf{p}}^*)=\sqrt{n_{\mathbf{p}}+1}
\langle0_+|\{n\}+1_{\mathbf{p}}\rangle^{K_1}.
\ee
Therefore, for a weak source,
\be
\langle \{n+1_{\mathbf{p}}\}|\{n\}\rangle^K=-iK_{\mathbf{p}}
\sqrt{n_{\mathbf{p}}+1}.
\ee
The corresponding probability of creating 1 more particle with momentum $\mathbf
{p}$ is
\be
|\langle \{n+1_{\mathbf{p}}\}|\{n\}\rangle^K|^2=|K_{\mathbf{p}}|^2
(n_{\mathbf{p}}+1).
\ee
In the last factor, the 1 represents spontaneous emission, and the $n_{\mathbf
{p}}$ is the enhancement effect of stimulated emission.

The corresponding analysis in the absorption case gives
\be
\langle \{n\}|\{n+1_{\mathbf{p}}\}\rangle^K=-iK_{\mathbf{p}}^*\sqrt{n_{\mathbf
{p}}+1},
\ee
or
\be
\langle \{n-1_{\mathbf{p}}\}|\{n\}\rangle^K=-iK_{\mathbf{p}}^*
\sqrt{n_{\mathbf{p}}},
\ee
that is, the probability of absorbing one particle is proportional to the
incident intensity.

\begin{figure}
\centering
\includegraphics{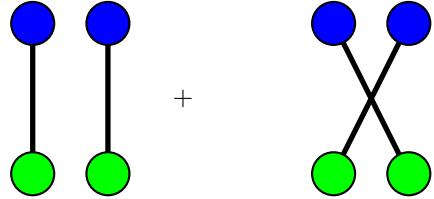}
\caption{\label{fig:s6} Exchange of two particles between spatially and
temporally separated sources.  Bose-Einstein symmetry implies that the
particles are exchanged between  either spatially separated source.  No 
interaction is to be inferred where the lines cross.}\end{figure}

Let us come back to space and time.  Adopting a more telegraphic notation,
we can write the vacuum persistence amplitude as
\bea
&&\langle 0_+|0_-\rangle^K=e^{-i\int K^*GK}=1-i\int d1\,d1' K^*(1)G(1-1')
K(1')\nn\\
&&\qquad\mbox{}-\frac12\int d1\,d1'\,d2\,d2'\,K^*(1)K^*(2)G(1-1')G(2-2')K(1')
K(2')+\dots.\label{spex}\nn\\
\eea
Here the numbers represent space-time points, $1=\mathbf{r}_1, t_1$, etc. 
Because the product of sources is symmetrical, we can replace
\be
G(1-1')G(2-2')\to\frac12[G(1-1')G(2-2')+G(1-2')G(2-1')],
\ee
Diagrammatically, the two terms can be represented as in Fig.~\ref{fig:s6}.
This builds in the symmetry in the labels.  The third term in Eq.~(\ref{spex})
is
\be
-\frac12\int d1\,d2\, K^*(1)K^*(2)\psi(1)\psi(2),\quad
\psi(1)\psi(2)=\psi(1,2)=\psi(2,1),
\ee
which exhibits another characteristic of Bose-Einstein statistics.  The field
$\psi$, the generalization of the wavefunction, must be totally symmetric
under interchange of the particles.

\section{Interactions}

\begin{figure}
\centering
\includegraphics{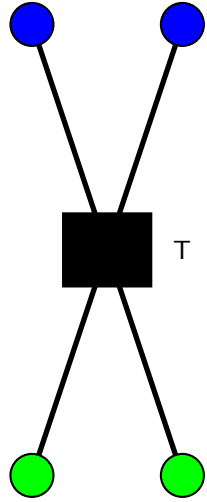}
\caption{\label{fig:s7} Two particles produced by the two earlier 
sources scatter by the processes labeled $\tens{T}$ 
and the scattered particles are detected by the two later sources.}
\end{figure}

\begin{figure}
\centering
\includegraphics{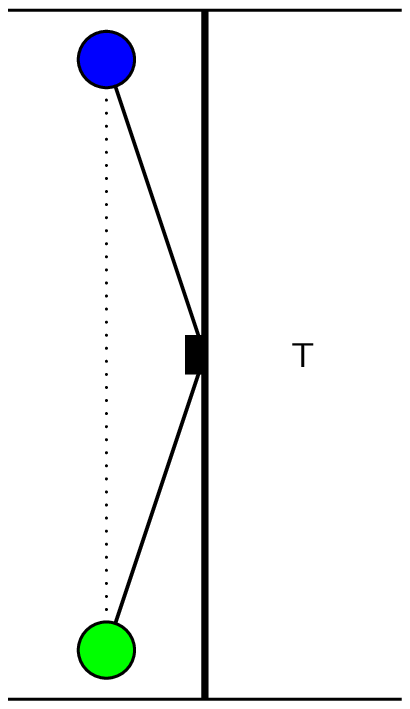}
\caption{\label{fig:s8} Diagram representing scattering of a particle
off a fixed center.  Here the heavy vertical line represents the fixed center
or nucleus, the dotted line represents the exchange of a particle directly
between the sources, and the solid lines represent the production of a
particle by the first source, its scattering off the fixed center, by
the process represented by the black box $\tens{T}$, 
and finally its absorption by the second source.}  
\end{figure}

Real particles interact with each other.  Thus we should have processes
such as sketched in Fig.~\ref{fig:s7}.
The particle emitted by the sources are scattered by interactions
represented by the black box labeled $\tens{T}$.
Let us begin with a simpler situation, scattering from a fixed center, for
example, a heavy nucleus.  This process can be represented with only two
sources, as shown in Fig.~\ref{fig:s8}.  
This diagram schematically represents the tracks of particles as seen in
a detector.  In the absence of interactions
\be
W=-\int K^* G^0 K,
\ee
where now we have used the superscript 0 to designate the free particle 
propagator or Green's function previously denoted simply by $G$.  This 
process in represented by the dotted line in Fig.~\ref{fig:s8}.  Now we
want to add something to this, the scattering process, represented generically
by $\tens{T}$.  
From a source point of view, both processes contribute to the vacuum
amplitude
\be
W=-\int K^*G^0K-\int K^*G^0\tens{T}G^0K,
\ee
where the last term means
\bea
&&-\int K^*(1)G^0(1-1')\tens{T}(1',1'')G^0(1''-1''')K(1''')\nn\\
&&\qquad=-\int\psi^{0*}(1')
\tens{T}(1',1'')\psi^0(1'').
\eea
The vacuum persistence amplitude is expressed as
\be
\langle0_+|0_-\rangle=e^{iW}
\ee 
for the same reason as before.
Because we have a well-defined causal situation,
\be
\psi^0(1)=\int G^0(1-1')K(1'), \quad t_1>t_{1'},
\ee
we can write
\be
\psi^0(\mathbf{r},t)=\sum_{\mathbf{p}}\sqrt{\frac{(d\mathbf{p})}{(2\pi)^3}}
e^{i\mathbf{p\cdot r}-iEt}(-i)K_{\mathbf{p}}\equiv\sum_{\mathbf{p}}
\psi_{\mathbf{p}}(\mathbf{r},t)(-i)K_{\mathbf{p}}.
\ee
Similarly,
\be
\psi^{0*}(1)=\int K^*(1')G^9(1'-1),\quad t_1<t_{1'}
\ee
implies
\be
\psi^{0*}(\mathbf{r},t)=\sum_{\mathbf{p}}\sqrt{\frac{(d\mathbf{p})}{(2\pi)^3}}
e^{-i\mathbf{p\cdot r}+iEt}(-i)K_{\mathbf{p}}^*\equiv\sum_{\mathbf{p}}
\psi_{\mathbf{p}}^*(\mathbf{r},t)(-i)K_{\mathbf{p}}^*.
\ee
The term $iW$ is the only term which describes the process being considered.
The vacuum amplitude includes
\be
\sum_{\mathbf{p}}\langle 0_+|1_{\mathbf{p}}\rangle^{K^*}\langle 1_{\mathbf{p}+}
|1_{\mathbf{p}'-}\rangle\langle1_{\mathbf{p'}+}\rangle^K\approx
\sum_{\mathbf{p}}(-iK_{\mathbf{p}}^*)\langle 1_{\mathbf{p}
+}|1_{\mathbf{p'}-}\rangle(-iK_{\mathbf{p'}}).
\ee
So, picking out the coefficients of $-iK_{\mathbf{p}'}$, $-iK_{\mathbf{p}}^*$,
we infer
\be
\langle 1_{\mathbf{p}}|1_{\mathbf{p'}}\rangle=-i\int\psi_{\mathbf{p}}(1)
\tens{T}(1,1')
\psi_{\mathbf{p}'}(1').
\ee
so if we knew $\tens{T}$ we could compute the scattering amplitude by 
taking matrix elements in this way.

\begin{figure}
\centering
\includegraphics{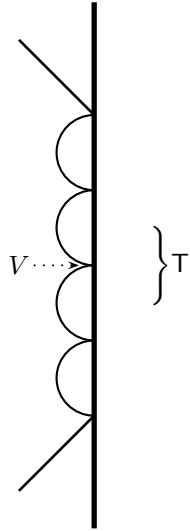}
\caption{\label{fig:s9} Multiple-scattering resolution of the scattering
operator $\tens{T}$.  
Here the diagram represents repetitions of elementary scattering
processes each described by a potential $V$.}
\end{figure}

Let us analyze the scattering process in more detail, by going into processes
which occur in a definite time.  That is, we break up the extended process,
represented by the black box in Fig.~\ref{fig:s8} by viewing it as a 
repetition of elementary processes represented by a potential $V$, 
as sketched in Fig.~\ref{fig:s9}.
To describe this, we extend the idea of the field. The vacuum amplitude is 
given in terms of
\be
W=-\int K^*[G^0K+G^0\tens{T}G^0K],
\ee where the quantity in square brackets is regarded as the field,
\be
\psi=\psi^0+\int G^0\tens{T}\psi^0,
\ee
where $\psi^0=\int G^0 K$ and $\tens{T}\psi^0$ 
may be thought of as an effective 
source.  In this way we get a superposition of effects.  Alternatively,
we may emphasize the last scattering act,
\be
\psi=\psi^0+\int G^0 V\psi,\label{psiofv}
\ee
where, as above, $\psi$ is due to an infinite number of elementary 
scattering acts.  This decomposition is self-consistent, because
we may write
\bea
\psi&=&\psi^0+\int G^0 V\psi\nn\\
&=&\psi^0+\int G^0V\psi^0+\int G^0VG^0V\psi^0+\int G^0VG^0VG^0V\psi+\dots.
\eea
So if the expansion makes sense
\be
\tens{T}=V+VG^0V+VG^0VG^0V+\dots.
\ee

The differential equation corresponding to Eq.~(\ref{psiofv}) is
\be
\left(i\frac\partial{\partial t}-T\right)\psi=K+V\psi,\ee
or
\be
\left(i\frac\partial{\partial t}-T-V\right)\psi=K.
\ee
So we see, indeed, that $V$ is the potential energy.  Thus, we indeed
have a multiple scattering process. 

This also generalizes the concept of the Green's function.  If we write
\be
\psi=\int G K,
\ee
where the Green's function satisfies
\be
\left(i\frac\partial{\partial t}-T-V\right)G(1,1')=\delta(1,1'),
\ee
and the action can be written as
\be
W=-\int K^*\psi-\int K\psi^*+\int\psi^*\left(i\frac\partial{\partial t}-T-V
\right)\psi,
\ee
which in value is equal to
\be
W=-i\int K^*GK.
\ee
The Green's functions include the possibility of bound states.

\begin{figure}
\centering
\includegraphics{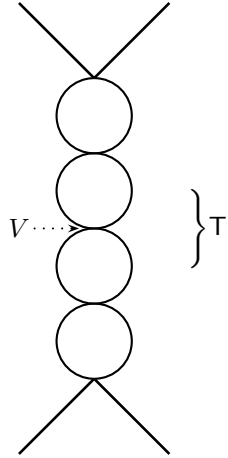}
\caption{\label{fig:s10} Scattering process described in terms of multiple
scattering.}
\end{figure}

More generally, think of 2-particle scattering, represented in 
Fig.~\ref{fig:s7}.  Single-particle exchange is represented by
\be
W_2=-\int K^*G^0K.  
\ee
But here, the particles do their thing, and the scattering is represented
by
\be
W_4=-\frac12\int (K^*G^0)_1(K^*G^0)_2 \tens{T}(12,1'2')(G^0K)_{1'}(G^0K)_{2'}.
\ee Again, the scattering amplitude $\langle 1_{\mathbf{p}_1}
1_{\mathbf{p}_1}|1_{\mathbf{p}_{1'}}1_{\mathbf{p}_{2'}}\rangle$ is
given simply in terms of $\tens{T}$. 
Moreover, we want to analyze the scattering in
terms of elementary processes, as sketched in Fig.~\ref{fig:s10}.
The terms in the vacuum amplitude of interest are contained in
$e^{i(W_2+W_4)}$ which when expanded contains the following terms with four
sources:
\bea
-\frac12W_2^2+iW_4&=&-\frac12\int K^*\psi^0K^*\psi^0-\frac{i}2\int 
(K^*G^0)(K^*G^0)\tens{T}(G^0K)(G^0K)\nn\\
&=&-\frac12\int K^*(1)K^*(1')\psi(1,2),
\eea
where
\be
\psi(1,2)=\psi^0(1)\psi^0(2)+i\int G_1^0G_2^0 \tens{T}\psi^0\psi^0.
\ee
This says that the source detects what comes directly, and what comes from
the scattering process.  On the other hand, we can look at this from the
last scattering point of view,
\be
\psi(1,2)=\psi^0(1)\psi^0(2)+i\int G_1^0G_2^0V\psi(1',2'),\label{2pfield}
\ee
where $V$ is the measure of the single-scattering process.  Again one
can iterate.  Look at the corresponding differential equation,
\be
\left(i\frac\partial{\partial t_1}-T_1\right)\left(i\frac\partial{\partial t_2}
-T_2\right)\psi(1,2)=K(1)K(2)+i(V\psi)(1,2),
\ee
where a matrix notation is used in the last term.  This can be written as
\be
\left[\left(i\frac\partial{\partial t_1}-T_1\right)
\left(i\frac\partial{\partial t_2}-T_2\right)-iV\right]\psi(1,2)=K(1)K(2).
\ee
Now make explicit the time localization of the potential:
\be
V(1,2;1',2')=\delta(t_1-t_2)\delta(t_{1'}-t_{2'})\delta(t_1-t_{1'})
V_{12}(\mathbf{r}_1-\mathbf{r}_2,\mathbf{r}_{1'}-\mathbf{r}_{2'}).\label{inst}
\ee
It might also be that the potential is localized in space:
\be
V_{12}(\mathbf{r}_1-\mathbf{r}_2,\mathbf{r}_{1'}-\mathbf{r}_{2'})
=\delta((\mathbf{r}_1-\mathbf{r}_2)-(\mathbf{r}_{1'}-\mathbf{r}_{2'}))
V(\mathbf{r}_1-\mathbf{r}_2),
\ee but we won't assume this in the following.  Then the Green's function
defined by
\be
\psi(1,2)=\int G(1,2;1',2')K(1')K(2),
\ee
satisfies
\be
\left[\left(i\frac\partial{\partial t_1}-T_1\right)
\left(i\frac\partial{\partial t_2}-T_2\right)-iV\right]G=\left(\delta(1-1')
\delta(2-2')\right)_{\rm sym},
\ee
where the subscript denotes symmetrization.
Now because of Eq.~(\ref{2pfield}), the integral equation for the Green's 
function is (symmetrization suppressed)
\be
G=G_1^0G_2^0+iG^0G^0VG. \label{2pinteq} 
\ee
The assumption of an instantaneous $V$ means that we can look at the equal-time
Green's function,
\be
G(\mathbf{r}_1,\mathbf{r}_2,t;\mathbf{r}_1',\mathbf{r}_2',t'),\quad
t=t_1=t_2, \quad t'=t_1'=t_2'.
\ee
Appearing in Eq.~(\ref{2pinteq}) is
\bea
&&iG^0(\mathbf{r}_1,t;\mathbf{r}'_1,t')G^)(\mathbf{r}_2,t;\mathbf{r}_2',t')
=-i\eta(t-t')\int\frac{(d\mathbf{p}_1)}{(2\pi)^3}\frac{(d\mathbf{p}_2)}
{(2\pi)^3}\nn\\
&&\quad\times \exp[i(\mathbf{p}_1\cdot(\mathbf{r}_1-\mathbf{r}_1')
+i(\mathbf{p}_2\cdot(\mathbf{r}_2-\mathbf{r}_2')-i(E_1+E_2)(t-t')],
\eea
a Green's function with energy $E_1+E_2$.  This obeys the equation
\be
\left(i\frac\partial{\partial t}-T_1-T_2\right)(iG^0G^0)=\delta(t-t')
\delta(\mathbf{r}_1-\mathbf{r}_{1}')\delta(\mathbf{r}_2-\mathbf{r}_{2}'),
\ee
the conventional Green's function equation.  Then
\be
\left(i\frac\partial{\partial t}-T_1-T_2\right)iG=\delta(t-t')
\delta(\mathbf{r}_1-\mathbf{r}_{1}')\delta(\mathbf{r}_2-\mathbf{r}_{2}')
+V_{12}(iG),\label{expected}
\ee
which is what we would expect.

Now we write the vacuum amplitude as
\bea
\langle 0_+|0_-\rangle^K&=&1-i\int K^*G^0 K-\frac12\int K^*K^* \psi(12)+\dots
\nn\\
&=&e^{-i\int K^*G^0K-\frac12\int K^*K^*[\psi(12)-\psi(1)\psi(2)]}+\dots,
\label{vacampscatt}
\eea
where in the last three-particle and higher interactions have been omitted.
Here we include in the second term only the effects of interaction, through
\be
\chi(1,2)=\psi(1,2)-\psi(1)\psi(2).
\ee

The integral equation (\ref{2pinteq}) written in terms of the last interaction
can also be written in term of the first interaction,
\be
G=G^0G^0+GiVG^0G^0.
\ee
Symbolically we can solve for $G$,
\be
G=\frac1{1-G^0G^0iV}G^0G^0=G^0G^0\frac1{1-iV G^0G^0},
\ee
which are formally identical.
This assumes the instantaneous interaction given by Eq.~(\ref{inst}).
Let us write
\be
G^0G^0=-iG^0_{1+2}, \quad iG=G_{1+2};
\ee
then our integral equation reads
\be
G_{1+2}=G^0_{1+2}+G^0_{1+2}V_{12}G_{1+2}.
\ee
which satisfies the differential equation (\ref{expected}), or
\be
\left(i\frac\partial{\partial t}-T_1-T_2-V_{12}\right)G_{1+2}=\delta(t-t')
\delta(\mathbf{r}_1-\mathbf{r}_{1'})\delta(\mathbf{r}_2-\mathbf{r}_{2'}).
\label{g1+2}\ee

We might be interested in non-instantaneous initial and final states, even
though we are assuming an instantaneous interaction.  Then we could write
\be
G=i G^0(t_>-t_<)G_{12}(t_<-t'_>)G^0(t'_>-t'_<),
\ee
where 
\be
G^0(t-t')\to -i\delta(\mathbf{r-r}') \quad\mbox{as}\quad \mbox t\to t'+0,\ee
hence the factor of $i$ is supplied by comparison with the instantaneous
limit, $t_1=t_2$, $t_1'=t_2'$. 

\section{Bound States}
Sources must be able to create composite structures.  We could have started
with composite particles, after all, so this is a aspect of self-consistency.
Write the integral equation for $G$ symmetrically,
\bea
G&=&G^0G^0+G^0G^0iV[G^0G^0+GiVG^0G^0]=G^0G^0+G^0G^0iVG^0G^0\nn\\
&&\quad\mbox{}+G^0G^0iVGiVG^0G^0.
\eea
We are still working with the instantaneous interaction approximation.  So
$G$ has the property of depending on two times.  Write again
\be
G=-iG_{1+2},\quad V\to V_{12}.
\ee
The description of the bound states is contained in $G_{1+2}$.  The part of the
vacuum amplitude involving $G_{1+2}$ is
\be
\langle 0_+|0_-\rangle=e^{-\frac{i}2\int K^*K^*G^0G^0V_{12}G^0_{1+2}V_{12}
G^0G^0 
KK}.\label{dbs}
\ee
This involves all interactions.

Now introduce the coordinates for the two-particle system, where we 
assume that the two particles have the same mass,
\be
\mathbf{R}=\frac12(\mathbf{r}_1+\mathbf{r}_2),\quad \mathbf{r}=\mathbf{r}_1
-\mathbf{r}_2.
\ee
The Green's function equation (\ref{g1+2}) becomes
\be
\left(i\frac\partial{\partial t}-\frac{\mathbf{P}^2}{2M}-\frac{\mathbf{p}^2}
{2\mu}-V(\mathbf{r})\right)G(\mathbf{r},\mathbf{R},t;
\mathbf{r}',\mathbf{R}',t')=\delta(t-t)\delta(\mathbf{r-r}')
\delta(\mathbf{R-R}'),\label{gfbs}
\ee
where we see the appearance of the total mass $M$, the reduced mass $\mu$,
the total momentum $\mathbf{P}$ and the relative momentum $\mathbf{p}$.
We are interested in the motion of the center of mass.  Let the internal
motion be described by an eigenfunction $\phi_k$ governed by a Schr\"odinger
equation,
\be
\left(E_k-\frac{p^2}{2\mu}-V\right)\phi_k(\mathbf{r})=0,
\ee
so multiplying Eq.~(\ref{gfbs}) by $\phi^*_n(\mathbf{r})$ and integrating over
$\mathbf{r}$ we obtain
\bea
&&\left(i\frac\partial{\partial t}-\frac{P^2}{2M}-E_n\right)\int(d\mathbf{r})
\phi^*(\mathbf{r})
G(\mathbf{r},\mathbf{R},t; \mathbf{r'},\mathbf{R}',t')\nn\\
&&\qquad=\delta(t-t')\delta(\mathbf{R-R}')\phi^*(\mathbf{r'}).
\eea
The Green's function can be expanded, therefore, in terms of a single-particle
Green's functions depending on the state $n$:
\be
G(\mathbf{r},\mathbf{R},t; \mathbf{r'},\mathbf{R}',t')=\sum_n\phi_n(\mathbf{r})
G_n(\mathbf{R-R}',t-t')\phi^*_n(\mathbf{r}').
\ee
Now from Eq.~(\ref{dbs}) we identify the effective source for an atomic state
\bea
K_n(\mathbf{R},t)&=&\frac1{\sqrt{2}}\int(d\mathbf{r})\phi^*(\mathbf{r})V_{12}
(\mathbf{r})G^0(\mathbf{R}+\frac{\mathbf{r}}2-\mathbf{r}_1,t-t_1)\nn\\
&&\quad\times G^0(\mathbf{R}-\frac{\mathbf{r}}2-\mathbf{r}_2,t-t_2)
 K(\mathbf{r}_1,t_1)K(\mathbf{r}_2,t_2)d1\,d2.\label{kn}
\eea
We must be very explicit that $E_n<0$, meaning that the sources are extended:
They must put out less energy than two free particles would have.  So $G^0$
does not refer to the propagation of a free particle, but rather it propagates
an excitation which does not get very far.  As a result, the Green's functions
become real in effect, by virtue of the integration, and
\be
K_n^*(\mathbf{R},t)=\frac1{\sqrt{2}}\int K^*K^* G^0G^0V_{12}\phi_n
\ee is actually the complex conjugate of Eq.~(\ref{kn}).  The consistency
of the  vacuum persistence amplitude  
\be
\langle 0_+|0_-\rangle=e^{-i\sum_n\int K_n^*(\mathbf{R},t)G_n(\mathbf{R-R}',
t-t')K_n(\mathbf{R}',t')}
\ee
demands that $K_n^*$ really be the complex conjugate of $K_n$,

Perhaps this remark is clarified by looking at the Fourier transform of
the free Green's function,
\be
\frac1{E-T+i\epsilon}.
\ee
When $E<0$ there is no singularity and the Green's function is real.

The formalism is flexible: it can deal with bound states whether analyzed or
not, which is especially useful in high energy physics.

Can we write down a formalism that expresses the dynamics and supplies the
field equations?  The affirmative answer is supplied by writing
\be
\langle 0_+|0_-\rangle=e^{iW},
\ee
with
\bea
W&=&\int\bigg\{\psi^*\left(i\frac\partial{\partial t}-T\right)\psi
-K^*\psi-\psi^*K-\frac12\psi^*\psi^*V\psi\psi\nn\\
&&\quad\mbox{}-\frac12\psi^*\psi^*V\chi-\frac12\chi^*V\psi\psi\nn\\
&&\quad\mbox{}-\frac{i}2\chi^*\left[\left(i\frac\partial{\partial t}-T\right)
\left(i\frac\partial{\partial t}-T\right)-iV\right]\chi\bigg\}.
\eea
Here we have introduce the 2-particle field $\chi$.  What does this imply
under field variations?  Varying with respect to $\psi^*$ gives
\be
\left(i\frac\partial{\partial t}-T\right)\psi=K+\psi^*V(\psi\psi+\chi),
\ee where in the last we see a realistic representation of a single-particle
source in the interaction with the other fields.
Varying with respect to $\chi^*$ yields
\be
\left[\left(\frac\partial{\partial t}-T\right)
\left(i\frac\partial{\partial t}-T\right)-iV\right]\chi=iV\psi\psi.
\ee
This says that $\chi$ is the part of the two-particle field that has interacted
at least once, thus
\be
\chi=GiV\psi\psi.
\ee
If we use these equations to evaluate $W$ we recover the previous result 
(\ref{vacampscatt}).

We have here two independent field $\psi$ and $\chi$, coupled by interaction.
This is the essence of the many-particle situation in high energy physics.
Essentially, we may regard the $-\frac12\chi^*V\psi\psi$ as a phenomenological
coupling between particles.

This nonrelativistic development of source theory was written down after the
relativistic formulation was developed, which is explicated in great detail
in Schwinger's three volume treatise \cite{Schwinger1989}.
\chapter{Concluding Remarks}
\label{concl}
We have traced Schwinger's development of action formulations from classical
systems of particles and fields, to the description of quantum dynamics 
through the Quantum Action Principle.  In the latter, we have described
quantum mechanical systems, especially the driven harmonic oscillator.
This is ahistorical, since Schwinger first developed his quantum dynamical
principle in the context of quantum electrodynamics in the early 1950s, and 
only nearly a decade later applied it to quantum mechanics, which is field
theory in one dimension---time.  At roughly the same time he was thinking
about quantum statistical systems \cite{Martin1959}, and it was natural to
turn to a description of nonequilibrium systems, which was the motivation
of the time-cycle method, although Schwinger put it in a general, although
simplified, context.
The time cycle method was immediately applied to quantum field theory by
his students, K. T. Mahanthappa and P. M. Bakshi
 \cite{mahanthappa1962,bakshi1963}.  % But rather
%than here tracing the profound and growing influence of this great paper,
%as well as the deep underpinning still provided by  Schwinger's action
%principle, % (over 1300 citations currently in Web of Science), 
Then we give  a sketch of the application of these methods
to quantum field theory, based on Schwinger's 1956 Stanford lectures,
and to what Schwinger perceived as the successor to
field theory, Source Theory. The latter appeared shortly after he received
his Nobel prize in 1963.  In fact the present document, which includes
components of Julian Schwinger's thinking over nearly four decades, shows
that these developments proceeding organically, and that the action principle
and Green's functions\footnote{One of Schwinger's last publications
\cite{Schwinger1993}
described the centrality of Green's functions to his life work.} played
central roles throughout his remarkable career in physics.  In fact, it has
been argued \cite{mehra} that the first ``source theory'' paper was in fact his
most cited one, written in 1951 \cite{Schwinger1951a}.

% But we have now reached a appropriate point to
%pause.  In Part II of this paper we will provide that elaboration, and trace
%some of the vast influence that Schwinger's development of these powerful
%techniques have had in all branches of theoretical physics.
  
\begin{acknowledgement}
I thank the Laboratoire Kastler Brossel, ENS, UPMC, CNRS, for its hospitality
during the completion of the first part of
 manuscript.  I especially thank Astrid 
Lambrecht and Serge Reynaud. The work was supported in part
with funding from the Simons Foundation, CNRS, and the Julian
Schwinger Foundation.  I thank my many students
at the University of Oklahoma, where much of the material reported here
was used as the basis of lectures in electrodynamics, quantum mechanics, 
and quantum field theory.  I especially thank Herb Fried for permission
to use his transcription of Schwinger's 1956 lectures as the basis for Chapter
6 here, and Walter Becker for his \LaTeX\ conversion of those notes.
\end{acknowledgement}

\backmatter%%%%%%%%%%%%%%%%%%%%%%%%%%%%%%%%%%%%%%%%%%%%%%%%%%%%%%%
%\include{glossary}
%\include{solutions}
%\printindex

%%%%%%%%%%%%%%%%%%%%%%%%%%%%%%%%%%%%%%%%%%%%%%%%%%%%%%%%%%%%%%%%%%%%%%

\end{document}